\documentclass{aa}  

\usepackage{graphicx}
\usepackage{txfonts}
\usepackage{caption}
\usepackage{hyperref}
\usepackage{url}
\usepackage{natbib}
\bibpunct{(}{)}{;}{a}{}{,} 

\begin{document} 
\title{Radiative transfer with POLARIS:\\I. Analysis of magnetic fields through synthetic dust continuum polarization measurements}
\author{S. Reissl\inst{\ref{inst1},\ref{inst2}}, S. Wolf\inst{\ref{inst1}} \and  R. Brauer\inst{\ref{inst1}}  }

\institute{\centering Institut für Theoretische Physik und Astrophysik, Christian-Albrechts-Universit\"{a}t zu Kiel, Leibnizstraße 15, 24098 Kiel, Germany \\
\href{mailto:rbrauer@astrophysik.uni-kiel.de}{rbrauer@astrophysik.uni-kiel.de,} 
\href{mailto:wolf@astrophysik.uni-kiel.de}{wolf@astrophysik.uni-kiel.de}\label{inst1}
\and
\centering Universit\"{a}t Heidelberg, Zentrum f\"{u}r Astronomie, Institut f\"{u}r Theoretische Astrophysik,
Albert-Ueberle-Stra{\ss}e 2, 69120 Heidelberg, Germany \\
\href{mailto:reissl@uni-heidelberg.de}{reissl@uni-heidelberg.de}\label{inst2}
}
                                                
\abstract
   {}
{We present POLARIS (\textbf{POLA}rized \textbf{R}ad\textbf{I}ation \textbf{S}imulator), a newly developed three-dimensional Monte-Carlo radiative transfer code. POLARIS was designed to calculate dust temperature, polarization maps, and spectral energy distributions. It is optimized to handle data that results from sophisticated magneto-hydrodynamic simulations. The main purpose of the code is to prepare and analyze multi-wavelength continuum polarization measurements in the context of magnetic field studies in the interstellar medium. An exemplary application is the investigation of the role of magnetic fields in  star formation processes.}
   {We combine currently discussed state-of-the-art grain alignment theories with existing dust heating and polarization algorithms. We test the POLARIS code on multiple scales in complex astrophysical systems that are associated with different stages of star formation. POLARIS uses the full spectrum of dust polarization mechanisms to trace the underlying magnetic field morphology.}
{Resulting temperature distributions are consistent with the density and position of radiation sources resulting from magneto-hydrodynamic (MHD) - collapse simulations. The calculated layers of aligned dust grains in the considered cirumstellar disk models are in excellent agreement with theoretical predictions.
Finally, we compute unique patterns in synthetic multi-wavelength polarization maps that are dependent on applied dust-model and grain-alignment theory in analytical cloud models.}
 {}
   \maketitle
%
\section{Introduction}
It is well established that giant molecular clouds within the interstellar medium (ISM) are the birthplace of stars and later planetary systems. However, understanding star formation remains one of the outstanding remaining challenges of modern astrophysics. In spite of significant progress in recent years \citep[][]{2010HiA....15..438C}, fundamental questions about the basic physics of star formation remain unanswered.\\
The importance of magnetic fields in regulating the star formation process is still a topic of ongoing research. Self-gravitating clouds have long been believed to reduce support against collapse by magnetic support and triggering star formation \citep[][]{1999osps.conf..305M} with time scales of collapse of  up to 10 Myr, assuming ambipolar diffusion  \citep[][]{2008A&A...487..247F,2009Sci...324.1408G,2010ApJ...725..466C} indicating a significant influence of magnetic fields on the star formation process. Observations, however, suggest that molecular clouds are shorter lived and the cloud collapse is controlled by turbulence with negligible magnetic contribution \citep[][]{2000ApJ...530..277E, 2004ASPC..322..299K}.\\
Observations of magnetic fields in molecular clouds are essential for understanding their role in the evolution from dense clouds to the formation of stars. One way to improve our knowledge about the involvement of magnetic fields in the star formation process is to trace its morphology with the help of dust grains. In addition to magnetic fields, dust grains also play a central role in the ISM -- from heating and cooling processes, astro-chemistry, to the polarization effects of light \citep[][]{2010MNRAS.404..265D}. Non-spherical rotating dust particles have the tendency to align with their shorter axis along the local magnetic field direction \citep[][]{1974ApJ...187..461M,2012JQSRT.113.2334V}. Therefore, mapping of light that is polarized by elongated and aligned dust grains in the near-infrared to sub-mm \citep[][]{2004MNRAS.352.1347L, 2005AA...430..979G,2011AA...535A..44F} offers one way to obtain information about the magnetic field morphology in regions associated with star formation.  Additionally, at optical to near infrared wavelengths, light from embedded stars scattered as dust grains enables us to investigate the spatial structure of star-forming regions and can also constrain the dust properties \citep[][]{1997MNRAS.286..895L}.\\
However, to make comparisons between star formation theories with observational data, one requires an expedient understanding of the intrinsic radiative transfer (RT), the dust properties, and the physics of grain alignment.
Unfortunately, the 3D dust RT problem is nonlinear, which makes it one of the hardest challenges in computational astrophysics. In idealized scenarios, the problem can be solved by using analytical radiative transfer techniques \citep[e.g.][]{1997MNRAS.291..121I,2002A&A...395..373V,2006ApJ...645..920S}. Environments with ongoing star formation, however, are regions where gas and dust density and magnetic fields form complex structures and their physical quantities cover several orders of magnitude. The Monte-Carlo (MC) method is a powerful approach to perform radiative transfer simulations independent of complexity. Here, nature is mimicked by sending photon packages along probabilistic pathways to obtain numerical solutions to RT problems. A broad variety of codes has been developed to cover problems of astrophysical importance such as the heating of dust, photoionization, and the calculation of polarization of scattered radiation {\citep[][]{1999pcim.conf..220J, 1999A&A...349..839W,2002ApJ...574..205W,2001sf2a.conf...65N,2003MNRAS.340.1136E, 2003CoPhC.150...99W,2003ApJ...583L..35S, 2009A&A...497..155M, 2011BASI...39..101W,2011ApJS..196...22B,2012ascl.soft02015D,2013prpl.conf2S001R,2014ascl.soft04006H}}.  \\
So far, vital questions concerning dust composition, alignment efficiency, and size distribution are unknown. With forthcoming observational  equipment, e.g. SOPHIA/HAWC+ \cite{2013AAS...22134514D} or ALMA \cite{2004...34.555}, the need for accurate and expedient modeling of observational polarimetry data is increasing. Hence, we developed a new RT code from scratch and fully incorporated radiative transfer, dust heating, and polarization algorithms, as well as state-of-the-art dust grain alignment theories. Here, we implemented the classical imperfect Davis-Greenstein (IDG) alignment owing to paramagnetic relaxation \citep[][]{1983ApJ...272..551A,1967ApJ...147..943J}, the radiative torque alignment (RAT) because of radiation-dust interaction \citep[][]{2007AAS...210.7904H}, and considered the effects of mechanical alignment (GOLD) \citep[][]{1952MNRAS.112..215G,1994MNRAS.268..713L,1995ApJ...451..660L} as a result of gas streams. \\
This paper is structured as follows: First, we present the POLARIS code and its basic RT and dust heating algorithm in Sect. \ref{code}. Then, we briefly discuss the applied dust grain model and the implemented polarization and grain alignment features in Sect \ref{sect:dust}. The optimization methods that were considered in the POLARIS code are listed in Sect. \ref{optim}.
We describe the challenges of the scattering process on partially aligned dust grains in Sect \ref{sect:scatter}. To demonstrate the functionality and predictability of POLARIS, we calculate synthetic intensity and polarization maps in Sect. \ref{sect:application}. Here we considered both analytical models and complex MHD-simulations of different stages of the star formation process. Finally,  we summarize our results and give a short outlook to future projects in Sect. \ref{sect:sum}.

\section{The code}
\label{code}
POLARIS is a three-dimensional  Monte-Carlo (MC) continuum radiative transfer (RT) code dedicated to post-process MHD simulations to determine the observability of the physical quantities. The code can make use of the full parameter set (density, temperature, velocity, and magnetic field distributions) delivered by MHD simulations to perform dust heating and complex polarized radiative transfer (RT) calculations.  All RT simulations are performed on an adaptive octree grid, which is optimized to maintain the data structure of certain MHD codes, e.g. FLASH \cite{2010ascl.soft10082F}. The grid allows a time efficient propagation of photon packages and provides a dynamical spatial resolution to resolve regions of high optical depth and simultaneously reduces the amount of required memory. The code is platform independent and completely written in $\rm{C++}$ in a strict object orientated architecture making it easy to develop classes for future purposes such as photoionization, line transfer, or to implement further grid geometries.\\
The code works in three independent main operations modes:
\begin{enumerate}
        \item MC simulations of self-consistent heating of the dust grains by an arbitrary number of radiation sources.
        \item MC calculation of the mean energy density and anisotropy by the radiation field for radiative torque (RAT) alignment (see Sect. \ref{sect:RAT}).
        \item Calculation of spectral energy distributions (SEDs) and synthetic multi-wavelength intensity and polarization maps for statistical analysis.
\end{enumerate} 
To access its full functionality, the code POLARIS parses script files. Predefined commands in a pseudo XML-style that is implemented in the code allows the user to create scripts with sequences of operation modes. 

\subsection{Radiative transfer}
\label{sq:rad}
Monochromatic photon packages with a fixed energy $\epsilon_{\rm{0}}$ from distinct sources are emitted into the model space. To cover the various radiation-dust interactions in the ISM, we provide classes of point (stars), diffuse (star field), thermal dust re-emission and external (interstellar radiation field and background objects) sources of radiation. Each source has its characteristic spatial emission, energy spectrum, and degree of linear and circular polarization. A higher amount of photon packages leads to an increase in calculation time but also results  in a better  S/N inherent in the MC process.\\
In the dust heating mode we use a combined algorithm of continuous absorption \citep[][]{1999A&A...344..282L} and immediate temperature correction \citep[][]{2001ApJ...554..615B}. First, the mass specific emissivity 
\begin{equation}
j(a,T_{\rm{d}})=\int{C_{\rm{abs,\lambda}}(a)B_{\rm{\lambda}}(T_{\rm{d}})d\lambda}
\label{eq:LucyJ}
\end{equation}
of the entire ensemble of dust grain sizes is pre-calculated and interpolated as a function of dust temperature for a user-defined temperature range using pre-tabulated dust grain properties (see Sect. \ref{sect:application}). Here, $B_{\rm{\lambda}}(T)$ is the Planck function, $T_{\rm{d}}$ is the temperature of the dust, and $C_{\rm{abs,\lambda}}$ the cross-section of absorption. Afterwards, the photon package propagates {a distance $l_{i}$} in the grid between two points of radiation-dust interactions. At each cell wall, the absorption rate per time unit 
\begin{equation}
\dot{E} = \frac{\epsilon_{\rm{0}}}{\Delta t} \sum_{\rm{i}}{C_{\rm{abs,\lambda_i}}(a)\times l_{i}} + \Delta \dot{E}
\label{eq:LucyE}
\end{equation}
is updated as presented in \cite{1999A&A...344..282L}, assuming constant temperatures and densities in each cell. {We extended the presented algorithm by an additional therm $\Delta \dot{E}$ to incorporate  temperatures provided by MHD simulations into the RT simulations (see Sect. \ref{setupMHDHeat} for detail}).   {Since we assume local thermodynamic equilibrium (amount of absorbed energy equals amount of re-emitted energy), the updated dust temperature is provided by solving}
\begin{equation}
        j(a,T_{\rm{d}}) = \frac{\dot{E}}{4 \pi V_{\rm{cell}}} 
.\end{equation}
The optical depth $\tau$ is integrated over the passed distance $l$ up to the next interaction ($\tau>\tau_{\rm{max}}$). The interaction distance is determined by $\tau_{\rm{max}} = -\log_{10}(1-z)$ \citep[see][]{2003CoPhC.150...99W}. In this paper, $z$ donates a randomly generated number with $z \in [0;1[$. At each point of interaction the scattering-absorption probability is determined by the albedo $C_{\rm{sca}}/C_{\rm{ext}}$. In the case of scattering ($C_{\rm{sca}}/C_{\rm{ext}}$<z), a new direction $(\vartheta', \varphi')$ is sampled from the phase function of the dust grain. POLARIS uses optionally isotropic scattering, full Mie-scattering or, alternatively, the inverted Henyey-Greenstein phase function
\begin{equation}
        \cos(\vartheta') = \begin{cases} \frac{1}{2g} \left[1+g^2-\left(\frac{1-g^2}{1-g+2g\times z}\right)\right]   &\mbox{if } g \neq 0 \\  1-2\times z & \mbox{if } g=0  \end{cases} 
        \label{eq:HG}
\end{equation}
with
\begin{equation}
        \varphi' = 2\pi \times z
,\end{equation}
 as presented in \cite{1941ApJ....93...70H}. For the latter, the probability distribution of scattering is described by one parameter, the anisotropy $g=\left\langle cos(\vartheta')\right\rangle \in [-1, 1]$, $g=0$ means isotropic scattering, where $g=-1$ backward scattering, and $g=1$ forward scattering, respectively. The anisotropy parameter $g$ and the full scattering matrix (see \ref{sect:scatter}) are pre-calculated with DDSCAT (\cite{2013arXiv1305.6497D}, see Sect. \ref{sect:RAT} and Sect. \ref{sect:application}) and is a function of grain size, grain orientation, and wavelength. Although alternative phase functions are currently discussed \citep[see][]{2008ApJS..177..546S}, the HG phase function still provides a good approximation for the scattering probability. During all dust-heating calculations, the dust grain orientation is assumed to be random.\\
The dust temperature $T_{\rm{d}}$ is the essential parameter for thermal radiation. Hence, the case of an absorption and immediate thermal re-emission event is treated by the instantaneous temperature correction technique of \cite{2001ApJ...554..615B}. A photon package absorbed by dust will be immediately  re-emitted, assuming a blackbody spectrum. The new wavelength $\lambda_{i}$ of the thermal dust photon emitted in the $i - th$ cell is determined by
\begin{equation}
        z = \frac{\int_{\lambda_{\rm{min}}}^{\lambda_{i}}{C_{\rm{abs,\lambda}} \frac{dB_{\rm{\lambda}}(T_{\rm{d,i}})}{dT}  d\lambda }}{\int_{\lambda_{\rm{min}}}^{\lambda_{\rm{max}}}{C_{\rm{abs,\lambda}} \frac{dB_{\rm{\lambda}}(T_{\rm{d,i}})}{dT}  d\lambda}}
        \label{eq:BW}
\end{equation}
to ensure thermal equilibrium. This process is repeated until the photon package reaches the boundary of the grid.Once the correct dust temperature is achieved by the dust heating algorithm, intensity and polarization maps can be calculated. \\
The multi-wavelength MC RT polarization simulation, including scattering, are calculated similarly to  dust heating and can be performed in a monochromatic or multi-wavelength mode. In the case of polarization calculations, the dust temperature now remains  constant and different dust-grain alignment mechanisms can be applied. 

\subsection{Dust modeling and alignment}
\label{sect:dust}
\begin{figure}[]   
 \centering
        \includegraphics[width=0.45 \textwidth]{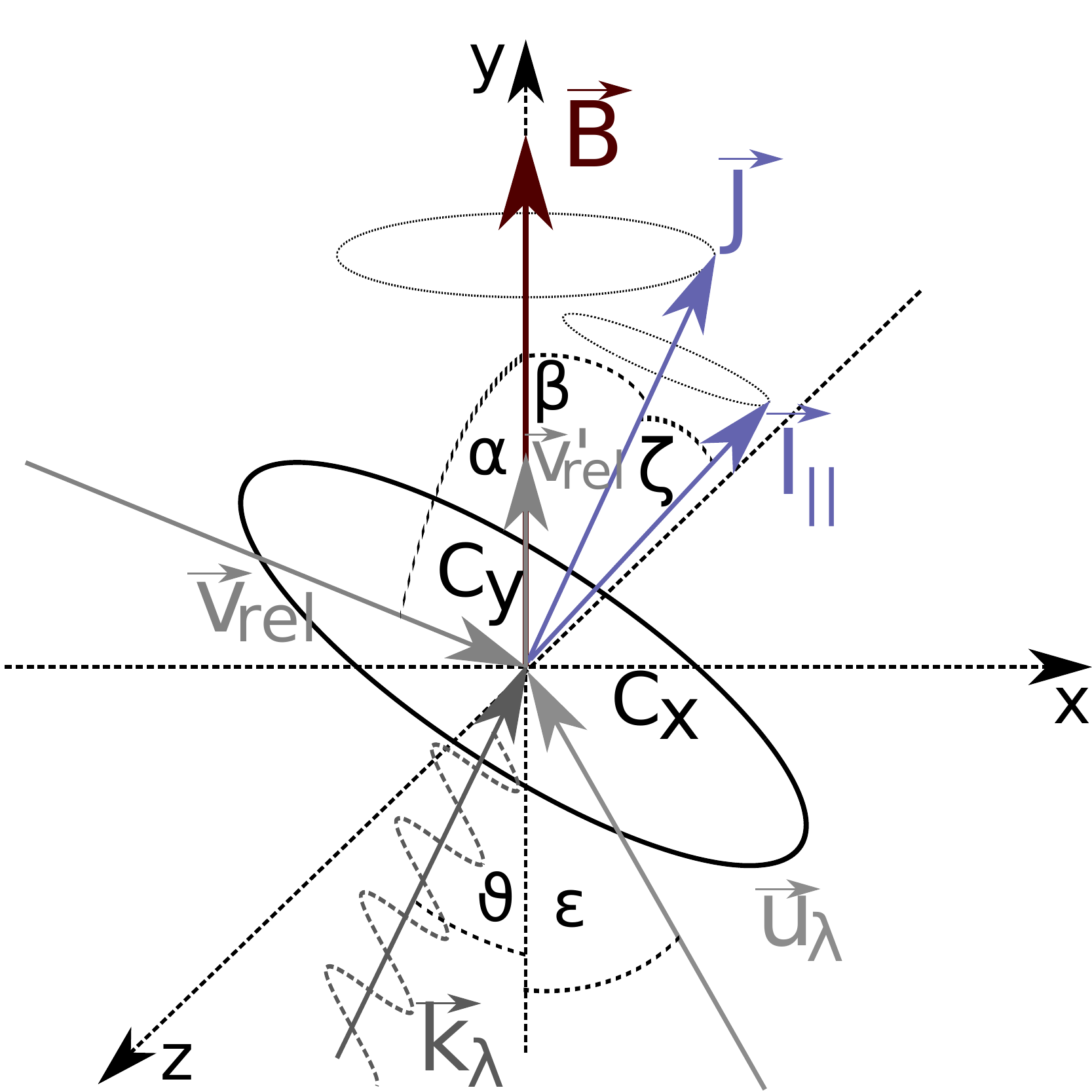}
        \caption{\small Geometrical configuration of a dust grain partially aligned with its angular momentum $\vec{J}$ to the magnetic field direction $\vec{B}$. The precession of $\vec{J}$ around $\vec{B}$ defines the cone angle $\beta,$ and the precession of maximum moment of inertia $\vec{I}$ around $\vec{J}$ defines the angle of internal alignment $\zeta$. Here, the angle $\vartheta$ is defined as being between the direction of incident light $\vec{k}$ and $\vec{B}$. The angles $\alpha$ and $\epsilon$ are between the direction of the supersonic velocity stream $\vec{v_{\rm{g}}}$ and $\vec{B}$ and between the anisotropy $\vec{u}_{\rm{\lambda}}$ in the radiation field and $\vec{B}$, respectively. The vector $\vec{v'_{\rm{rel}}}$ is the projection of $\vec{v_{\rm{g}}}$ on the direction of the magnetic field direction $\vec{B}$. }
        \label{Fig:Ang}
\end{figure}

Interstellar dust grains consist of silicate and carbon compounds and make up only about $1\%$ of total mass of the ISM. However, the extinction of dust grains can lead to significant degrees of linear and circular polarization from the optical to millimeter regime because of scattering, dichroic extinction, and thermal re-emission. \\
POLARIS considers the dust as a mixture of different dust components. Each component consist of its own material, size distribution, sublimation temperature, and alignment behavior. When the dust temperature exceeds the sublimation temperature of a certain material, this component will be removed from the cell. {For the mixing of different dust materials, the distinct cross-sections can simply be multiplied with their ratio of abundance and added up \citep[see e.g.][]{2010MNRAS.404..265D}.}\\
Non-spherical spinning dust grains align with their shorter axis, which is parallel to the local magnetic field direction. Thus, they preferentially block light perpendicular to the magnetic field direction, which results in a polarization  parallel to the magnetic field while re-emitted light is polarized perpendicularly. The radiation emitted by dust will mostly be found  in the mid-infrared, for a given dust temperature. Hence, the orientation of polarized light is related to the direction of the projected underlying magnetic field morphology, leading to a characteristic polarization pattern.\\
The implemented RT equations for extinction and polarization  \citep[][]{2002ApJ...574..205W} require the sum and difference of cross-sections ($C=\pi a^2 \times Q$) parallel $C_{\rm{ext,||}}$ and perpendicular $C_{\rm{ext,\bot}}$ to the direction of the magnetic field \citep[see][for detail]{2014A&A...566A..65R}. Here, $a$ is the effective radius of the spherical grain with equivalent volume to that of a spheroid grain and $Q$ is the efficiency of interaction.\\
We use the Stokes vector $ S = (I,Q,U,V)^T$ to describe the change in intensity and degree of polarization where $I$ is intensity, $Q$ and $U$ quantify the linear polarization, and $V$ the circular polarization. {Writing the radiative transfer equation in the Stokes vector formalism allows to calculate the effects of dichroic extinction, thermal re-emission (\cite{1974ApJ...187..461M}) with:}

\begin{equation}
\frac{\rm{d}}{n_{\rm{d}}ds}\begin{pmatrix} I\\ Q \\ U \\ V \end{pmatrix}=-\begin{pmatrix} \overline{C}_{\rm{\rm{ext}}} &  \overline{C}_{\rm{pol}} & 0 & 0 \\  \overline{C}_{\rm{pol}} & \overline{C}_{\rm{\rm{ext}}} & 0 & 0\\ 0 & 0 & \overline{C}_{\rm{\rm{ext}}} & - \overline{C}_{\rm{circ}} \\ 0 & 0 & \overline{C}_{\rm{circ}} &  \overline{C}_{\rm{pol}} \end{pmatrix}\begin{pmatrix} I\\ Q \\ U \\ V \end{pmatrix} + B_{\rm{\lambda}}(T_{\rm{d}})\begin{pmatrix} \overline{C}_{\rm{abs}}\\ \Delta \overline{C}_{\rm{abs}} \\ \Delta \overline{C}_{\rm{abs}} \\ 0 \end{pmatrix},
\label{eq:radtrans}
\end{equation}
where $T_{\rm{d}}$ and $n_{\rm{d}}$ are the temperature and number density, respectively, of the dust. In this paper $\overline{X}$ denotes the averaging over grain size distribution and $\left\langle X \right\rangle$ over orientation. So far, POLARIS is optimized for oblate dust grains but adjustments for arbitrary shapes can easily be applied.
In Eq. \ref{eq:radtrans}, the required matrix of cross-sections  depends on grain size, orientation, and wavelength. The matrix elements are determined by the values along the minor axis ($C_{\rm{ext,||}}$) and major axis ($C_{\rm{ext,\bot}}$), and are calculated for each dust grain size in each RT simulation with

\begin{equation}
        C_{\rm{ext,x}}=  \left\langle C_{\rm{ext}} \right\rangle+\frac{1}{3} R\times \left( C_{\rm{ext,||}}-C_{\rm{ext,\bot}} \right) 
\end{equation}
\begin{equation}
        C_{\rm{ext,y}}=  \left\langle C_{\rm{ext}} \right\rangle+\frac{1}{3}R \times \left( C_{\rm{ext,||}}-C_{\rm{ext,\bot}} \right)\left( 1- 3 \sin^2(\vartheta) \right)
        \label{eq:cross}
.\end{equation}
The angle $\vartheta$ is between incident light and magnetic field direction (see Fig. \ref{Fig:Ang}). For oblate dust grains
\begin{equation}
        C_{\rm{ext}}=0.5(C_{\rm{ext,x}}+C_{\rm{ext,y}}),
        \label{eq:Cext}
\end{equation}
\begin{equation}
        C_{\rm{pol}}=0.5(C_{\rm{ext,x}}-C_{\rm{ext,y}}),
        \label{eq:Cpol}
\end{equation}
and
\begin{equation}
        \left\langle C_{\rm{ext}} \right\rangle = \left(2C_{\rm{ext,||}}+C_{\rm{ext,\bot}}\right)/3.
\end{equation}
The Rayleigh reduction factor $R$ \citep[][]{1968nim..book..221G} is introduced to take care of imperfect grain alignment and is defined as
\begin{equation}
        R=\left\langle G\left(\cos^2(\beta)\right) G\left(\cos^2(\zeta)\right) \right\rangle
.\end{equation}
Here, $\beta$ is the alignment cone angle between the angular momentum $\vec{J}$ and magnetic field $\vec{B}$ and $\zeta$ is the internal alignment angle between the axis of the largest moment of inertia $\vec{I_{\rm{||}}}$ and $\vec{J}$, while $G(x)=1.5x - 0.5$. The Rayleigh reduction factor is defined as $R \in [-0.5;1],$ where positive values correspond to an alignment with the longer grain axis that is perpendicular to the magnetic field direction and vice versa for negative values. We note that the Rayleigh reduction factor can also be a function of wavelength and effective dust grain radius. Consequently, grain alignment and, subsequently, polarization is completely determined by first order moments weighted over a distribution function defined by $\left\langle \cos^2(x) \right\rangle = \int{f(x)\sin(x)dx}$. Unfortunately, alignment and internal alignment do not work independently. For exact solutions, a simultaneous integration of both distribution functions $f(\beta)$ and $f(\zeta)$ over cone angle $\beta$ and angle $\zeta$, respectively, is required. Depending on the considered grain alignment mechanism, the distribution functions, in turn, are also functions of density, temperature, magnetic field strength, velocity, and direction of the incident light (see Sect. \ref{sect:IDG}, \ref{sect:GOLD} and \ref{sect:RAT}). Because of the enormous parameter space, the reduction in polarization cannot be pre-calculated without losing precision. The exact calculations that are required, however,  for each photon package-dust interaction results in a further burden to 3D MC-RT simulations. Therefore, in POLARIS we approximate 
\begin{equation}
        \left\langle G\left(X\right) G\left(Y\right) \right\rangle \approx \left\langle G\left(X\right)  \right\rangle \times \left\langle G\left(Y\right) \right\rangle \left(1+f_{\rm{c}}\right)
\end{equation}
with a correlation factor $f_{\rm{c}}$ where $f_{\rm{c}} = 0$ stands for no correlation \citep[][]{2013ApJ...779..152H,2014ApJ...790....6H}. The correlation factor is of minor influence to the net polarization, meaning that this approach  is well justified.\\
Once, the Rayleigh reduction factor is known, the exact cross-sections for the RT simulation can be calculated with Eqs. \ref{eq:Cext} and \ref{eq:Cpol} for a given size distribution $n(a)$ between a minimal $a_{\rm{min}}$ and maximal $a_{\rm{max}}$ grain size with
        \begin{equation}
        \overline{C}_{\rm{ext}}=  \int_{a_{\rm{min}}}^{a_{\rm{max}}} {C}_{\rm{ext}}(a) n(a) da 
        \label{eq:avg}
\end{equation}
\begin{equation}
        \overline{C}_{\rm{pol}}=  \int_{a_{\rm{min}}}^{a_{\rm{max}}} {C}_{\rm{pol}}(a) n(a) da 
.\end{equation}
The same procedure is used for the cross-sections of scattering $\overline{C}_{\rm{sca}}$, absorption $\overline{C}_{\rm{abs}}$, and circular polarization $\overline{C}_{\rm{circ}}$ to obtain the matrix required in Eq. \ref{eq:radtrans}.

\subsubsection{Imperfect internal (II) alignment}
\label{sect:IDG}
The cause of internal alignment \citep[see][]{1979ApJ...231..404P} is the Barnett effect \citep[][]{B1915}. Any change in angular momentum (e.g. gas-dust collisions) forces free electrons in the paramagnetic dust grain material to align their spin parallel to the axis of grain rotation. This induces a net magnetic moment and, subsequently, grain alignment. However, this net magnetic moment is not constant because of thermal fluctuations inside the dust grain. Subsequently, thermal fluctuations are transferred into a wobbling motion, leading to a precession of the largest moment of inertia $\vec{I_{\rm{||}}}$ arround $\vec{J}$. The distribution function for imperfect internal alignment is 
\begin{equation}
        f(\zeta)=N\times \exp \left(-\frac{J_{eff}-^2}{2I_{\rm{||}}k_{\rm{B}} T_{\rm{d}}} \left[1+(h-1)\sin^2(\zeta)\right]\right)
        \label{eq.distr}
\end{equation} 
\citep[][]{1996ASPC...97..425L}, where $h=I_{\rm{||}}/I_{\rm{\bot}}$ is the ratio of the momenta of inertia, $J_{eff}$ is the rms value of the angular momentum, and $N$ is a normalizing constant (see App. \ref{apA} for details).

\subsubsection{Imperfect Davis-Greenstein (IDG) alignment}
\label{sect:IDG}
For the IDG, the alignment mechanism is paramagnetic relaxation \citep[][]{1951ApJ...114..206D,1967ApJ...147..943J,1979ApJ...231..404P,1979ApJ...231..417S}. In rotating non-spherical paramagnetic dust grains the electrons' spin axis follows the external magnetic field direction. However, for a sufficiently high angular velocity, the electrons cannot fully respond. Spin-lattice interaction transfers the component of angular velocity perpendicular to the magnetic field into internal heat. In the ISM, paramagnetic relaxation and subsequently perfect alignment of the dust grains is opposed by random gas bombardment.\\
The ratio of the characteristic timescales of paramagnetic relaxation and gas-dust collisions provides a threshold for the effective grain radius $a$ up to which dust grains align
\begin{equation}
        \delta_{\textit{0}} = 2.07\times 10^{20}\frac{B^2}{n_{\textit{g}} T_{\textit{d}} \sqrt{T_{\textit{g}}}}[m],
        \label{eq:delta}
\end{equation}
where $T_{\rm{g}}$ is the gas temperature and $n_{\rm{g}}$ is the number density of the gas.
Dust grains with radii $a$ above $\delta_{\textit{0}}$  no longer significantly contribute to polarization. For the IDG, an analytical function for the distribution of the opening angle $\beta$ \citep[see][]{1979ApJ...231..417S} allows us to calculate 
\begin{equation}
        \left\langle \cos^2(\beta) \right\rangle=  \left(1 - \sqrt{\frac{\xi^2(a)}{1-\xi^2(a)}}      \right) \arcsin{\frac{\sqrt{1 - \xi^2(a)}}{1 - \xi^2(a)}}
        \label{eq:IDGfinal}
\end{equation}
\citep[][]{1983ApJ...272..551A},  with 

\begin{equation}
        \xi^2(a) = \frac{a+\delta_ {\textit{0}}\times \frac{T_{\textit{d}}}{T_{\textit{g}}}}{a+\delta_{\textit{0}}}
        \label{eq:IDGzetha}
\end{equation}
and subsequently the Rayleigh reduction factor. To calculate the internal alignment, we use the approximation from \cite{1997ApJ...484..230L} for the angular momentum
\begin{equation}
 J_{eff}^2\approx \sqrt{\left(1+0.5 s^{-2}\right) \left(1+T_{\rm{d}} / T_{\rm{g}} \right)} \times J_{\rm{th}}^2
\end{equation}
where $J^2_{\rm{th}} \approx 2 I_{\rm{||}}k_{\rm{B}}T_{\rm{g}}$ is the thermal angular momentum obtained from gas-dust collisions and $s$ is the aspect ratio of the grains. This approach delivers reliable values in the limit of thermal equilibrium ($T_{\rm{d}}/T_{\rm{g}}=1$). However it slightly underestimates the internal alignment for $s<0.5$ and $T_{\rm{d}}/T_{\rm{g}}<0.2$.

\subsubsection{Gold (magneto mechanical) alignment}
\label{sect:GOLD}
The alignment of dust grains by purely mechanical alignment was proposed by \cite{1952MNRAS.112..215G}. This mechanism requires a predominant direction in the velocity field of the gas. The dust grains align with their longer axis parallel to the gas flow. In magneto-mechanical alignment \citep[][]{1994MNRAS.268..713L,1995ApJ...451..660L,1997ApJ...483..296L}, the alignment behavior depends on both gas flow and magnetic field. Here, the reference frame is determined by the Barnett effect and, subsequently, the magnetic field direction, but not the gas stream. In this paper we also refer to magneto-mechanical alignment  as GOLD alignment. The gas flow must be supersonic, otherwise the dust grain orientation remains randomized \citep[see][]{1952MNRAS.112..215G,1995ApJ...451..660L}.\\
A distribution function for the cone angle $\beta$ was derived by \cite{1976Ap&SS..43..291D}. For spheroidal dust grains, the first order moment $\left\langle \cos^2(\beta) \right\rangle$ is completely defined by the anisotropy in the gas stream
\begin{equation}
        s = -\frac{1}{2}\frac{ \left\langle \vec{v_{\rm{rel}}} \right\rangle -3 \left\langle \vec{v'_{\rm{rel}}} \right\rangle }{ \left\langle \vec{v_{\rm{rel}}} \right\rangle - \left\langle \vec{v'_{\rm{rel}}} \right\rangle }
,\end{equation}
where $\vec{v'_{\rm{rel}}}$ is the part of the relative drift velocity $\vec{v_{\rm{rel}}}$ projected on the magnetic field direction and $g = h^{-1}$ is the grain non-sphericity \citep[][]{1994MNRAS.268..713L}. In the case of oblate dust grains, $g$ is negative and 
\begin{equation}
        \left\langle \cos^2(\beta) \right\rangle = \begin{cases} \frac{\sqrt{-g} \arcsin(\sqrt{-s/(1 + g))}}{s \arctan{\sqrt{sg/(1 + s + g)}}} - 1/s &\mbox{if } s<0  \\
\frac{\sqrt{-g} \sinh^{-1}(\sqrt{s/(1 + g))}}{s \tanh^-{1}(\sqrt{-sg/(1 + s + g))}} - 1/s & \mbox{if } s > 0  \end{cases}
\label{eq:gold}
.\end{equation}
Here, the Rayleigh reduction factor is independent of wavelength and grain size. The polarization can change sign for an angle $\alpha$ between magnetic field and a predominant gas stream lower than $54^{\circ}$ (see Fig. \ref{Fig:Ang}). For other grain shapes we refer to \cite{1996ASPC...97..425L}.\\
For the internal alignment the angular momentum can be approximated \citep[][]{1997ApJ...483..296L} by
\begin{equation}
        J_{eff}^2 \approx k_{\rm{B}} I_{\rm{||}}(2/h+1)\left( \frac{T_{\rm{g}}+T_{\rm{d}}}{2} + \frac{\mu m_{\rm{H}} v_{\rm{rel}}^2}{6 k_{\rm{B}}}\right)
        \label{eq:goldJ}
,\end{equation}
with the molecular mass $\mu m_{\rm{H}}$ of the gas. For the required conditions of supersonic velocity we locally calculate the Mach-number in each cell with
\begin{equation}
        M=  \frac{v_{\rm{rel}}}{\sqrt{k_{\rm{B}}T_{\rm{g}} /\mu m_{\rm{H}}}}
        \label{eq:mach}
\end{equation}
and demand $M>1$. Otherwise, we consider the dust grains to be randomized.\\
Possible mechanical alignment of helical dust grains in sub-sonic environments is currently being discussed \citep[e.g.][]{2007ApJ...669L..77L} but is not yet  in POLARIS.

\subsubsection{Radiative torque (RAT) alignment}
\label{sect:RAT}
It was noticed by \cite{1976Ap&SS..43..291D} that radiation can both spin up and align dust grains. The alignment mechanism is also known as the Barnett effect.\\
Based on a numerical approach and the DDSCAT Code \citep[][]{1993ApJ...405..685D,1994ApJ...405..685D, 2000ascl.soft08001D,2013arXiv1305.6497D}. \cite{1996ApJ...470..551D,1997ApJ...480..633D,2003ApJ...589..289W} demonstrate that RATs can significantly align a number of different grain shapes. The spin up of dust grains is also supported by laboratory experiments \citep[see][]{2004ApJ...614..781A}. Later, the analytical framework for RAT alignment was developed in  \cite{2007AAS...210.7904H, 2007AAS...210.7901H}, where they reproduced the numerical findings from \cite{2003ApJ...589..289W}.\\
With increasingly effective dust grain radius $a$, RATs become more efficient and the dust grains get suprathermally spun up. Analytical models by \cite{2007AAS...210.7904H, 2007AAS...210.7901H} show grain alignment at attractor points with high angular momentum (high-J) and low angular momentum (low-J). The high-J attractor point is stable and, with supra-thermal rotation ($\omega_{\rm{rad}} > 3\times\omega_{\rm{th}}$), dust grains can be considered to be perfectly aligned. Most dust grains drift to the unstable low-J attractor, where they are prone to thermal fluctuations and are subsequently partially disaligned. To determine the effective radius ($a_{\rm{alg}}$) at which the dust grains start to align, we follow the procedure described in \cite{2007ApJ...663.1055B} with
\begin{equation}
        \left(\frac{\omega_{\rm{rad}}}{\omega_{\rm{gas}}}\right)^2 = \frac{ a \rho_{\rm{d}}}{\delta m_{\rm{H}}}\left[\frac{\int Q_{\Gamma}(\epsilon)\lambda\gamma_{\rm{\lambda}} \overline{u}_{\rm{\lambda}} d\lambda}{n_{\rm{g}}k_{B} T_{\rm{g}}}  \frac{1}{1+\frac{t_{\rm{gas}}}{t_{\rm{rad}}}} \right]^2.
        \label{eq:omega}
\end{equation}
Here, $Q_{\Gamma}(\epsilon)$ is the alignment efficiency of the RAT, which depends on the angle $\epsilon$ between the predominate direction in the radiation field and the magnetic field direction. The alignment efficiency $Q_{\Gamma}(\epsilon)$ is also a function of  grain size, wavelength, and grain composition, and can be pre-calculated. The shape-dependent parameter $\delta,$ as well as the characteristic timescales of gas bombardment $t_{\rm{gas}}$ and radiation $t_{\rm{rad}}$ acting on the dust grains are introduced in detail in \cite{1996ApJ...470..551D, 1997ApJ...480..633D}. Checking for supra-thermal conditions also requires  information about the local anisotropy $\gamma_{\rm{\lambda}}$ and the mean energy density $\overline{u}_{\rm{\lambda}}$ of the radiation field.\\
These quantities can be determined with POLARIS by a MC mode, where we store the energy density direction $\vec{u}_{\rm{\lambda}}$ in each cell. In this mode we consider all radiation sources including thermal dust re-emission. Following \cite{1999A&A...344..282L} for the mean energy density per path length and wavelength we derive
\begin{equation}
  \vec{u}_{\lambda,i} d \lambda  =   \frac{\epsilon_{\rm{0}} }{c \Delta t V_{\rm{cell}}}     \frac{\vec{k}\times l_{\rm{i} }}{\left|\vec{k}\right|}d \lambda
\label{eq:LucyEnDens}
.\end{equation}
This allows us to calculate the mean energy density
\begin{equation}
  \overline{u}_{\rm{\lambda}} =  \sum_{i=1}{ \left| \vec{u}_{\rm{\lambda,i}}\right| }
\label{eq:EnDens}
\end{equation}
and the anisotropy parameter determined by
\begin{equation}
        \gamma_{\rm{\lambda}} =   \frac{1}{\overline{u}_{\rm{\lambda}} } \left|\sum_{i=1} \vec{u}_{\rm{\lambda,i}}\right| \in \left[0; 1\right] .
\end{equation}
With RATs, gas bombardment can even increase grain alignment by lifting grains from the low-J to the high-J attractor point. At the current stage of its development no analytical function for the distribution of the cone angle $\beta$ is available. While perfect alignment can be assumed at both high-J and low-J attractor points, internal aligned is just given for suprathermally rotating grains. At low-J the angular momentum  can be assumed to be thermal \citep[][]{2009ApJ...697.1316H} and  the distribution of internal alignment (Eq. \ref{eq.distr}) becomes a function of grain geometry alone because of $h$. Now, for the Rayleigh reduction factor, only the moment $\left\langle G_{\rm{low-J}}\left(\cos^2(\zeta)\right) \right\rangle$ remains of relevance. When we consider $f_{\rm{high-J}}$ to be the fraction of dust grain aligned at the high-J attractor point the Rayleigh reduction factor is\begin{equation}
        R = \begin{cases} f_{\rm{high-J}}+(1-f_{\rm{high-J}})\left\langle G_{\rm{low-J}}\left(\cos^2(\zeta)\right) \right\rangle &\mbox{if } a\geq a_{\rm{alg}}  \\0 & \mbox{if } a < a_{\rm{alg}}  \end{cases} 
        \label{eq:Rrat}
.\end{equation}
So far, $f_{\rm{high-J}}$ cannot be exactly  determined \citep[][]{2009ApJ...697.1316H} and remains a free parameter.\\
Once, a dust grain starts to tumble in the presence of a magnetic field a Larmor precession acts on the grains because of the Barnett effect. For the RATs to work, the magnetic field strength has to be sufficiently high to resist disaligment by gas-dust interactions. This requires the characteristic timescale for Larmor precession ($t_{\rm{L}}$) to be lower than the gas dumping time ($t_{\rm{gas}}$): $t_{\rm{L}}<t_{\rm{gas}}$ \citep[][]{2009ApJ...704.1204H, 2009ApJ...697.1316H}, which gives us a limit of 
\begin{equation}
        \left| \vec{B} \right| > 4.1\times 10^{-9} a \frac{n_g \sqrt{T_{\rm{g}}} T_{\rm{d}}}{s^2} [T]
        \label{eq:larm}
\end{equation}
for the magnetic field strength. This limitation is similar in nature to that of  Eq. \ref{eq:delta}. Since the coupling with the magnetic field is also the Barnett effect, the same limitation for the field strength is applied to RT simulations with GOLD alignment.

\subsection{Scattering on partially aligned dust grains}
\label{sect:scatter}
For each scattering event the coordinate system of the dust grain is determined by the magnetic field direction. The coordinate system of the Stokes vector must be transformed into the dust reference frame.  For scattering from one direction $(\vartheta, \varphi)$ to a new direction $(\vartheta', \varphi')$, the Stokes vector $S$ is modified via a 4x4 scattering matrix $\hat{M}$ by $S' = \hat{M}S$. Non-spherical, partially aligned dust grains can require a full 16-element matrix and an asymmetry factor $g$, with  both dependent on direction of inclination, scattering direction, wavelength, and grain size, which makes scattering a multi-dimensional problem that is fully supported by POLARIS.\\
The scattering matrix $\hat{M}$  and asymmetry factor $g$ can be pre-calculated with DDSCAT. For imperfectly aligned wobbling dust grains the incident $(\vartheta, \varphi)$ and scattering angles $(\vartheta', \varphi')$ become functions of alignment angles $\beta$ and $\zeta$, respectively. To obtain an exact scattering matrix for an average-sized dust grain one has to integrate the alignment distribution functions $f(\beta)$ and $ f(\zeta)$, respectively, as well as the grain size distribution $n(a)$ simultaneously. Here, we face the same problem as for the calculation of cross-sections in Sec. \ref{sect:dust}. In the current stage of development, the exact distribution function for RAT alignment is still missing. Solving the exact integrals for GOLD or IDG alignment would push the timescales of 3D MC-RT simulations beyond accessible computational equipment. To handle the scattering on non-spherical imperfectly aligned dust grains, we apply the following approximation:
\begin{equation}
\begin{split}
\hat{M}_{ij}(\lambda, R,a, \vartheta, \varphi, \vartheta', \varphi') \approx \\
           R(a)\times \hat{M_{ij}}(a,\lambda, \vartheta,\varphi, \vartheta', \varphi') + \left(1-R(a)\right) \left\langle \hat{M_{ij}}(a,\lambda, \vartheta', \varphi')\right\rangle .
\end{split}
\label{eq.aprox}
\end{equation}
The same approximation is used for the orientation and size-dependent asymmetry parameter $g$ of the HG-Phase function (see Eq. \ref{eq:HG}):
\begin{equation}
\begin{split}
g(\lambda, R,a, \vartheta, \varphi) \approx \\
           R(a)\times g(a,\lambda, \vartheta,\varphi) + \left(1-R(a)\right) \left\langle g(a,\lambda, \vartheta, \varphi)\right\rangle
\end{split}
\label{eq.aproxG}
,\end{equation}
where $\left\langle \hat{M_{ij}}(a,\lambda, \vartheta', \varphi')\right\rangle$ and $\left\langle g(a,\lambda, \vartheta, \varphi)\right\rangle$ are the entries of the scattering matrix and the asymmetry factor, respectively, for randomly aligned dust grains. With increasing efficiency in dust grain alignment the Rayleigh reduction factor reaches unity and the dust grains and the contribution of randomly aligned dust grains diminishes.

\begin{figure*}
   \centering
         \includegraphics[width=1.0\textwidth]{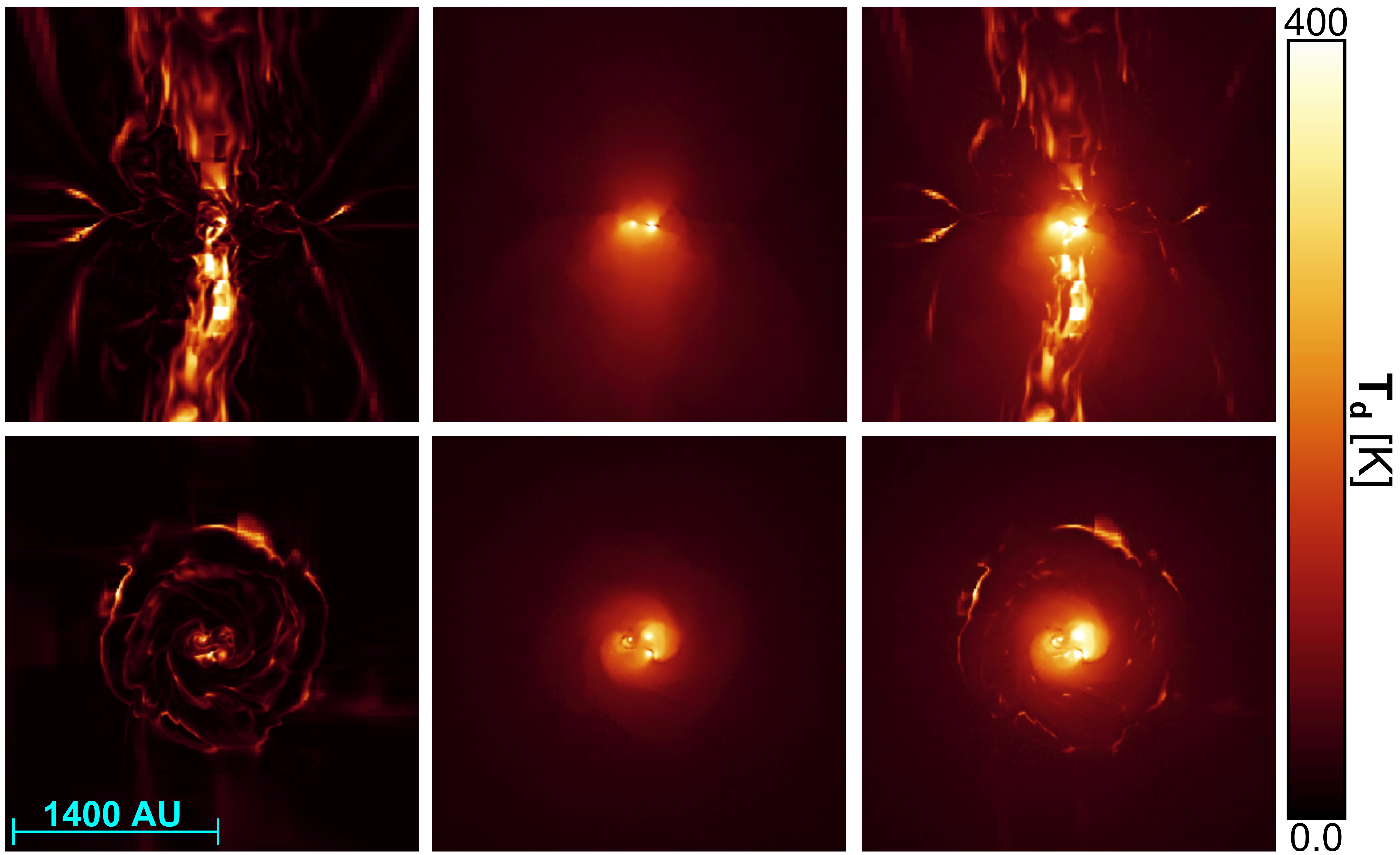}
        
   \caption{\small Dust temperature distribution in the mid-planes of a 3D MC-RT/MHD collapse simulation. Rows show the results of the same model, but of inclinations differing by $90^{\circ}$. In the left column, the dust was heated by 3D MC-RT simulation alone. The middle column shows the dust temperature distribution as a result of MHD shock heating. The dust temperature combined by offset dust heating, as described in Sect. \ref{setupMHDHeat}, is shown in the right column. The length of $1400\ \rm{AU}$ is for scale. We applied an upper cut-off of 400K (blue regions) to the mid-plane images for better illustration.}

\label{fig:heat}
\end{figure*}

\subsection{Radiative transfer with MHD data}
\label{setupMHDHeat}
The complexity of star formation can not be accurately mapped by simple analytical models. However, with the advantages of high-resolution MHD simulations, e.g. FLASH \cite{2010ascl.soft10082F}, it is possible to provide complex 3D physical data as an input for MC-RT simulations. Not only can this data   deliver density, magnetic field, and velocity distributions, but it can also provide the exact parameter of stars appearing in such simulations. This stars have to be considered as additional sources of radiation. For RT simulations with MHD  data, we have to take the following into account: For the implemented alignment mechanisms to work, the characteristic timescales have to be taken into account. By post-processing  MHD data, the shortest time steps in the MHD simulations must exceed the largest timescale of alignment mechanisms to ensure a steady-state of grain alignment \citep[see][]{1996ApJ...470..551D, 1997ApJ...480..633D, 2003ApJ...589..289W}. \\
Finally, a dust temperature can also be provided by the actual MHD simulations. With temperatures resulting from MHD simulations, we have to deal with two effects of dust heating: RT-heating because of radiation sources and  MHD-heating because of the compression and gas-dust interaction \citep[][]{2006MNRAS.373.1091B,2007ApJ...656..959K}. Both temperatures are equally physically well-motivated, however, they are rarely fully implemented in MHD codes. To account for all effects of dust heating, POLARIS optionally performs  an extended dust heating algorithm. Originally, the algorithm of \cite{1999A&A...344..282L} assumes a cell's energy content $\dot{E}$ to be empty at the beginning of the radiative dust heating process. Since we have already pre-calculated the emissivity of an ensemble of dust grains with a certain temperature, we can now perform  the reverse calculation. Using Eq. \ref{eq:LucyJ}, we determine the offset amount of energy $\Delta \dot{E}$ that corresponds to the dust temperature delivered by the MHD data.  We refer to this method as offset dust heating in the following.

\subsection{Code optimization}
\label{optim}
An 3D MC-RT simulation is challenging when it comes to run time and the capabilities of available computational equipment. The POLARIS code is parallelized to run on shared memory machines using the OpenMP library with the propagation  of several photon packages  simultaneously through the grid. The run-time scales well with the number of used processors. However, complex environments provided by MHD simulations, the physically well-motivated dust models, and alignment mechanisms requires a huge parameter space to solve the RT problem, including polarization in 3D. For this reason POLARIS is equipped with several optimization algorithms to perform RT calculations in a reasonable time. To propagate photon packages in regions with extreme optical depth more efficiently, we implemented the flowing algorithms: The modified random walk (MRW), a technique that uses a diffusion approximation instead of an MC approach  \citep[][]{2009A&A...497..155M,2010A&A...520A..70R} for optically thick regions and the forced first scattering  \citep[e.g.][]{1970A&A.....9...53M,1999ApJ...525..799W} that increases the signal-to-noise ratio in optically thin dust environments \citep[see][]{1984AIPC..115..665M}. For the calculation of the RAT alignment, the MRW cannot be applied. While the mean energy density is correctly approached, the MRW leads to a loss of information about the anisotropy of the radiation field. However, since the effective grain radii are pre-calculated and therefore discrete values, a far lower amount of photon packages is required to converge against a distinct value of $a_{\rm{alg}}$ in the RAT mode.\\
Furthermore, for synthetic intensity and polarization maps, the S/N can  be enhanced with the peel-off technique \citep[][]{1984ApJ...278..186Y} by sending a fraction of the radiation toward the observer at each scattering event. With longer wavelengths, scattering becomes increasingly insignificant and a ray-tracing algorithm can be applied. Here, we simply add up all contributions because of extinction and re-emission along the line of sight. Compared to a radiative transfer with scattering, the ray-tracing approach is more effective and allows us to achieve an excellent S/N \citep[][]{2010A&A...520A..70R,2012ascl.soft02015D}.\\
Radiation sources emit photons in the full wavelength range of its characteristic SED. However, certain wavelengths can be neglected since they do not contribute significantly to dust heating and alignment compared to the total emitted energy of the source. A wavelength range selection algorithm reduces the number of wavelengths considered and subsequently the computational effort.

\section{Applications}
\label{sect:application}
In this section, we show selected applications of the POLARIS code, demonstrating its features and predictive capability. In particular, we show RT simulations in astrophysical environments, representing different stages of star formation on multiple scales. We run 3D MC-RT simulations to calculate  temperature distributions and synthetic polarization maps. A benchmark of POLARIS' scattering calculations is provided in appendix \ref{apAA}.\\
The often used dust grain model of \cite{1977ApJ...217..425M} suggests an upper cut-off in grain radius at $a_{\rm{max}} = 250\ \rm{nm. }$ More recent models \citep[e.g.][]{2001ApJ...548..296W}, however, indicate an upper limit with $\rm{\mu m}$-radii. Since RAT alignment is heavily dependent on the maximum value of grain size, we pre-calculated a database for oblate dust grains with a fixed aspect,  $118$ effective radii, and $100$ wavelength logarithmically distributed between $a \in [5\ \rm{nm}: 2\   \rm{\mu m}]$ and $\lambda \in [90\ \rm{nm}:2\ \rm{mm}]$, respectively. Here, we used the program DDSCAT \citep[][]{2013arXiv1305.6497D} and approximated the oblate shape of the dust grains with $171 500$ distinct dipoles and an error tolerance of $8.5 \times 10^{-3}$. As materials, we considered astro-silicate and graphite, used the refraction indices of  \cite{2000AAS...197.4207W}, and applied the $1/3-2/3$ approximation to take care of the anisotropic graphite structure \citep[][]{1993ApJ...414..632D}. We weighted the efficiencies for each wavelength according to the power-law distribution to derive the average cross-sections over all effective radii (see Eq. \ref{eq:avg}).\\
Once the photon package escapes the grid, the degree of linear polarization $P_{\rm{l}}$ is determined
by the Stokes vector with
\begin{equation}
        P_{\rm{l}}=\sqrt{\frac{Q^2+U^2}{I^2}}
\end{equation}
and its projected position angle $\chi$

\begin{equation}
        \chi=\frac{1}{2} \arctan\left( \frac{U}{Q} \right)
.\end{equation}
The aspect ratio of minor axis to major axis remains $s = 0.5$ in all cases. Higher aspect ratios would
result in higher peak values and more contrast in the maps of linear and circular polarization. The dust heating is assumed to be independent of effective radius and aspect ratio, which is in good agreement with the findings of \cite{1980A&A....88..194H} and \cite{1999A&A...349L..25V} within the considered range of parameters. 

\subsection{Offset dust heating}
\label{setupHeat}
In this section, we post-processed an MHD collapse simulation with a total mass of $100\ \rm{M_{\odot}}$ \citep[see][for a detailed description]{2011MNRAS.417.1054S}. We took a snapshot after a simulation time of $5000\ \rm{yr}$ and consider just the inner most region of the MHD simulation with an accretion disc of about $1400\ \rm{AU}$. The dust temperature, as well as position, stellar radii, and surface temperatures of the stars were provided by the simulation. For dust material, we use a mixture of $62.5 \%$ astro-silicate and $37.5 \%$ graphite with an an upper cut-off for the size distribution at $a_{\rm{max}}=250\ \rm{nm}$. The highest temperature in these simulations is below the sublimation temperature of the silicate and graphite grains, respectively. First, we ignore the MHD dust temperature and performed a 3D MC-RT simulation, with the stars as radiation sources alone. Then, we adjust the temperature with the method described in Sect \ref{setupMHDHeat}.\\
In Fig. \ref{fig:heat}, we show the resulting dust temperature distributions in the center x-y planes and x-z planes. For the RT simulations with no offset energy (Fig. \ref{fig:heat} left panels), the dust temperature shows a diffuse distribution with the highest temperature near the stars and regions with lowest densities. The dust temperature distribution matches the density structure near the disk region, while the outflow regions remain hidden. Regions with high densities are shielded from stellar radiation and cannot be heated sufficiently. High density regions, however, gain temperature by the dynamical processes of shock heating in MHD simulations (Fig. \ref{fig:heat} middle panels). The MHD dust temperature resembles the dust density in all regions while underestimating the temperature towards the center region of the accretion disk. By considering  both RT and MHD leads to a shift in the profitability distribution of the wavelength of re-emission (see Eq. \ref{eq:BW}) towards shorter wavelengths. In this regime of wavelengths, absorption of radiation is more effective because of higher cross-sections of absorption $C_{\rm{abs,\lambda}}$. This results in a net temperature that is higher than  single temperatures alone, which would not have been the case by simply adding up the dust temperatures resulting from RT and MHD simulations.

\subsection{Disk models}
\label{setupDisk}
\begin{figure}[]
        \begin{minipage}[c]{1.0\linewidth}
                        \begin{center}
                                \includegraphics[width=1.0\textwidth]{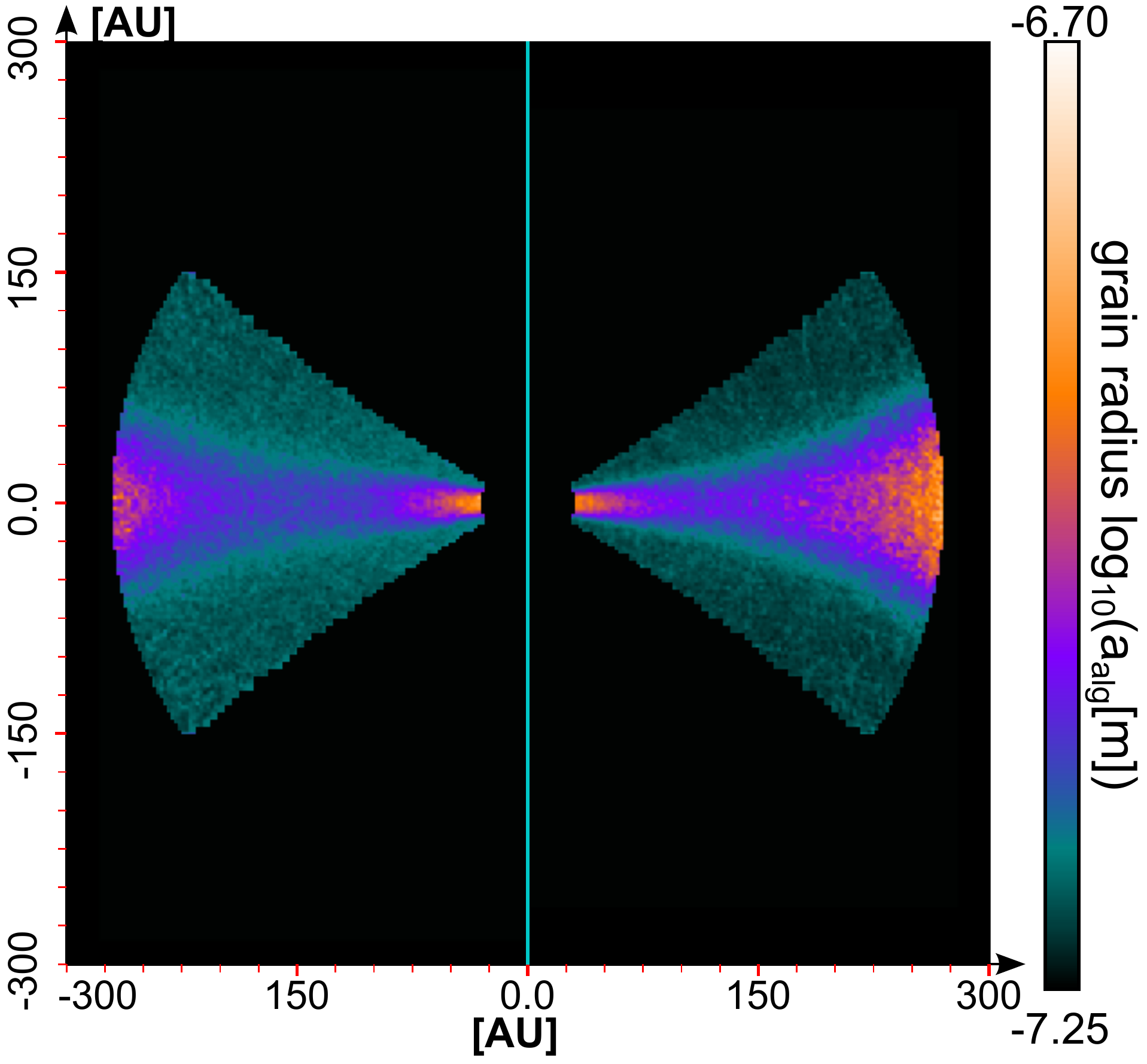}
                        \end{center}
                \end{minipage}  
                                
\caption{\small Distribution of grain sizes $a_{\rm{alg}}$, where dust particle start to align to the magnetic field direction $\vec{B}$ according to RAT theory. The  plane shown is perpendicular to the mid-plane of the disks. The disk model $D03$ with $10^{-3}\ \rm{M_{\odot}}$ is in the left panel and the model $D05$ with $10^{-5}\ \rm{M_{\odot}}$ is in the right panel. Lower values of $a_{\rm{alg}}$ result in higher degrees of linear polarization.}
\label{DiskLayer}
\end{figure}
In this section, we trace the magnetic field morphology in circumstellar disks of different mass with POLARIS.  The magnetic field plays an important role in the formation of stars, as well as in the formation and evolution of the circumstellar disk and, subsequently, in the formation of planets. Angular momentum of the gas component of the disk can, effectively, be transferred outwards by magnetic braking and interstellar outflows allow the fall of gas towards the center. Thus, to understand circumstellar disk and planet formation, we need to investigate the properties of the underlying magnetic field that is involved in the formation process.\\
By considering the dust grains to be partially aligned with the direction of the magnetic field, the field morphology may be investigated by polarization observations in the sub-mm regime of wavelength without the contribution of scattering in the polarization signal.
\subsubsection{Model setup}
The density distribution of our disk models are parametrized as presented in \cite{1973A&A....24..337S}. The model assumes the density distribution to follows:
\begin{equation}
        n_{\rm{g}}(r,z)=n_{\rm{0}}  \left(\frac{r_{\rm{0}}}{r}\right)^\alpha \exp\left( -\frac{1}{2}\frac{z}{h(r)} \right)^2
,\end{equation}
where
\begin{equation}
        h(z) = h_{\rm{0}}\left(\frac{r}{r_{\rm{0}}} \right)^\beta .
\end{equation}
Here, $r$ and $z$ are cylindrical coordinates, $h_{\rm{0}}$ is the scale height at $r_{\rm{0}}$ and the quantities $\alpha$ and $\beta$ are parameters that characterize the radial density profile and the disk flaring. We adjusted $n_{\rm{0}}$ for the entire disk mass to range between $10^{-3}\ [\rm{M_{\odot}}]$ and $10^{-6}\ [\rm{M_{\odot}}]$ for the different models.
\begin{table}[]
\begin{tabular}[]{ l | c }
Parameter & Value \\ \hline \hline
    $\alpha$ & $2.6$\\
                $\beta$ &       $1.25$\\
                $h_{\rm{0}}\ [\rm{AU}]$ & $10$\\
                $r_{\rm{0}}\ [\rm{AU}]$ & $100$\\
    $R_{\rm{in}}\ [\rm{AU}]$ & $25$\\
                $R_{\rm{out}}\ [\rm{AU}]$ & $280$\\
                $M_{\rm{disk}}\ [\rm{M_{\odot}}]$ & $10^{-6}$, $10^{-5}$, $10^{-4}$, $10^{-3}$\\
                $R_{\rm{*}}\ [\rm{R_{\odot}}]$ & $6$ \\
                $T_{\rm{*}}\ [\rm{K}]$ & $4000$ \\
                $T_{\rm{d0}}\ [\rm{K}]$ & $10$ \\
\end{tabular}
\caption{Table of physical quantities for the applied disk models. We consider different disk masses and refer to the models with a total disk mass of $10^{-3}\ \rm{M_{\odot}}$ as $D03$, with $10^{-4}\ \rm{M_{\odot}}$ as $D04$, with $10^{-5}\ \rm{M_{\odot}}$ as $D05$, and with $10^{-6} \rm{M_{\odot}}$ as $D06$, respectively.}
\label{tab:1}
 \end{table}
\begin{figure}[]
        \begin{minipage}[c]{0.49\linewidth}
                        \begin{center}
                                \includegraphics[width=0.9\textwidth]{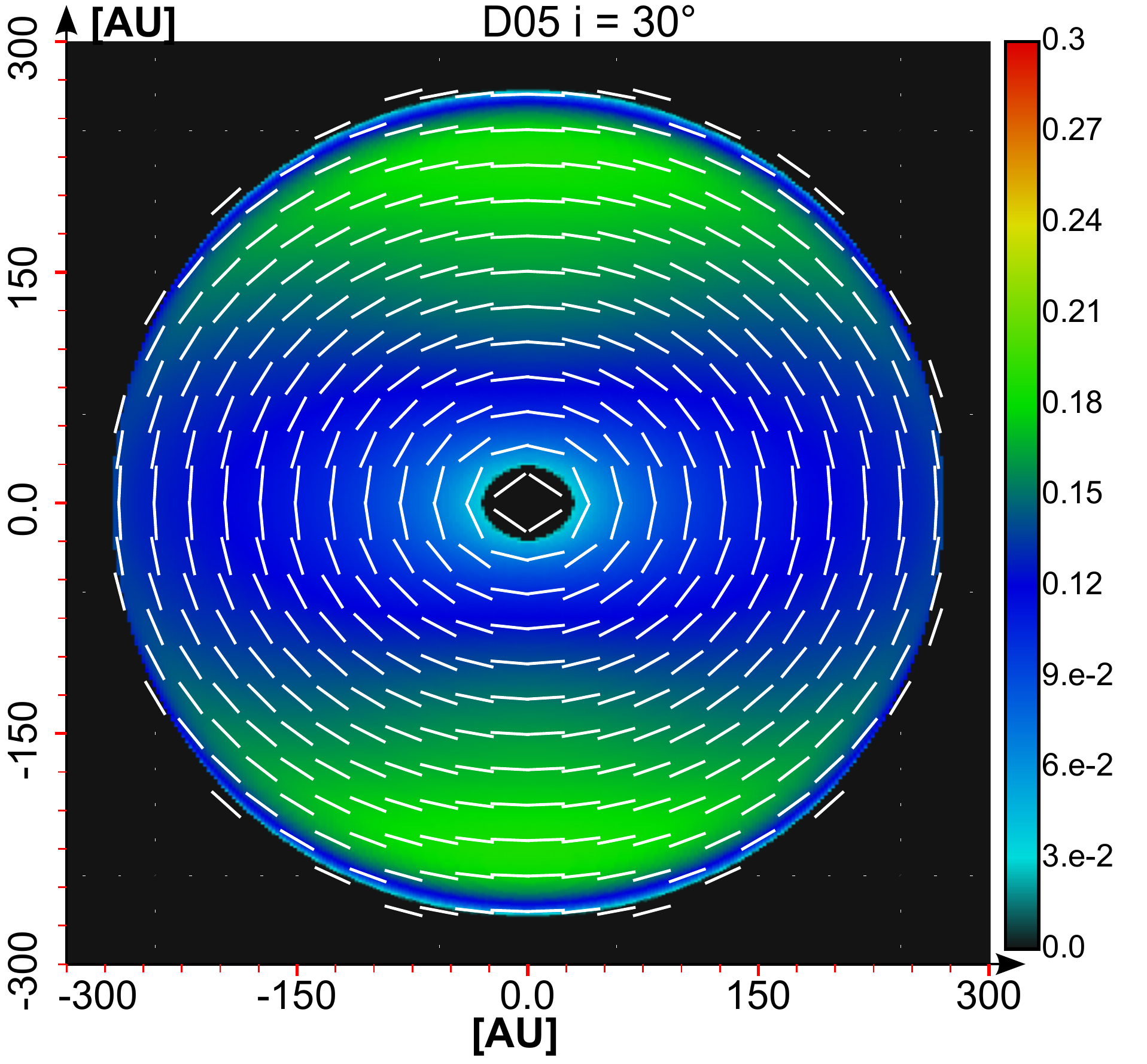}\\
                                \includegraphics[width=0.9\textwidth]{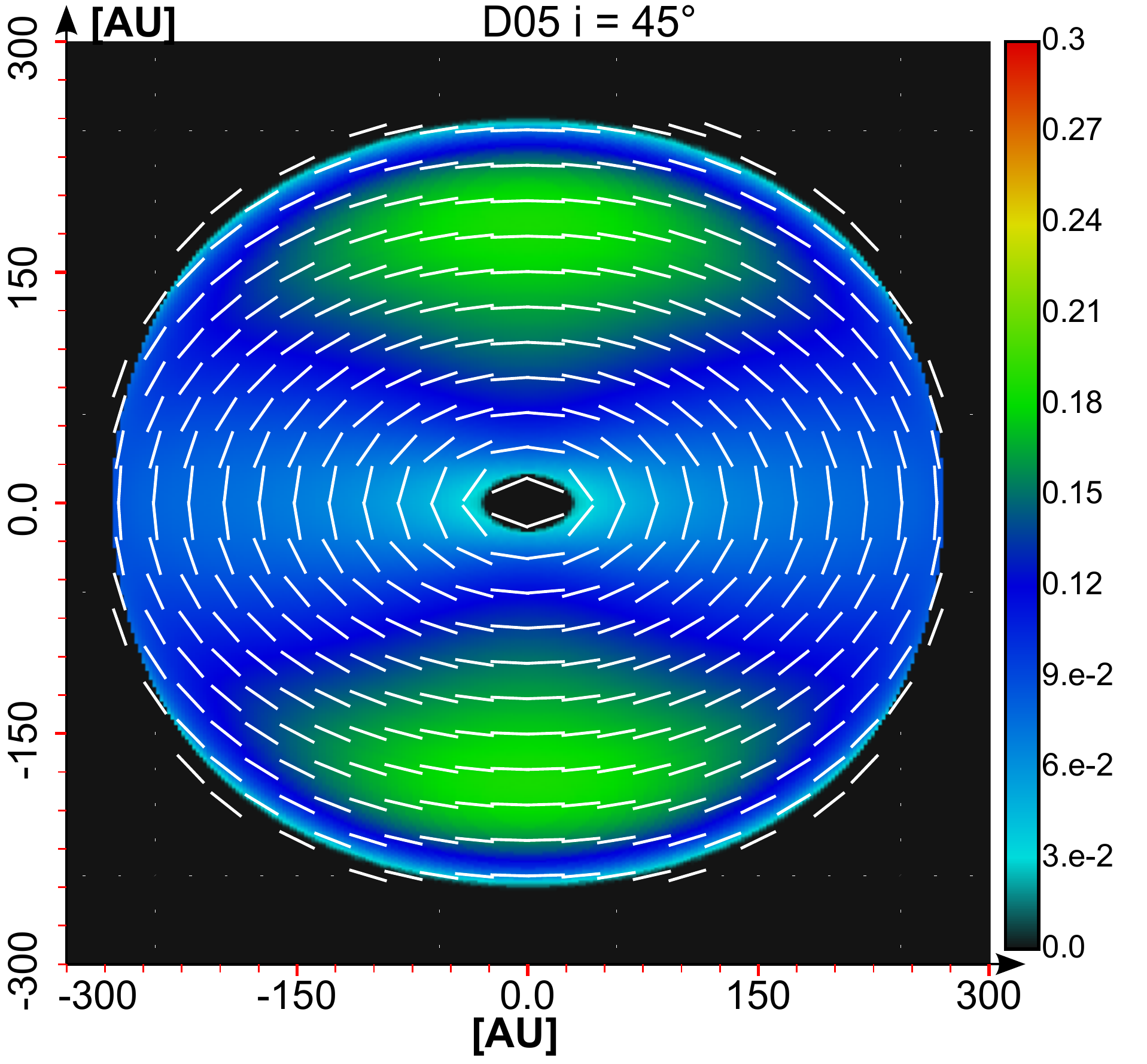}\\
                                \includegraphics[width=0.9\textwidth]{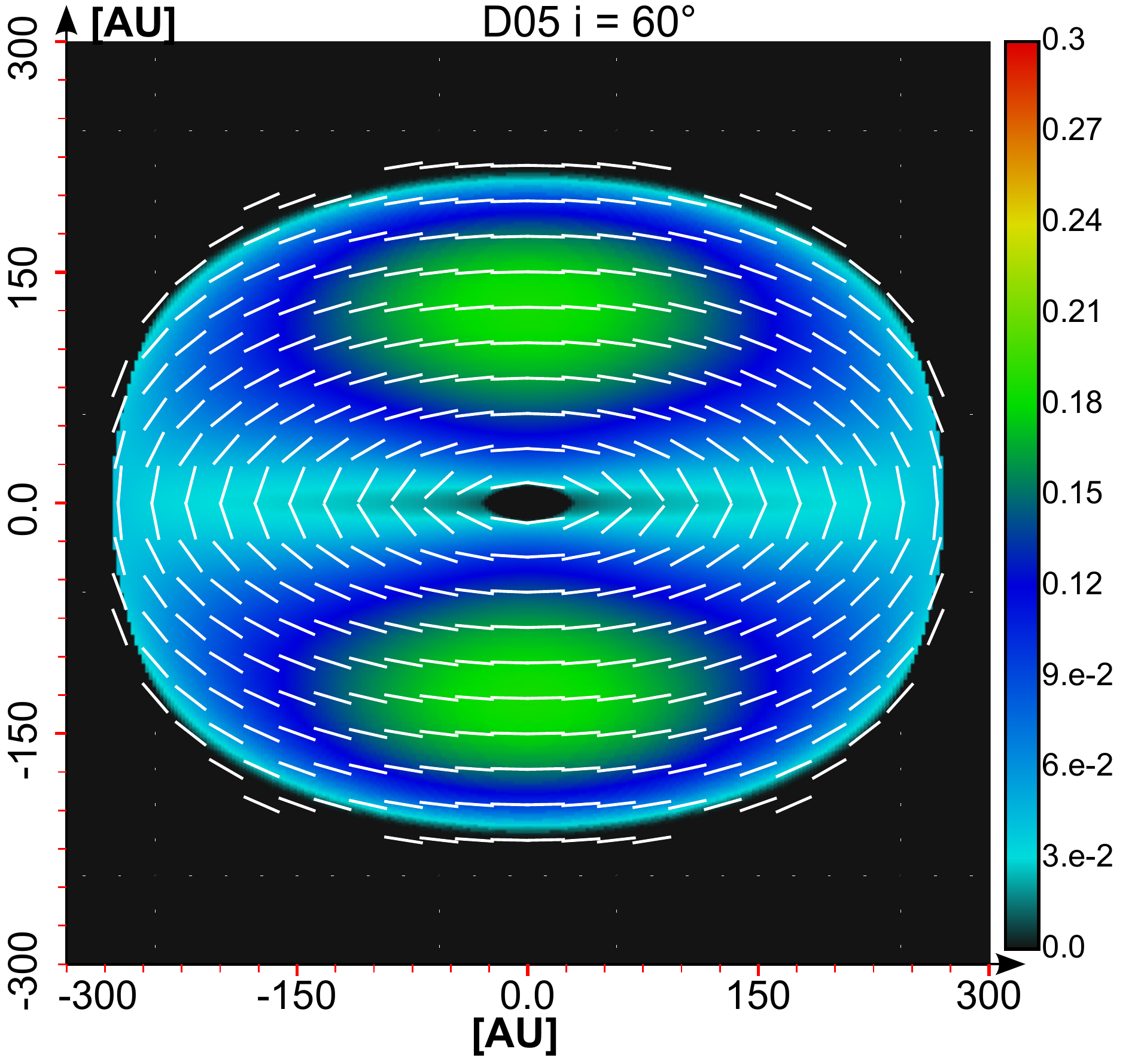}\\
                                \includegraphics[width=0.9\textwidth]{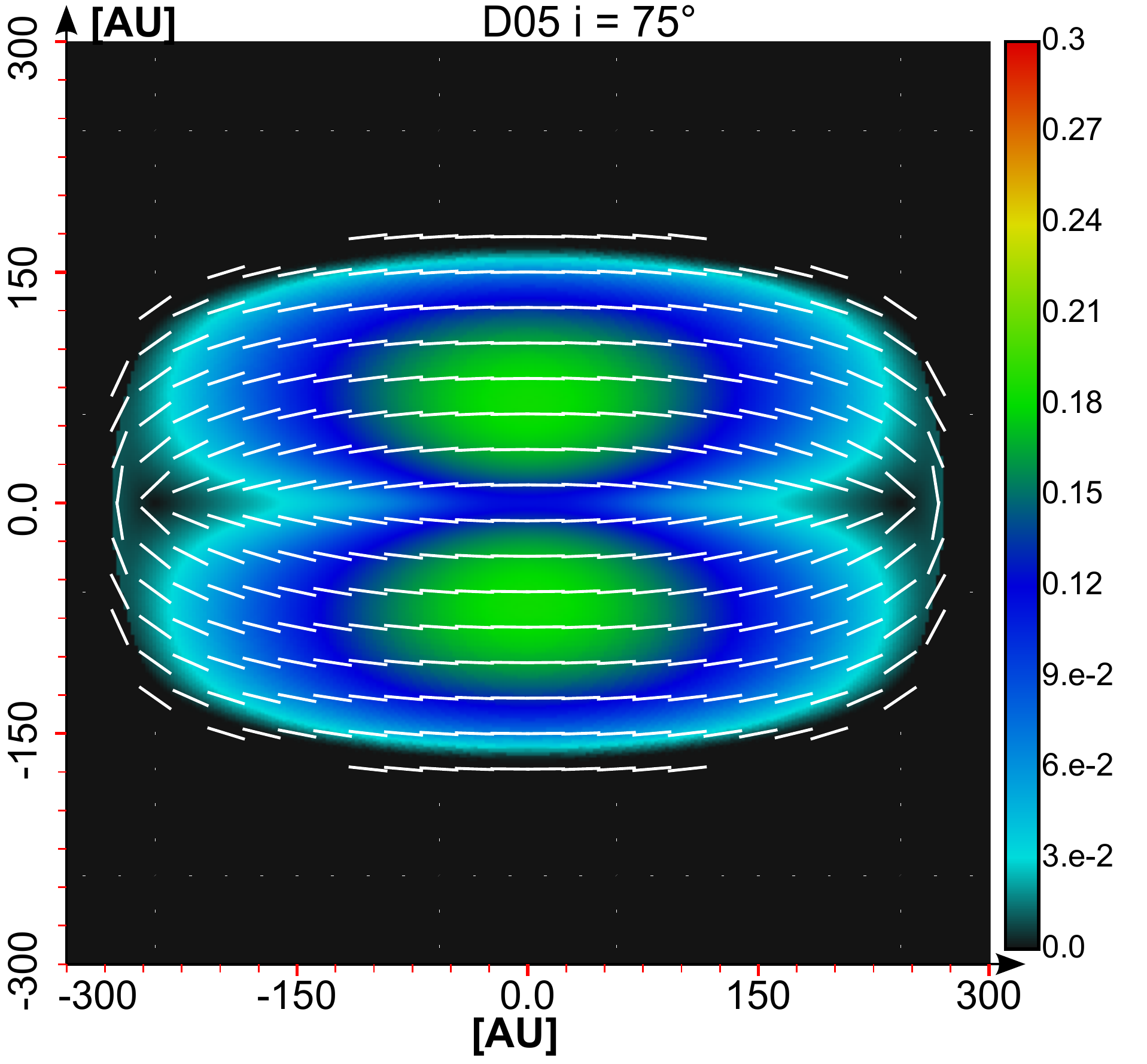}\\
                                \includegraphics[width=0.9\textwidth]{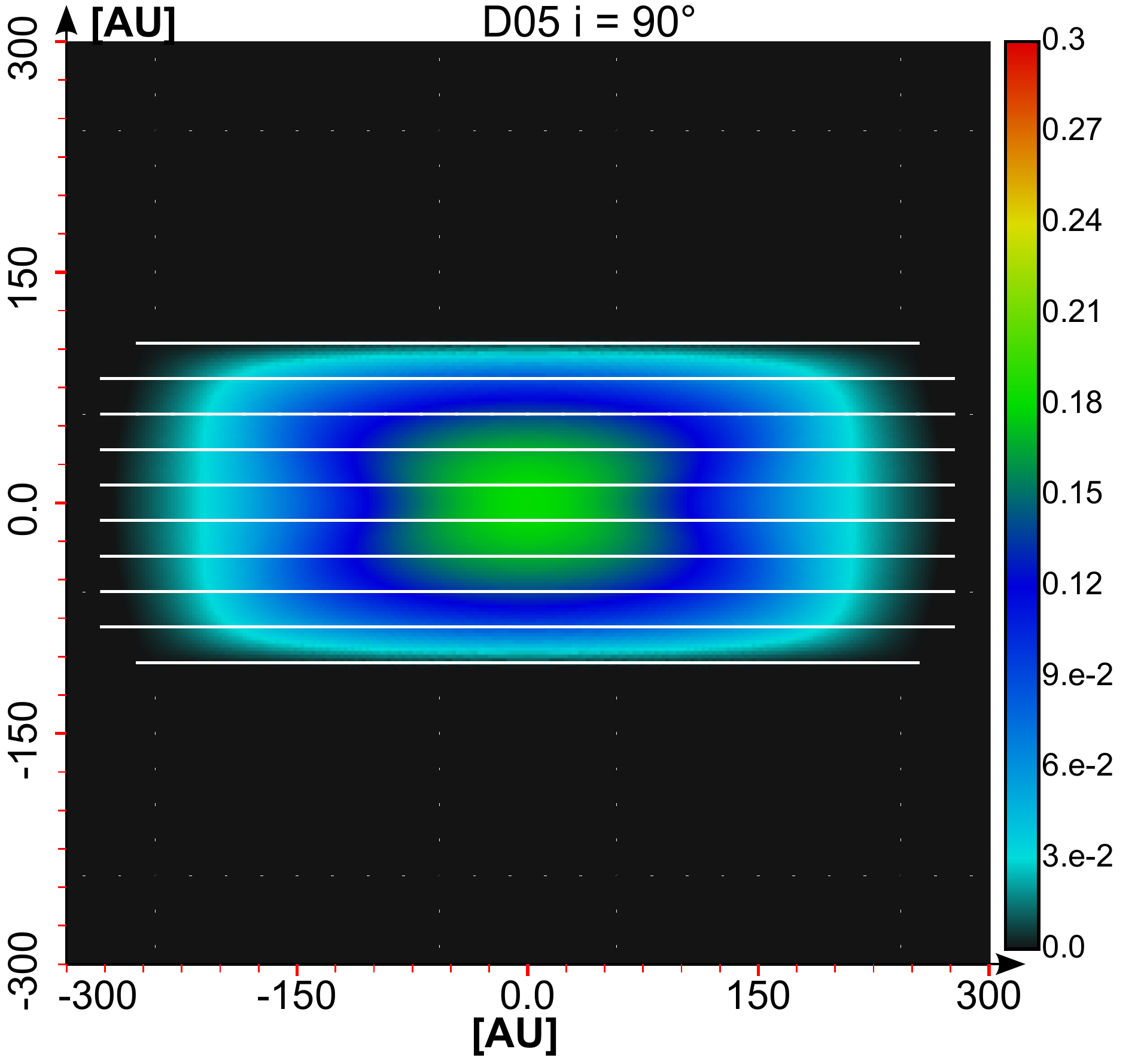}
                        \end{center}
                \end{minipage}
                \begin{minipage}[c]{0.49\linewidth}
                        \begin{center}
                                \includegraphics[width=0.9\textwidth]{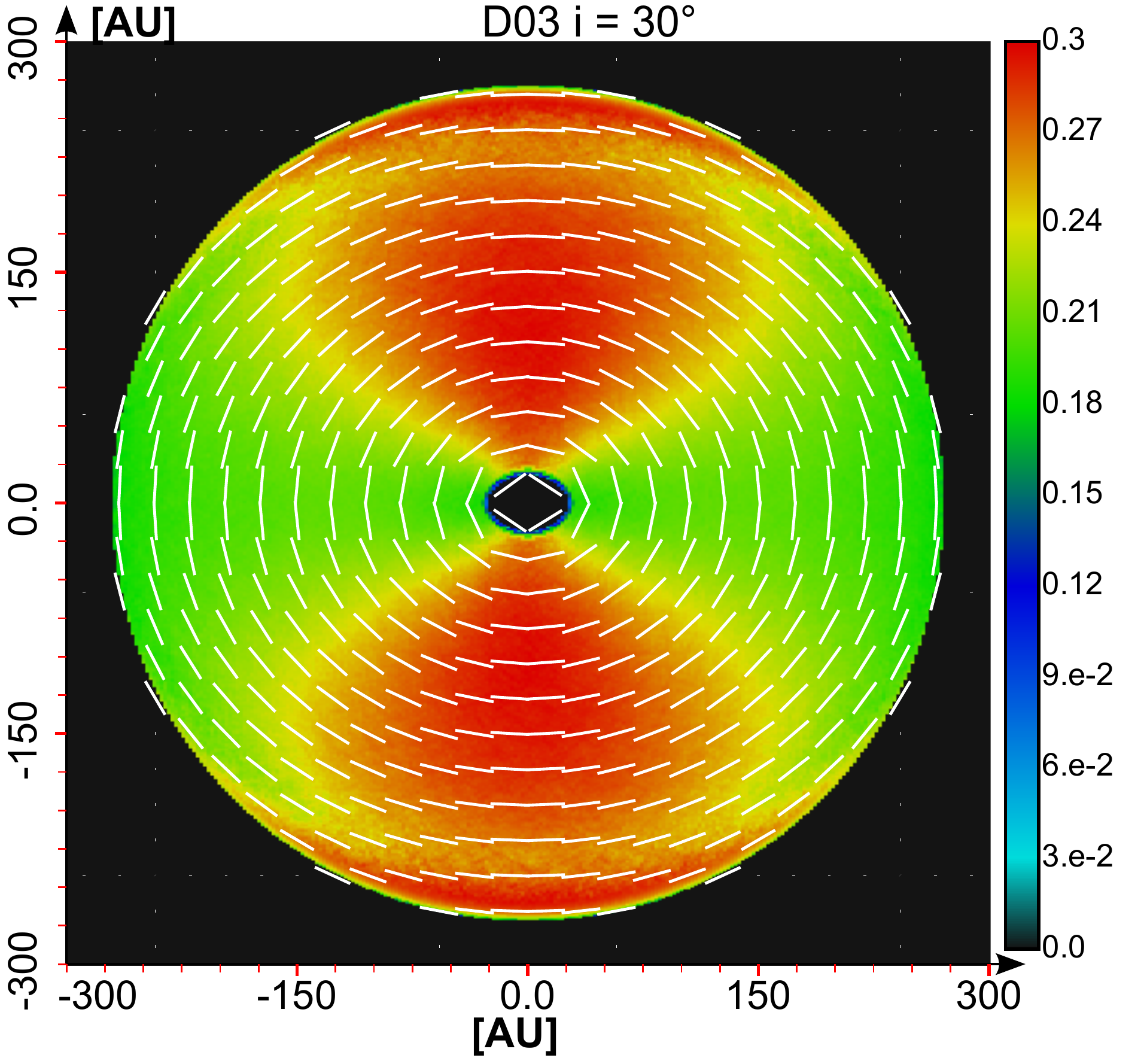}\\
                                \includegraphics[width=0.9\textwidth]{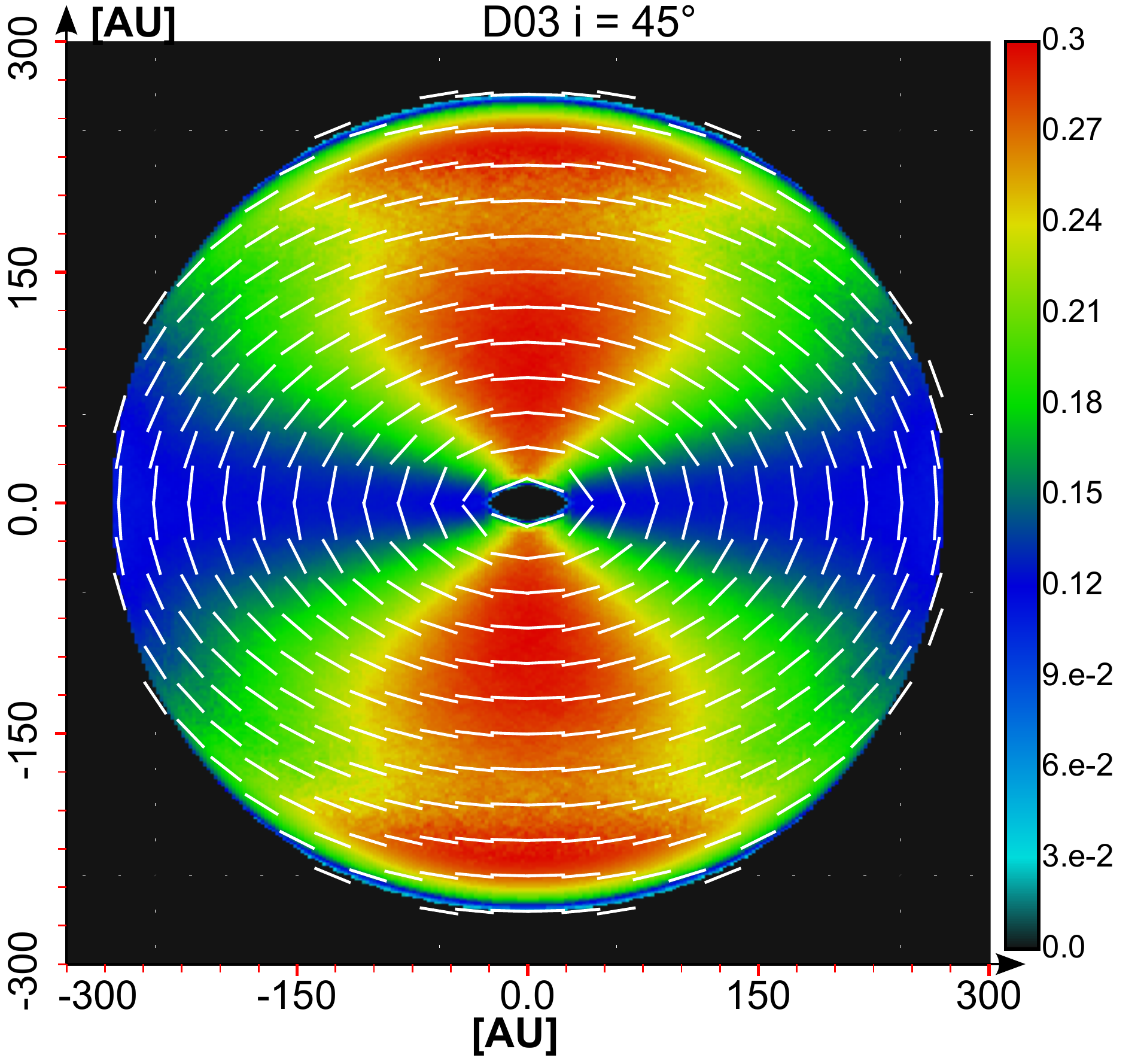}\\
                                \includegraphics[width=0.9\textwidth]{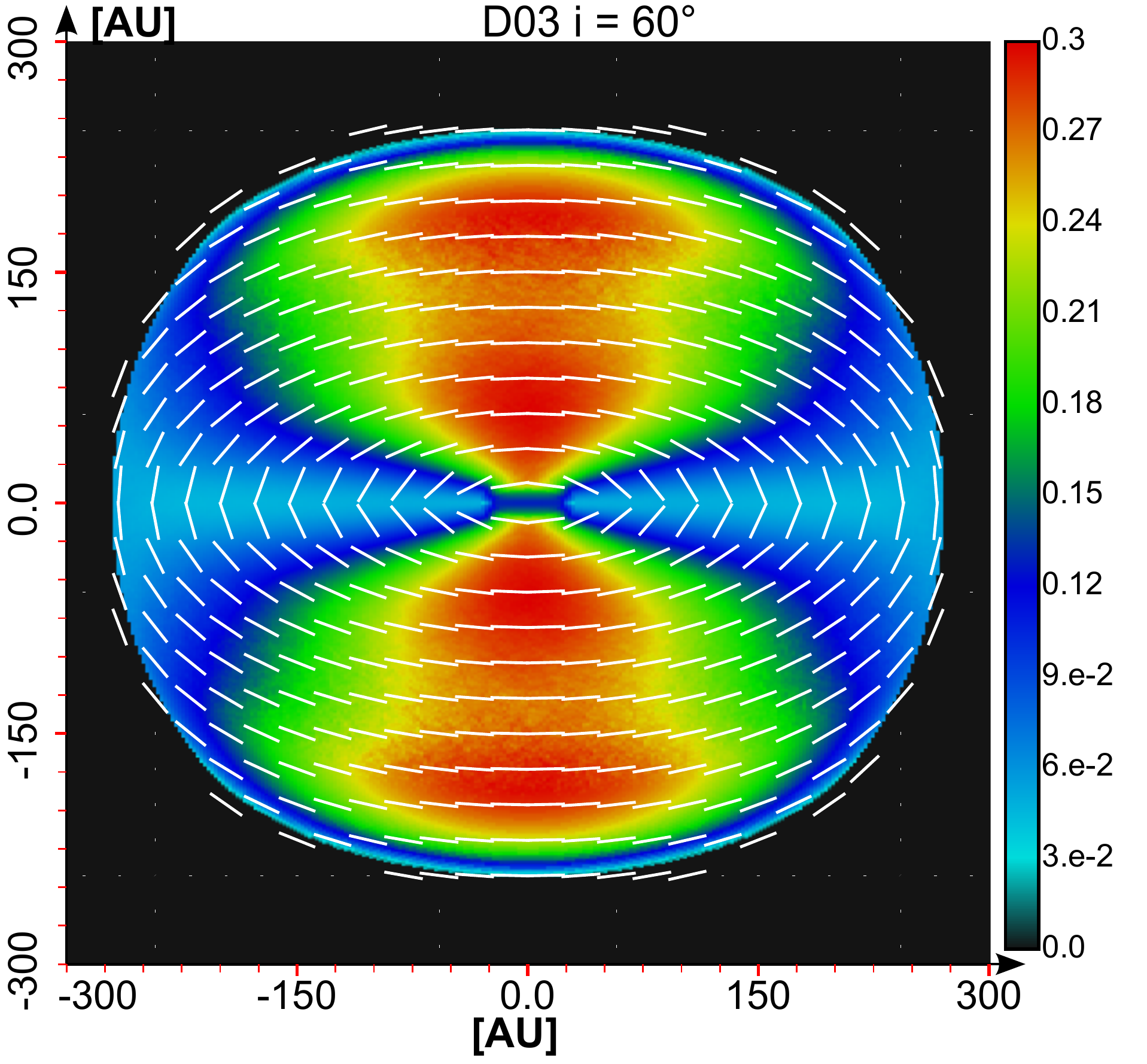}\\
                                \includegraphics[width=0.9\textwidth]{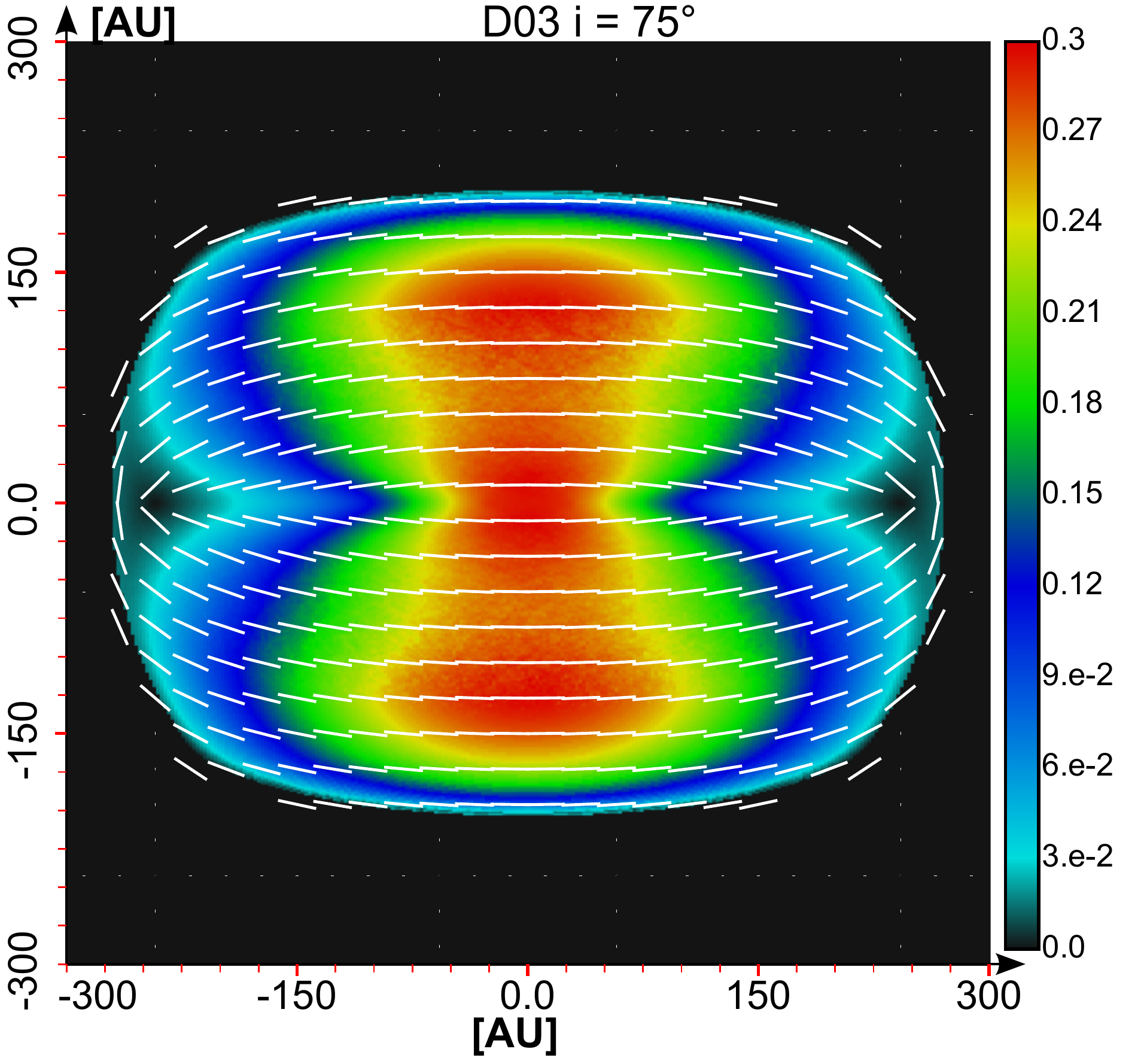}\\
                                \includegraphics[width=0.9\textwidth]{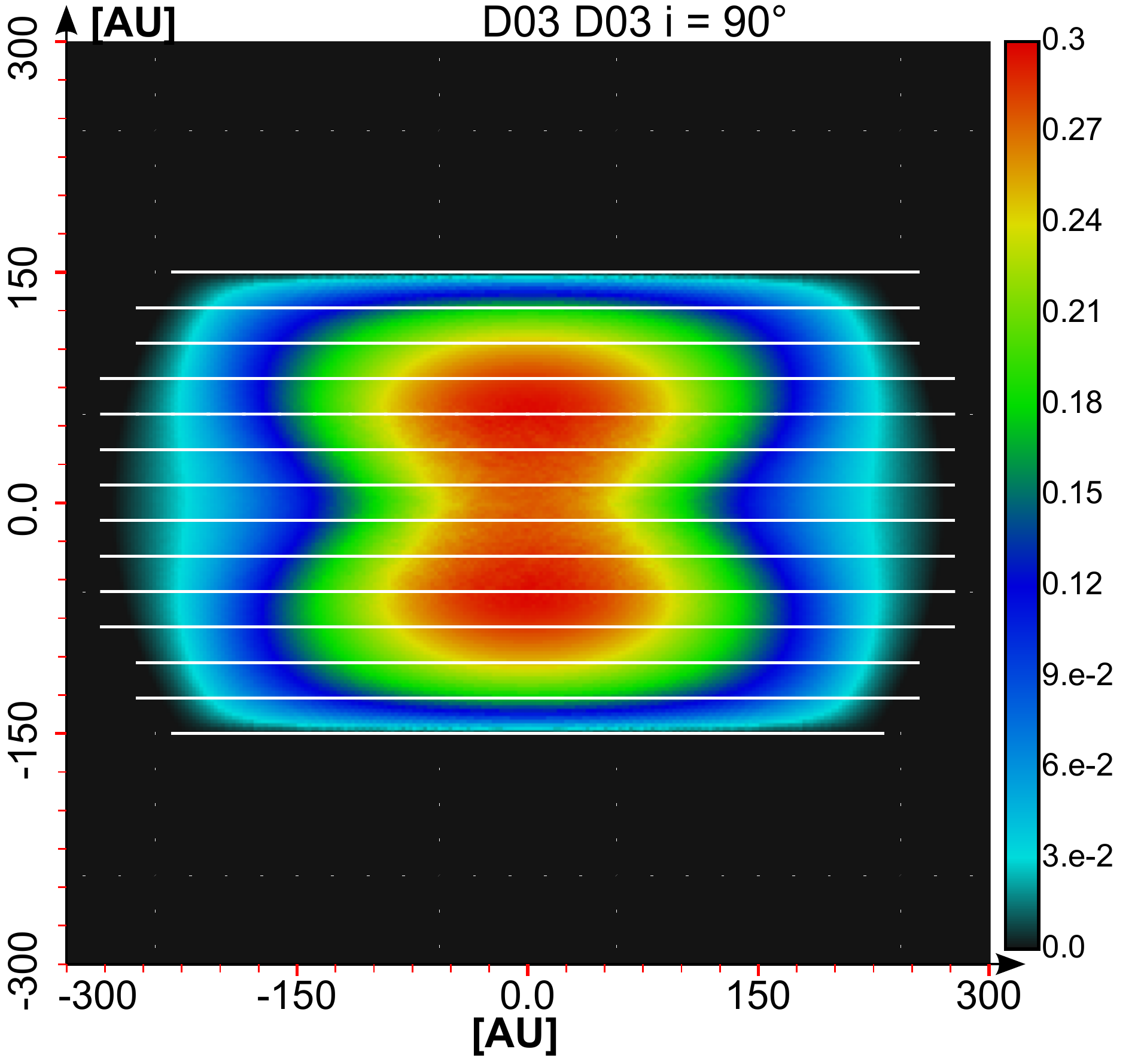}
                        \end{center}
                \end{minipage}

        \caption{\small{Maps with pattern of linear polarization overlaid with normalized orientation vectors at an exemplary wavelength of $\lambda= 515\ \rm{\mu m}$. We applied an offset of $90^{\circ}$ to the orientation vectors to match the projected toroidal magnetic field morphology. The model D03 with $10^{-3}\ \rm{M_{\odot}}$ is in the right column and model D05 with $10^{-5}\ \rm{M_{\odot}}$ is shown in the left column for inclination angles from $ i=30^{\circ}$ (top row) to $i=90^{\circ}$ (bottom row) in steps of $15^{\circ}$.}}
        \label{DiscRot}
\end{figure}
Theory and observations show a strong correlation with $n\propto B^{-\kappa}$ between magnetic field strength and density over several orders of magnitude \citep[][]{2011MNRAS.417.1054S,1966MNRAS.133..265M,1999ApJ...520..706C}. Here, we assume $\kappa = 0.6$ to keep the magnetic field strength between $10^{-8}\ \rm{T}$ in the center and $10^{-17}\ \rm{T}$ at the edges of the model space. The magnetic field is assumed to be dragged with the rotation of the disk because of ionized gas. Therefore, the field morphology can be approximated by a toroidal field \citep[see][]{2014A&A...566A..65R}.\\
Here, the considered dust consists  of only astro-silicate as material with a cut-off in grain size at $a_{\rm{min}}=5\ \rm{nm}$ and $a_{\rm{max}}=2\ \rm{\mu m}$, respectively with a power-law size distribution of $dn(a)\propto a^{-3.5} da$.\\
For the dust heating, we consider an offset dust temperature of $10\ \rm{K}$ within the boundaries of the disk and post-process the temperature with POLARIS and a central star as a radiation source for all disk models. The offset dust temperature is of minor influence to the following calculations but ensures a non-zero temperature in each cell in the denser regions of the model space, which enables us to reduce the number of photon packages. The model of \cite{1973A&A....24..337S} assumes the dust to be in thermal equilibrium with the gas ($T_{\rm{g}}=T_{\rm{d}}$). In this case, Eq. \ref{eq:IDGzetha} reaches unity and the IDG Rayleigh reduction factor (see Eq. \ref{eq:IDGfinal}) is undefined. Subsequently, the IDG mechanism cannot be applied in the disk models. Hence, in this section, we only consider   the effects of grain alignment according to RAT theory and assume the internal alignment to be perfect. As a second step, we apply the RAT mode of POLARIS to calculate the anisotropy and energy density of the radiation field in the disk, as described in Sect. \ref{sect:RAT}, to determine the minimal dust grain radius $a_{\rm{alg}}$ at which the dust grains start to align.\\
\subsubsection{Results}
Fig. \ref{DiskLayer} shows the resulting distribution of the $\rm{D03}$ and $\rm{D05}$ disk models. The distribution of aligned radii is layered with smaller aligned dust grains, which are starting to align at the disk's surface with larger grains towards the center plane. The model with the higher disk mass $\rm{D03}$ on the left hand side of Fig. \ref{DiskLayer} shows a distinct distribution of imperfectly aligned dust grains. In contrast to model $\rm{D03}$ for model $\rm{D05}$ on the right hand side of Fig. \ref{DiskLayer}, smaller dust grains are better aligned at the surface of the disk. However, with decreasing mass, the alignment of dust grains increases quickly towards the center plane and the outer regions. This is because of different dust and stellar contributions to the radiation field as well as a local dust temperature.\\
The optical depth decreases with lower disk mass. Therefore, the surface is more efficiently heated in the disks with higher mass. Whereas the grain alignment is suppressed near the sublimation radius for all models as a result of high temperatures in this region, the stellar radiation becomes more irrelevant in the outer regions of the disk. Consequently, the $\rm{D03}$ model is less influenced by the central star. The radiation in the innermost regions in each disk originates from thermal photons emitted in the hot surface layers and not the star itself. Subsequently, the higher mass model $\rm{D03}$ has aligned dust grains in much deeper layers as the lower mass model $D6$. The resulting layered distribution in dust grain alignment is consistent with the results of \cite{2014MNRAS.438..680H}.\\
In Fig. \ref{DiscRot}, we show the effects of thermal dust re-emission, which is dependent on the total disk mass. We compare the resulting polarization maps for the disk models $\rm{D03}$ and $\rm{D05}$ as a function of inclination angle at an exemplary wavelength of $\lambda = 515\ \rm{\mu m}$. The pattern and orientation of linear polarization in Fig \ref{DiscRot} are in good agreement with the predictions by \cite{2005ApJ...631..361C} and \cite{2007ApJ...669.1085C}. Since all disk models are optically thin at that wavelength, the synthetic polarization maps appear to be symmetric.\\
Because lower values of the characteristic size of aligned grains $a_{\rm{alg}}$ are associated with a higher degree of linear polarization (see Eq. \ref{eq:Rrat}), the $\rm{D03}$ disk model shows a more distinct pattern of linear polarization. A lower degree of linear polarization in low mass disks is consistent with the results shown in Fig. \ref{DiskLayer}. This would also hold even if we considered imperfect internal alignment since the internal distribution function for RATs is independent of the local physical conditions of the disk models and just a function of grain geometry (see Sect. \ref{eq:Rrat}). The pattern in the normalized polarization vectors are identical for all models, which reveals the projected toroidal magnetic field geometry at all inclination angles since the influence of scattering can be neglected at $\lambda = 515\ \rm{\mu m}$. However, just for the $\rm{D03}$ model, the degree of linear polarization ($P_{l}>0.5\%$) should be detectable in the entire disk, even for low inclination angles. \\
The patterns of linear polarization have sharp cut-offs with no polarization at the borders of the model. This is a result of the applied power-law correlation between density and magnetic field strength ($n\propto B^{-\kappa}$). Hence, the magnetic field  no longer satisfies the minimal required field strength (see Eq. \ref{eq:larm}) in the thinner disk regions.
\subsubsection{Discussion}
Aligned dust grains as an explanation of polarized radiation in the FIR and submm from disks has already been indicated by the observations of \cite{1999ApJ...525..832T}. However,  the extent  to which the observation of polarized light  reveals the underlying magnetic field morphology is still a matter for debate. Numerical calculations  (\cite{2007ApJ...669.1085C,2014MNRAS.438..680H}) predict that polarized radiation should emerge from the circumstantial disk region because of RAT alignment.
In contrast to predictions, observations of circumstellar systems \citep[][]{2009ApJ...704.1204H} show no detectable degree of linear polarization at all. The overestimation of RAT alignment, because of perfect internal alignment in earlier studies  \citep[][]{2007ApJ...669.1085C}, cannot account for this finding alone. An internal alignment with a conservative chosen parameter $f_{\rm{high-J}} = 0.25$ would still result in a detectable degree of linear polarization. However, $f_{\rm{high-J}}$ can still not be analytically determined and might also vary depending on the predominant local conditions of each disk. Another explanation might be the shape and composition of the dust grains. We cannot  expect  a similar amount of non-spherical dust grains that can align to be equally present in all circumstantial disks since this may vary significantly from disk to disk.\\
The distribution of dust grain sizes is also not expected to be equal in the entire disk. Dust grain growth and dust settling are processes that redistribute dust grains of different sizes in the disk \citep[see e.g.][]{2005ApJ...625..414T}. As a result, larger dust grains are more likely near the disk plane. Larger dust is also more likely to be aligned by RATs than  smaller dust grains on the disk surface. Dust settling in disks might also help to explain the  low amount of linear polarization and need to be considered in future studies with a more sophisticated dust model.

\subsection{Cloud model}
\label{setupCloud}

\begin{table*}[ht]
\begin{center}
\begin{tabular}[l]{ l | c | c | c | c | c | c | p{6.5cm} }
 Parameter & $n_{g1}\ [\rm{m^{-3}}]$ & $R_{c1}\ [\rm{ly}]$ & $n_{g2}\ [\rm{m^{-3}}]$ & $R_{c2}\ [\rm{ly}]$ & $R_{\rm{*}}\ [\rm{R_{\odot}}]$ & $T_{\rm{*}}\ [\rm{K}]$  & position of the star  $[ly]$ \\ \hline
 Value & $10^{13}$ & $0.2$ & $10^{12}$ & $0.1$ & $8$ & $6000$ & C01(-0.53,-0.0,-0.0), $\quad$ C02(0.0,0.0,0.13) \\
\end{tabular}

\end{center}
\caption{Table of physical parameter for the applied cloud models.} 
\label{tab:CM}
 \end{table*}

 \begin{figure*}[]
\begin{minipage}[c]{0.329\linewidth}
                        \begin{center}
                                \includegraphics[width=0.99\textwidth]{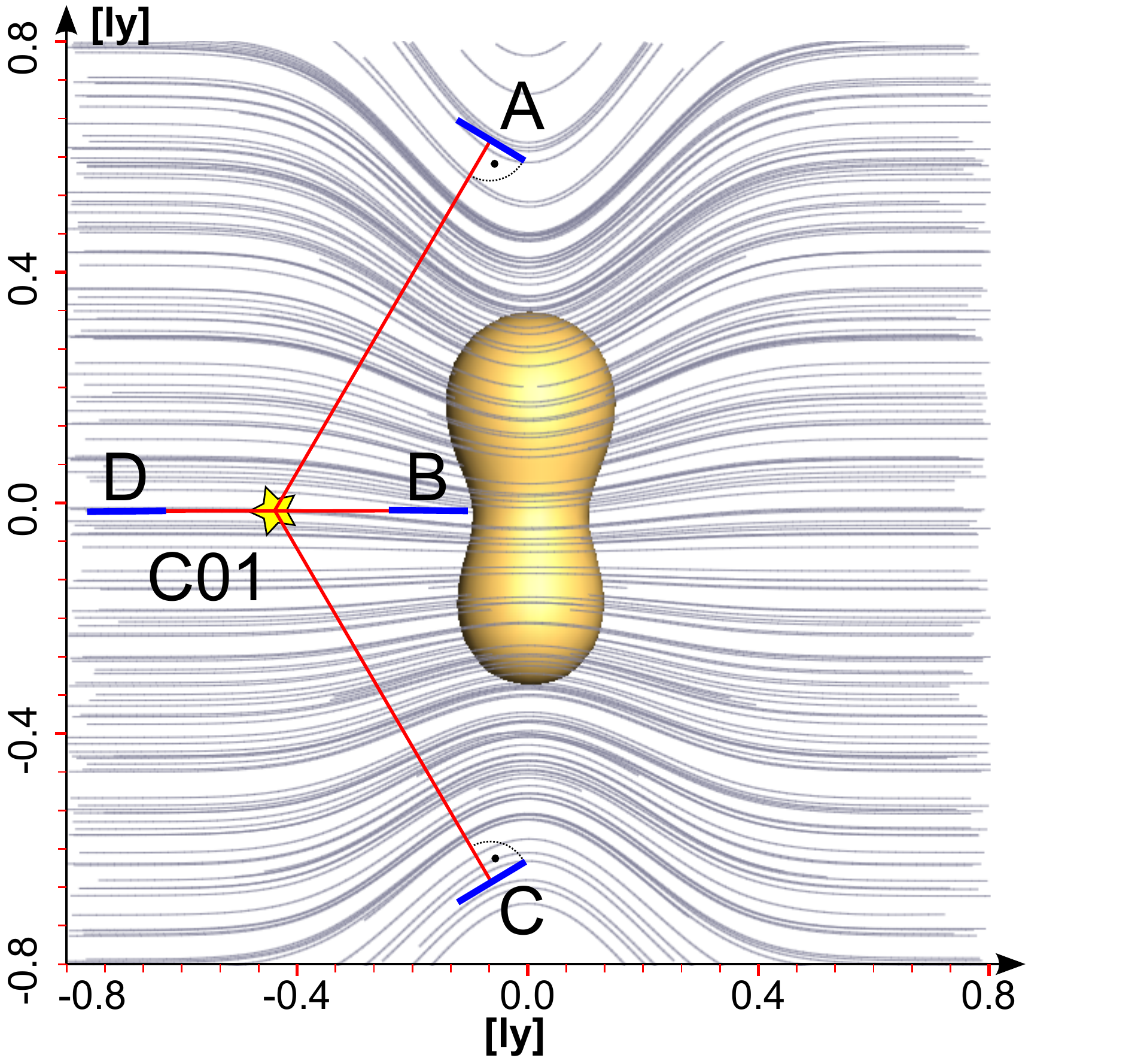}
                        \end{center}
                \end{minipage}
                \begin{minipage}[c]{0.329\linewidth}
                        \begin{center}
                                \includegraphics[width=0.99\textwidth]{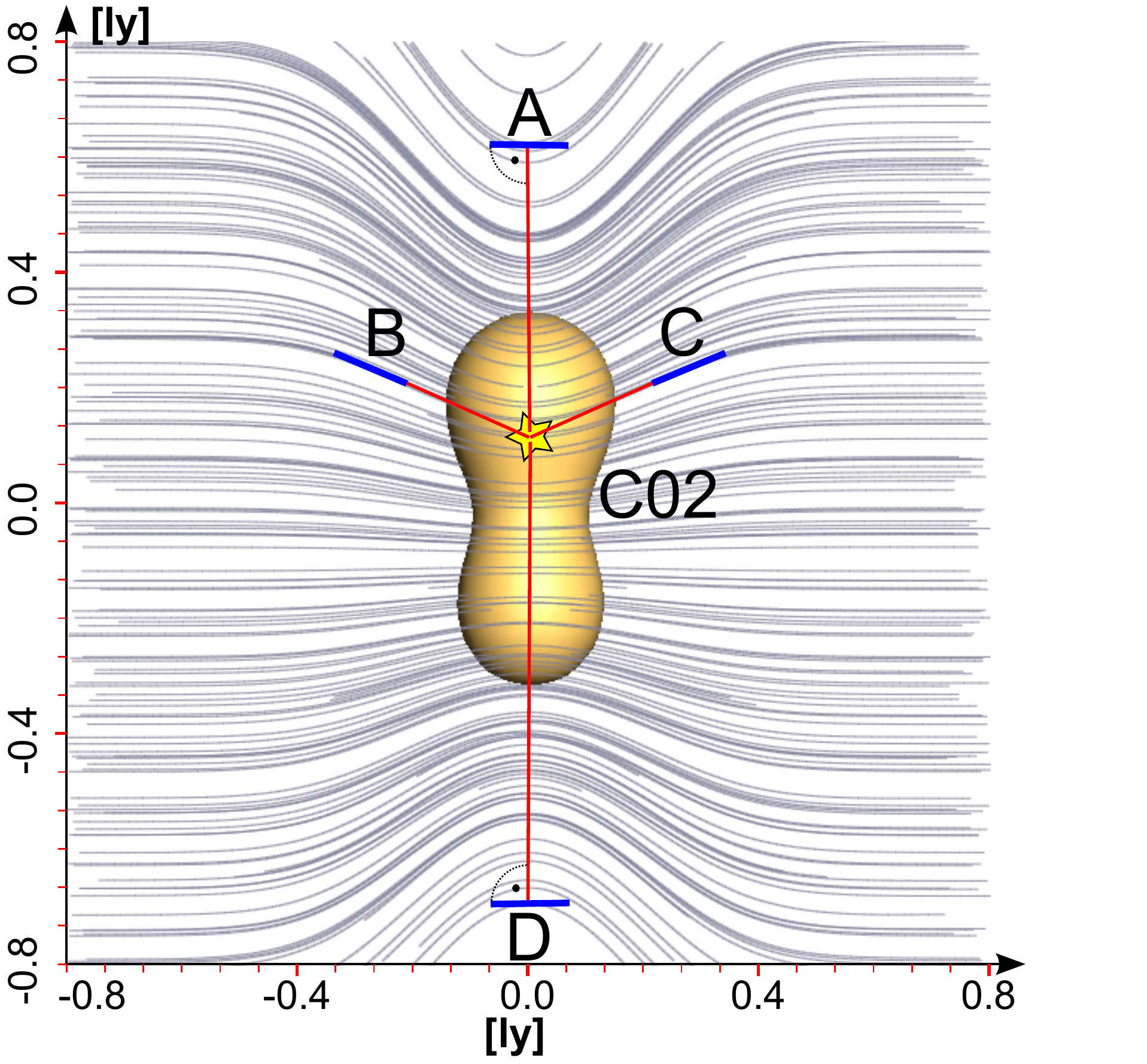}
                        \end{center}
                \end{minipage}  
                                \begin{minipage}[c]{0.329\linewidth}
                        \begin{center}
                                \includegraphics[width=0.99\textwidth]{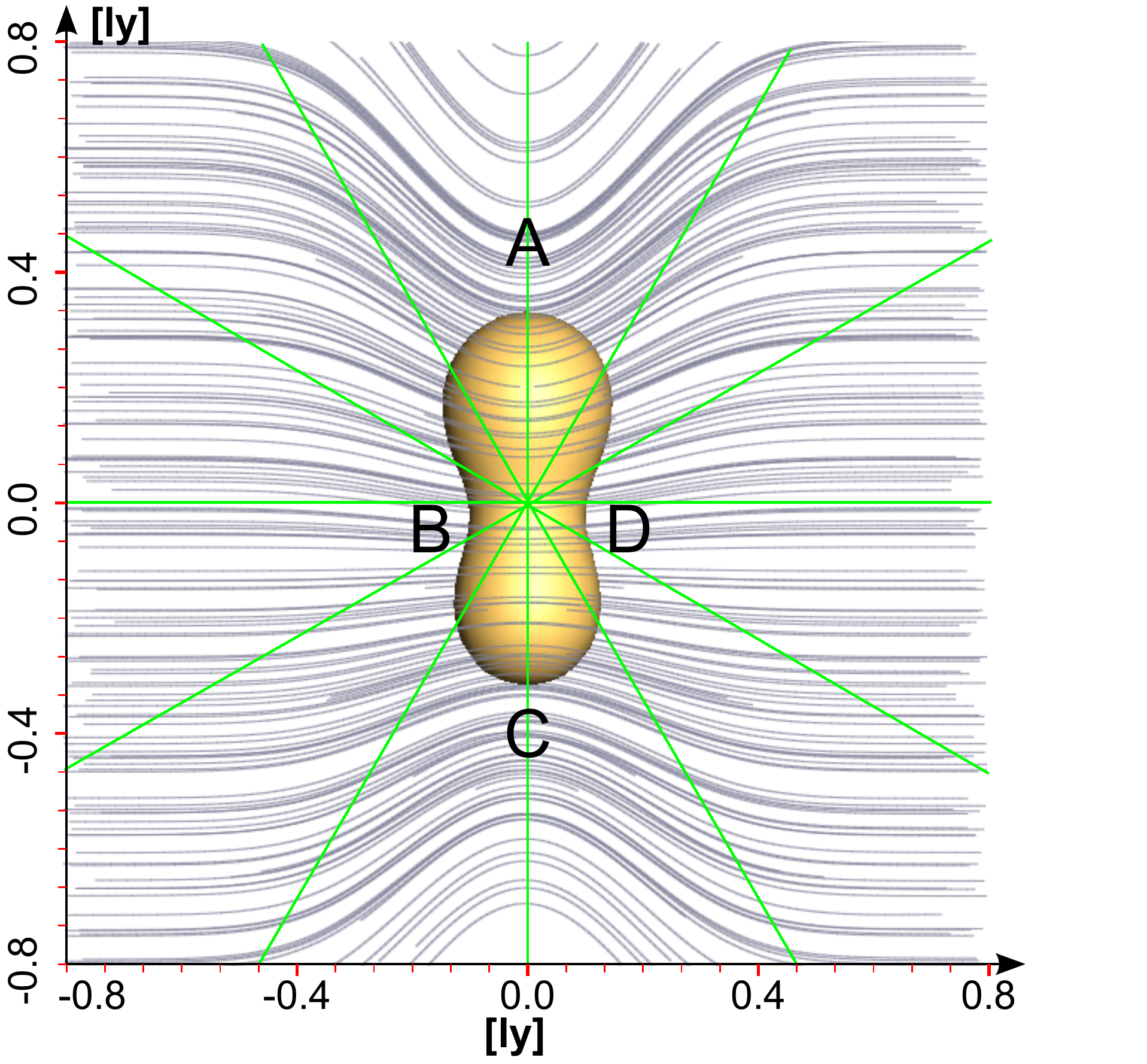}
                        \end{center}
                \end{minipage}  
                                
        \caption{{Schematic illustration of the cloud models $\rm{C01}$ (left panel) and $\rm{C02}$ (right panel).} Both models differ only by the position of a single star. The isocontour surface at $log(n_{\rm{d}})=3.27\ \rm{m^{-3}}$ is in dark yellow. The direction of the common gas stream $\vec{v}$ (left panel) toward the center is shown in green lines and the  analytical hourglass field morphology is shown in grey. Red lines and blue bars indicate the different angles between the radiation field and the magnetic field direction. The letters $A - D$ indicate areas with expected characteristic features for RAT and GOLD alignment, respectively.}
        \label{fig:globModel}
\end{figure*}

\begin{figure*}[]
\begin{center}

        \begin{minipage}[c]{0.32\linewidth}
                        \begin{center}
                                \includegraphics[width=1.0\textwidth]{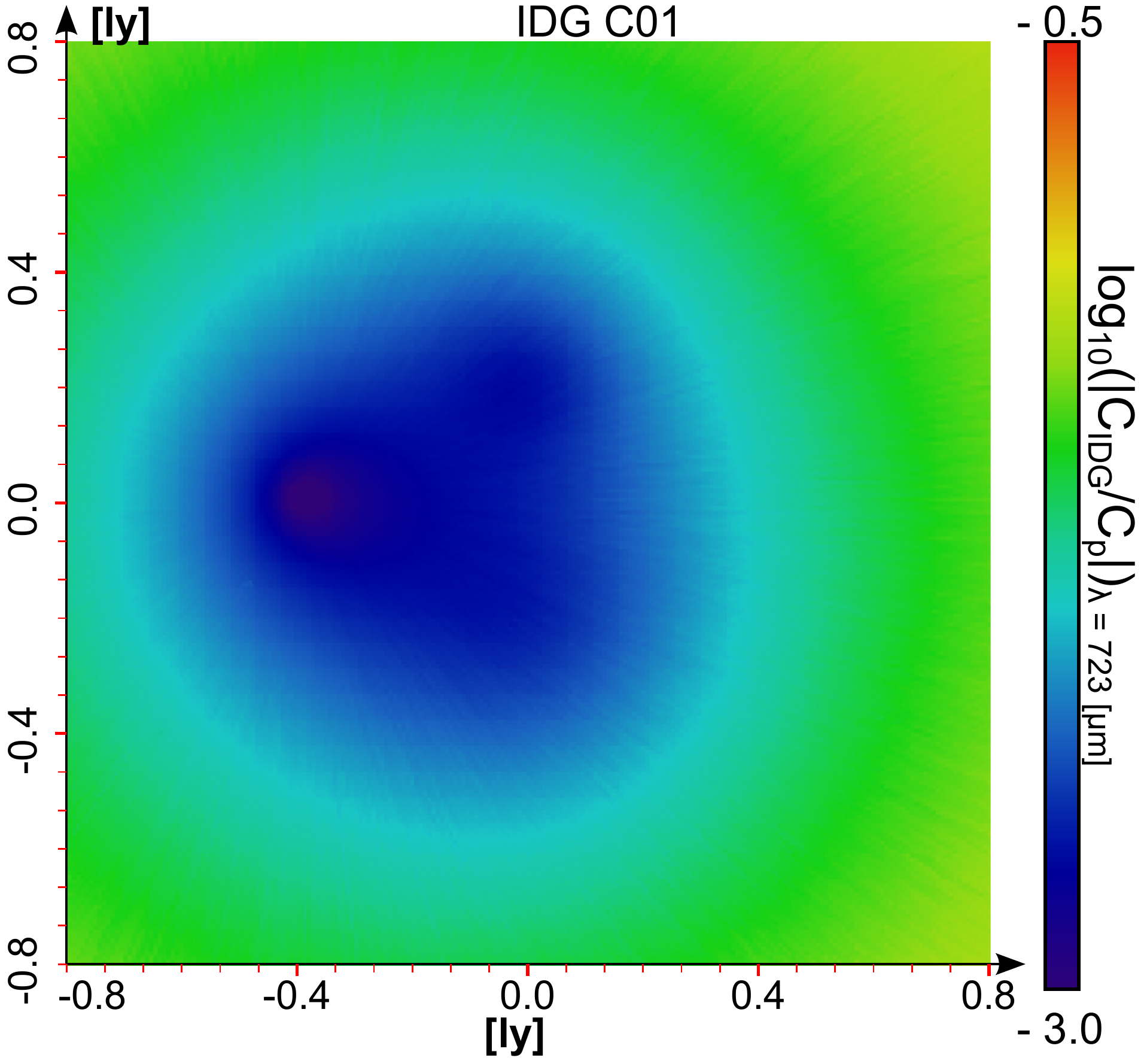}\\
                                \includegraphics[width=1.0\textwidth]{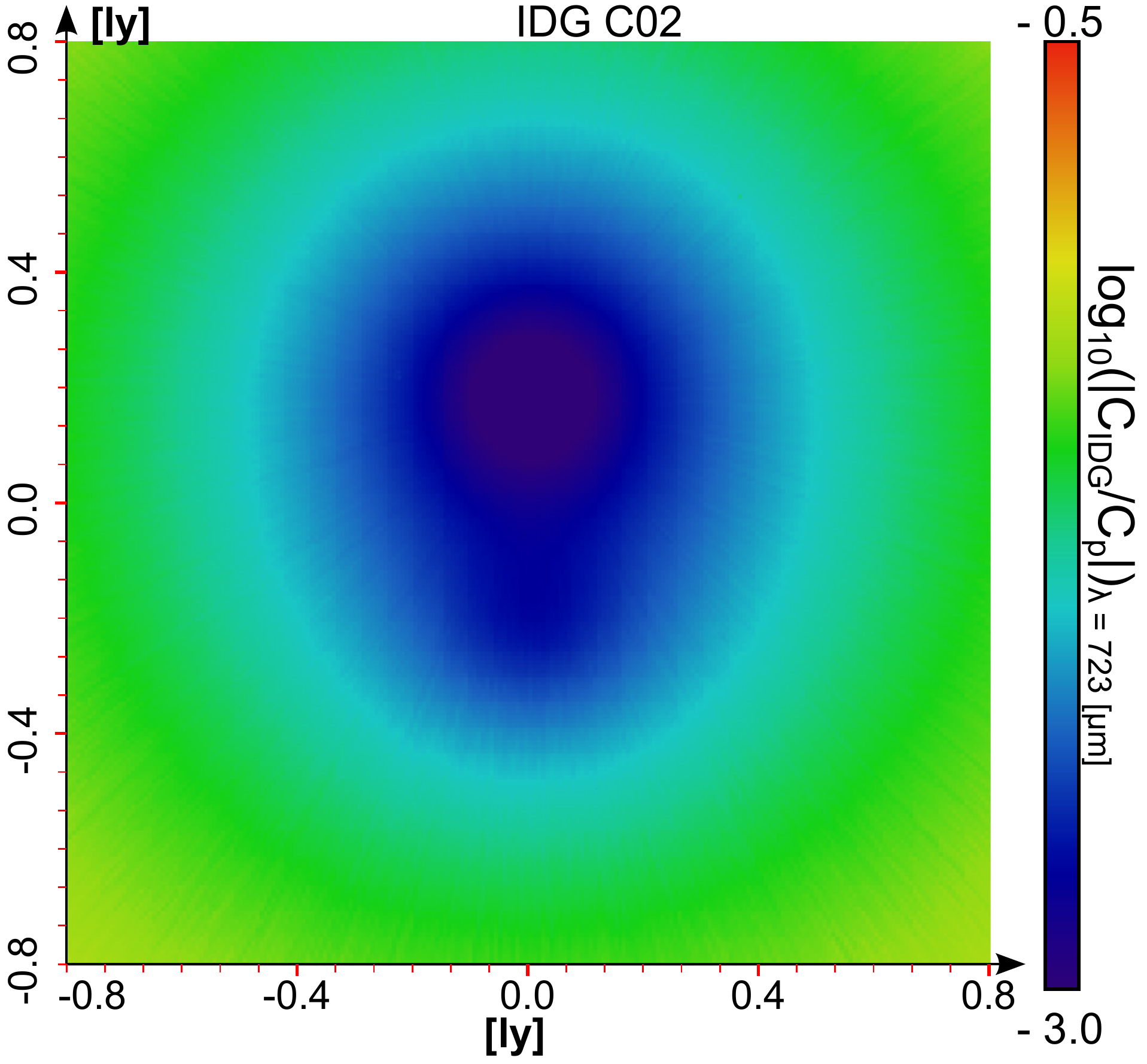}
                        \end{center}
                \end{minipage}
                                \begin{minipage}[c]{0.32\linewidth}
                        \begin{center}
                                \includegraphics[width=1.0\textwidth]{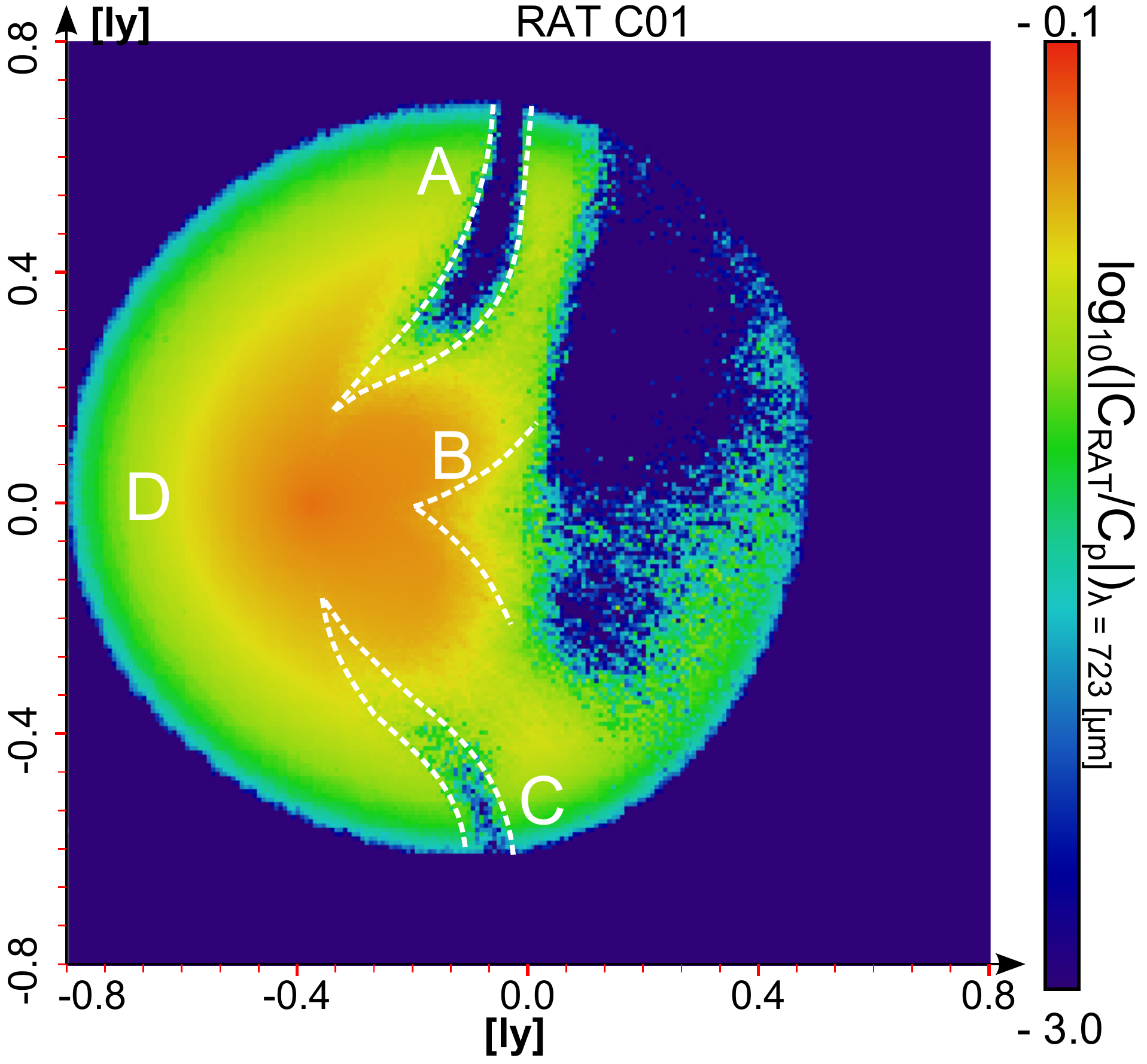}\\
                                \includegraphics[width=1.0\textwidth]{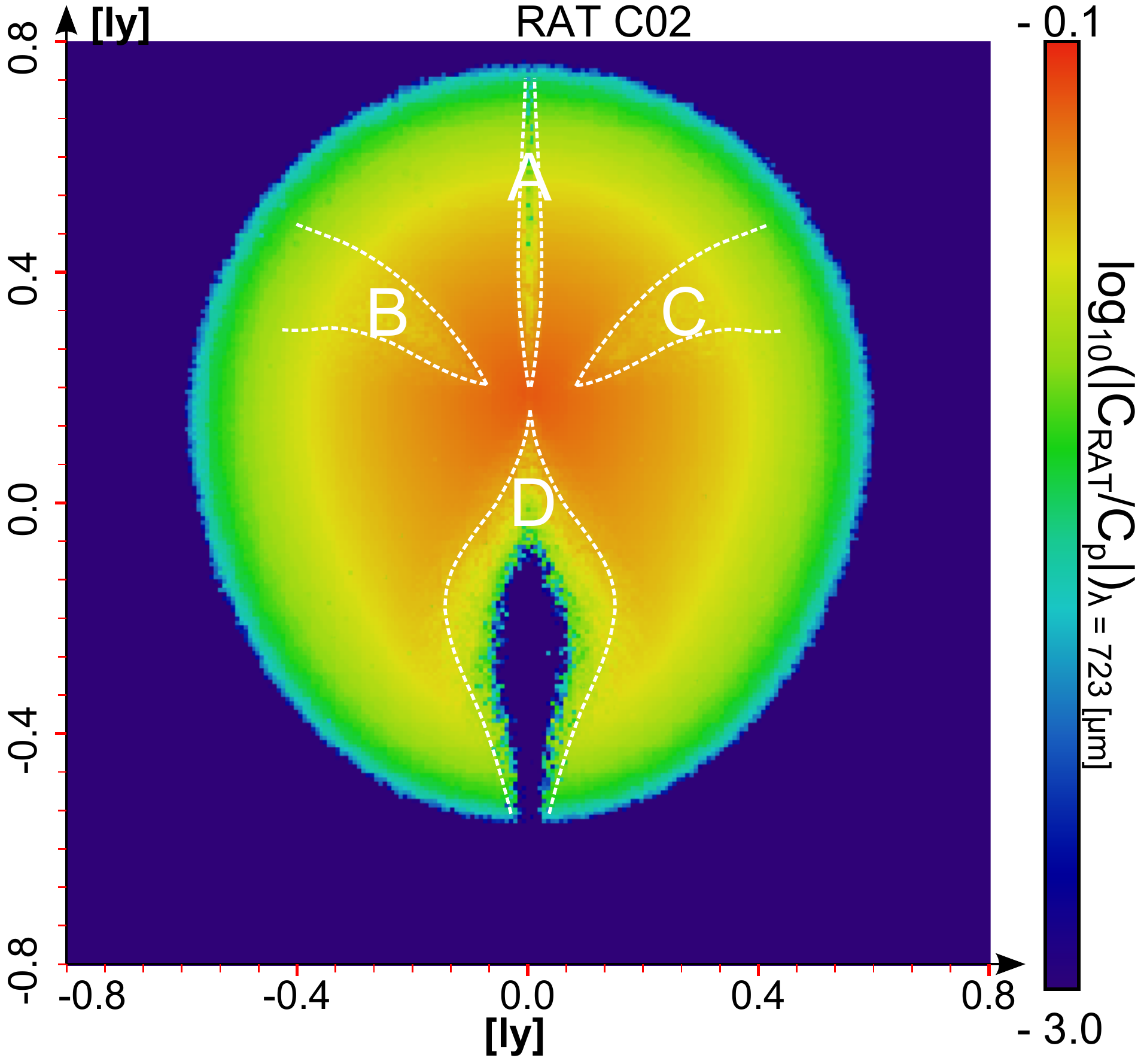}
                        \end{center}
                \end{minipage}  
                \begin{minipage}[c]{0.32\linewidth}
                        \begin{center}
                                \includegraphics[width=1.0\textwidth]{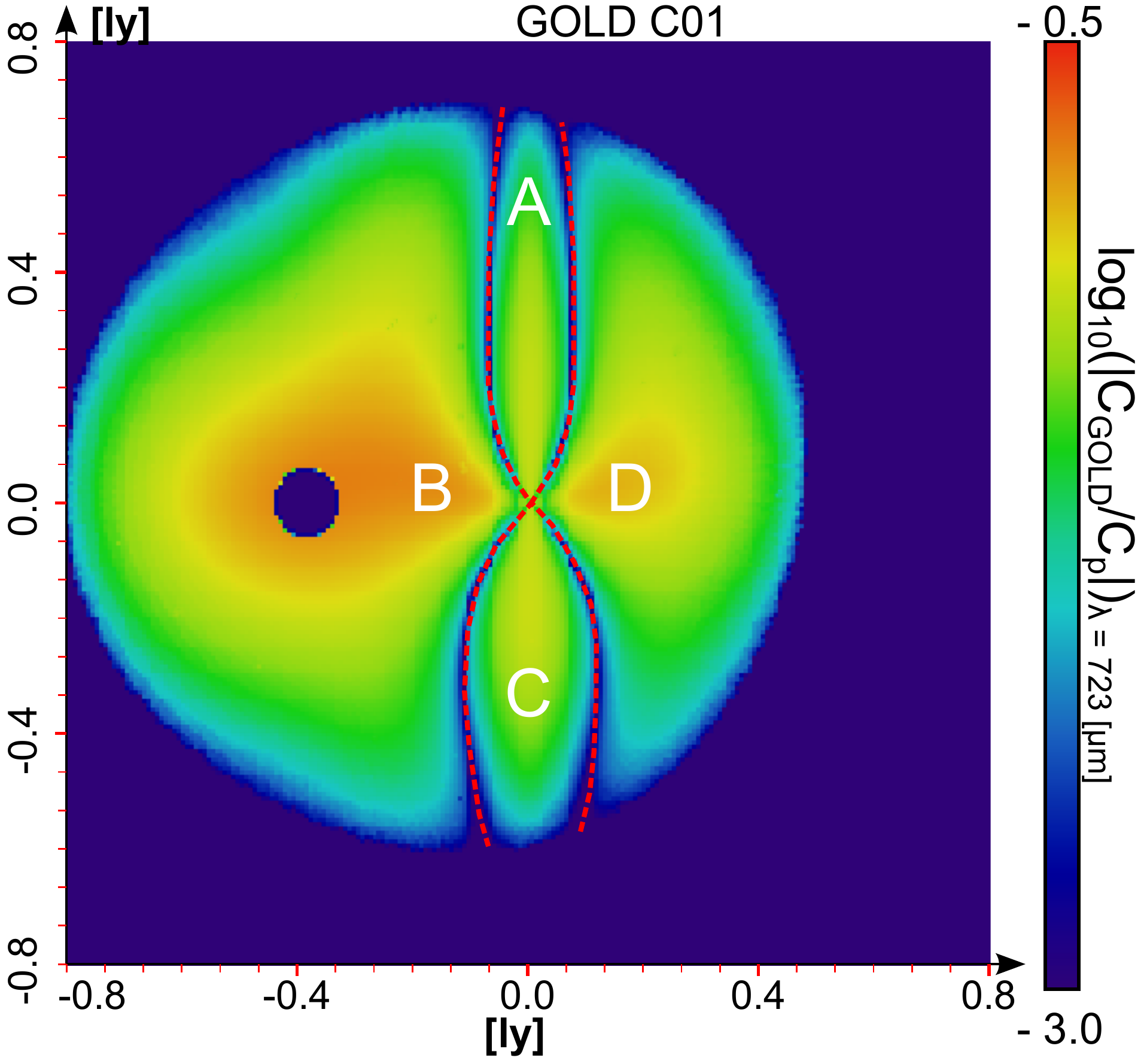}\\
                                \includegraphics[width=1.0\textwidth]{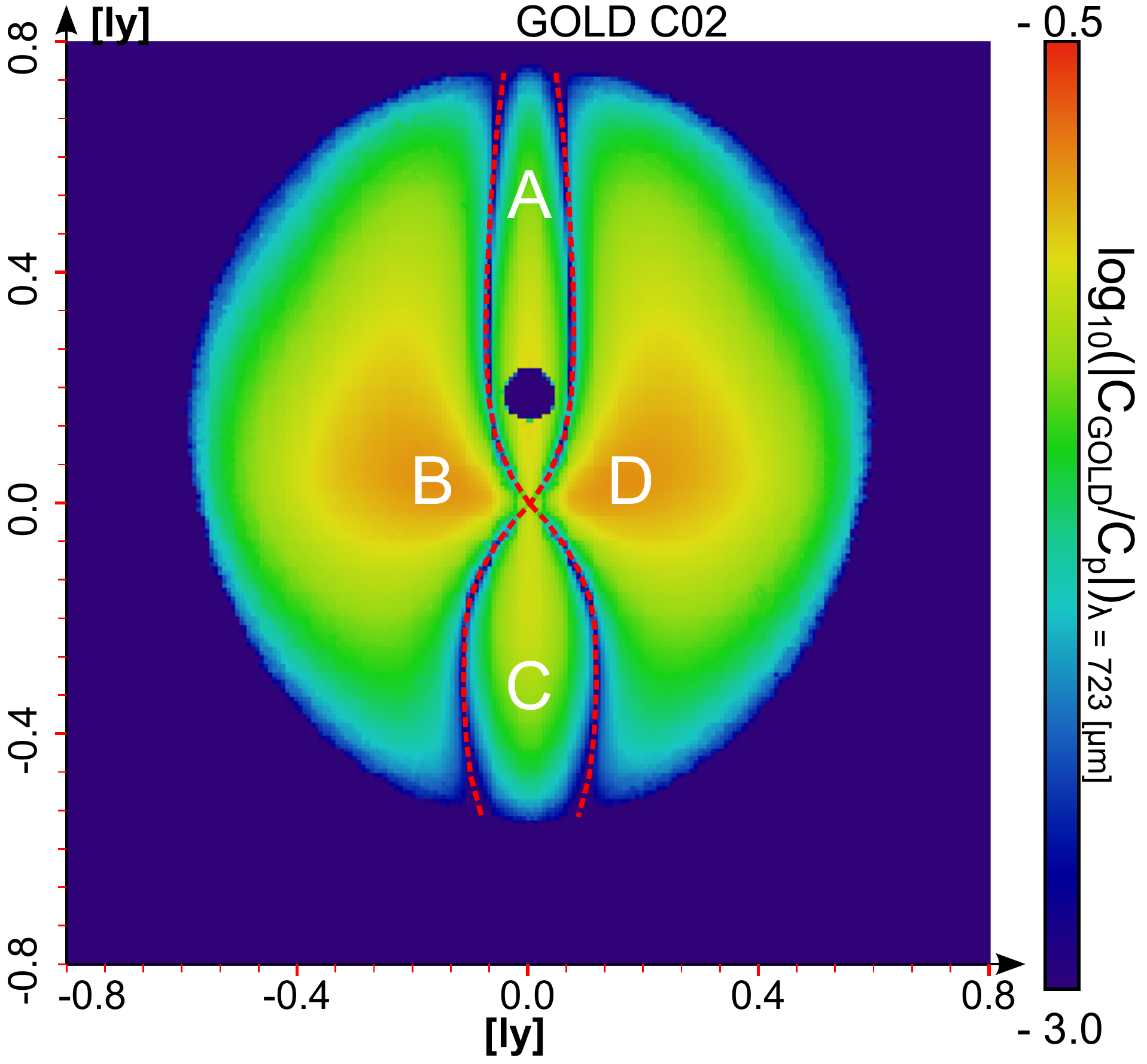}
                        \end{center}
                \end{minipage}
\end{center}
        
\caption{\small Ratio of polarization cross-sections $\log_{10}\left(  \left|  C_{\rm{x}} / C_{\rm{p}} \right| \right)_{\lambda = 723\ \rm{\mu m}}$ of imperfectly aligned dust grains $C_{\rm{x}}$, to perfectly aligned dust grain $C_{\rm{p}}$ in the mid-plane of the model space for the $C01$ model (top row) and $C02$ model (bottom row). The $x$ stands for IDG alignment (left columns), RAT alignment (middle columns), and GOLD alignment (right columns). In the middle columns, dotted white lines indicate the regions where the alignment efficiency $Q_{\Gamma}(\epsilon)$ is at its minimum. The dotted red lines in the right columns show the transition where GOLD alignment  changes the sign of linear polarization. The areas $A - D$ correspond to that of Fig. \ref{fig:globModel}.}
\label{fig:Cross}
\end{figure*}
In this section, we examine the pattern of linear polarization emerging from different grain alignment theories separately. The main focus here is on the ambiguities associated with different grain alignment theories used to infer the underlying magnetic field morphology on a predefined cloud model.\\
Our cloud model consists of two spheres following a Bonnor-Ebert density profile \citep[][]{1955ZA.....37..217E,1956MNRAS.116..351B} with a distance of $0.375\ \rm{ly}$ from each other. The characteristic radii of the density distribution are $0.1\ \rm{ly}$ and $0.2\ \rm{ly}$, with a number density for the gas of $10^{12}\ \rm{m^{-3}}$ and $10^{13}\ \rm{m^{-3}}$, respectively. Density and temperature are chosen to be consistent with theoretical models and observations \citep[][]{1997A&A...326..329L,2011A&A...535A..49S}.
\subsubsection{Model setup}
We modeled an hourglass-shaped magnetic field geometry by using an analytical function \citep[see][]{2014A&A...566A..65R}, with its center in the denser sphere as expected by observations \cite[e.g.][]{1999ApJ...525L.109G,2011AA...535A..44F}. The field strength was assumed to be constant with $\left|\vec{B}\right|=1.5\times 10^{-8} \rm{T}$ \citep[][]{2008A&A...487..247F,2010ApJ...724L.113B}. To compare calculations performed by POLARIS with the expected predictions for the GOLD alignment mechanism, we also assume a supersonic velocity stream $\vec{v}$ with an in-fall direction to the common center of both Bonnor-Ebert spheres. In this section, we also consider the effects of the internal alignment with a correlation factor of $f_{\rm{c}} = 0.6$ and assume $f_{\rm{high-J}}=0.5$. The initial dust temperature is adjusted with a single star as radiation source (see Tab. \ref{tab:CM}). We consider two different configurations concerning the position of the star. In model $C01,$ the star is embedded in the center of the denser Bonnor-Ebert sphere, and in model $C02,$  the star is $0.53\ \rm{ly}$ away from the common center of both spheres. A schematic illustration of the cloud models provided in Fig. \ref{fig:globModel}.\\
For the dust model, we apply a range of effective radii of $ a \in[5\ \rm{nm}: 250\ \rm{nm}]$ with a size distribution of $n_{\rm{d}} \propto a^{-3.5}$ and consider a mixture of   $62.5\%$ silicate and $37.5\%$ graphite \citep[][]{1977ApJ...217..425M,2001ApJ...551..807D}. However, just the silicate grains align with the direction of the magnetic field while the graphite remains randomized \citep[][]{1977ApJ...217..425M}. An initial sphere temperature of $T_{\rm{d}} = 10\ \rm{K}$ was used for both cloud models, $C01 $ and $C02,$ and we corrected the temperature according to the method described in Sect. \ref{setupMHDHeat}. Here, we assume dust and gas not to be  in thermal equilibrium and assume a correaltion of $T_{\rm{g}} = 10 \times T_{\rm{d}}$. In a final step, we applied the RAT mode of POLARIS to calculate the distribution of the size cutoff $a_{\rm{alg}}$ of dust grains to align. With this configuration the models allow the calculation of the characteristic linear polarization pattern for IDG, RAT, and GOLD alignment mechanism.
\subsubsection{Results}
Fig. \ref{fig:Cross} shows the ratio of cross-sections of partially aligned dust grains to the cross-sections of perfectly aligned dust grain in the mid-plane of both models separately for all alignment mechanisms at an exemplary wavelength of $\lambda = 723\ \rm{\mu m}$. In both cloud models, the surrounding regions of the stars are also the regions of highest dust temperature. The IDG mechanism reacts sensitively to both values, the density, and the temperature, because of the threshold $\delta_{\rm{0}}$ (see Eq. \ref{eq:delta}). Therefore the efficiency of grain alignment is lowest towards the center of the clouds and near the star (see Fig. \ref{fig:Cross} left panels).\\
The expected alignment pattern according to RAT theory is contrary to that of the IDG alignment. The emission of heated dust and stellar radiation contribute to the local energy density $u_\lambda$ and the anisotropy parameter can reach a maximum value up to $\gamma_\lambda = 0.9$ in close proximity to the star. The characteristic dust grain radius $a_{\rm{alg}}$ decreases and dust grains with radii above $a_{\rm{alg}}$ become aligned. Since, the process of determining  RAT alignment quantities requires two separate MC simulations (dust heating and calculation of the radiation field), the grain alignment of RATs has an inherent higher amount of noise. This becomes apparent in Fig. \ref{fig:Cross} (middle panel) in the inner cloud regions. Besides the amount of energy density and anisotropy, the grain alignment also highly depends   on the RAT alignment efficiency $Q_{\rm{\Gamma}}(\epsilon)$. This parameter, in turn, depends on the chosen dust grain geometry and has, in our dust model of oblate dust grains, its minimum at angles of $\epsilon = 0^{\circ}$ and $\epsilon = 90^{\circ}$ between radiation and magnetic field direction. This leads to areas of reduced grain alignment, indicated by white dotted lines in Fig.\ref{fig:Cross} (middle panel). In Fig.\ref{fig:Cross} (middle top panel), the low degree of grain alignment in the areas $A$ and $C$ is a direct result of the angle dependency of RAT theory because of $Q_{\rm{\Gamma}}(\epsilon)$. However, the low grain alignment on the left half of the panel is a result of the shielding of this region by the dense  clouds from stellar radiation and thus the overall energy density $\overline{u}$ remains low. The expected low grain alignment in the area $D$ cannot be found. This is because the overall energy density $\overline{u}$ outweighs the angle dependency in this very thin region. In Fig.\ref{fig:Cross} (middle bottom panel), the low degree of grain alignment in the areas $A - D$ follows   the angle between predominant direction of the radiation field and the magnetic field direction more clearly. This result is in excellent agreement with the expectations of the model setup shown in Fig. \ref{fig:globModel} and RAT theory.\\
\begin{figure}[]
        \begin{minipage}[c]{0.49\linewidth}
                        \begin{center}
                                \includegraphics[width=1.0\textwidth]{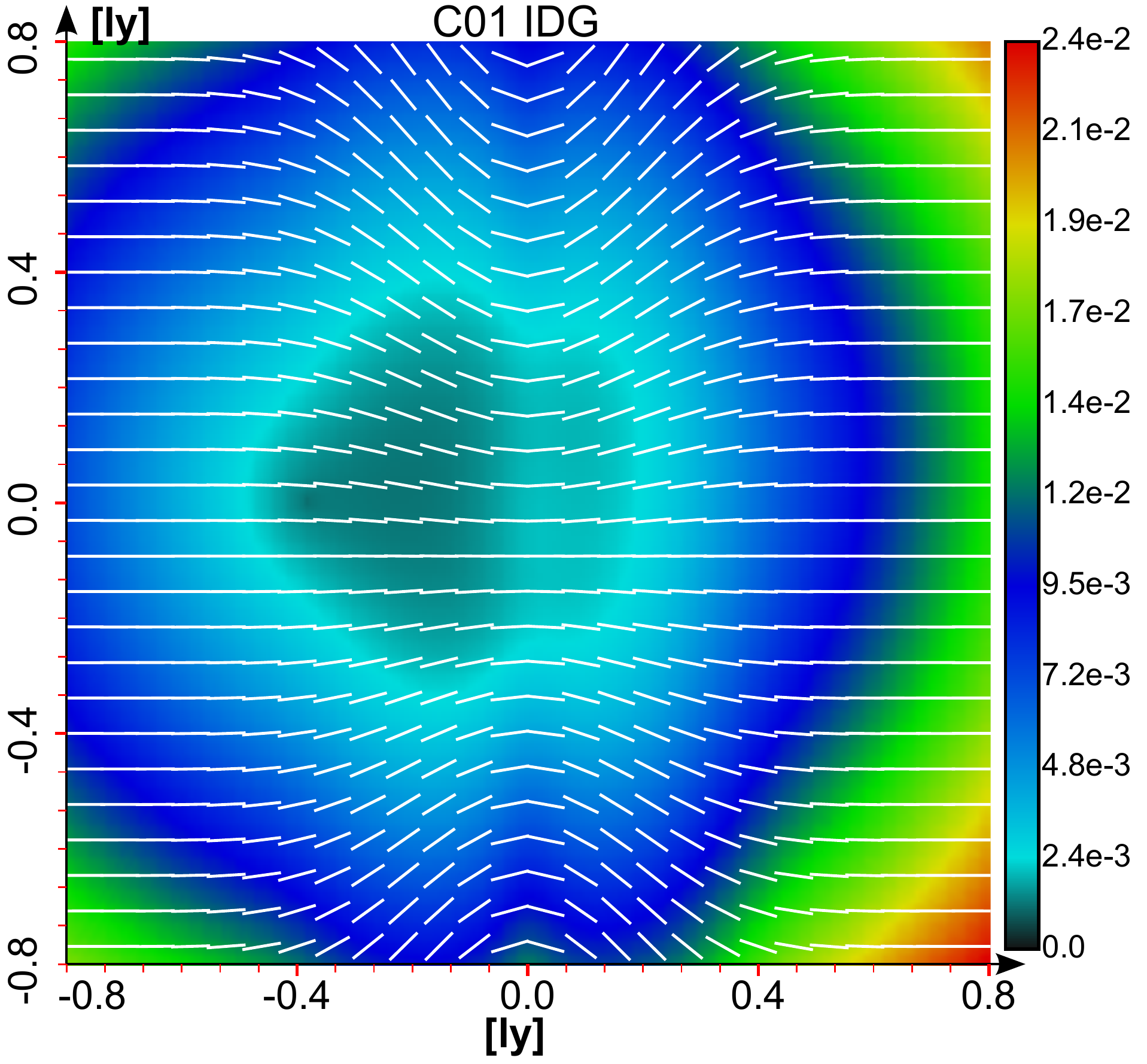}\\
                                \includegraphics[width=1.0\textwidth]{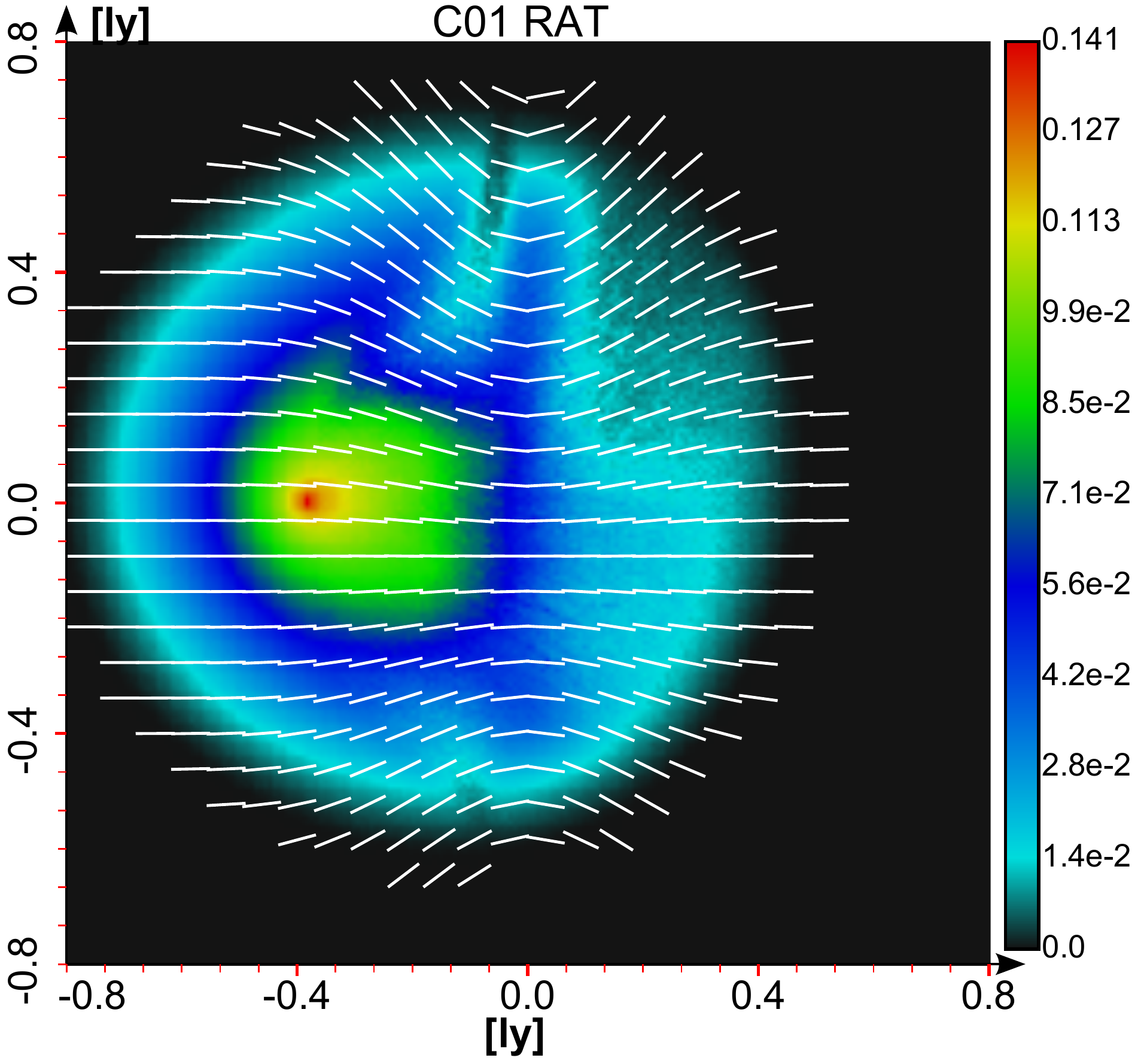}\\
                                \includegraphics[width=1.0\textwidth]{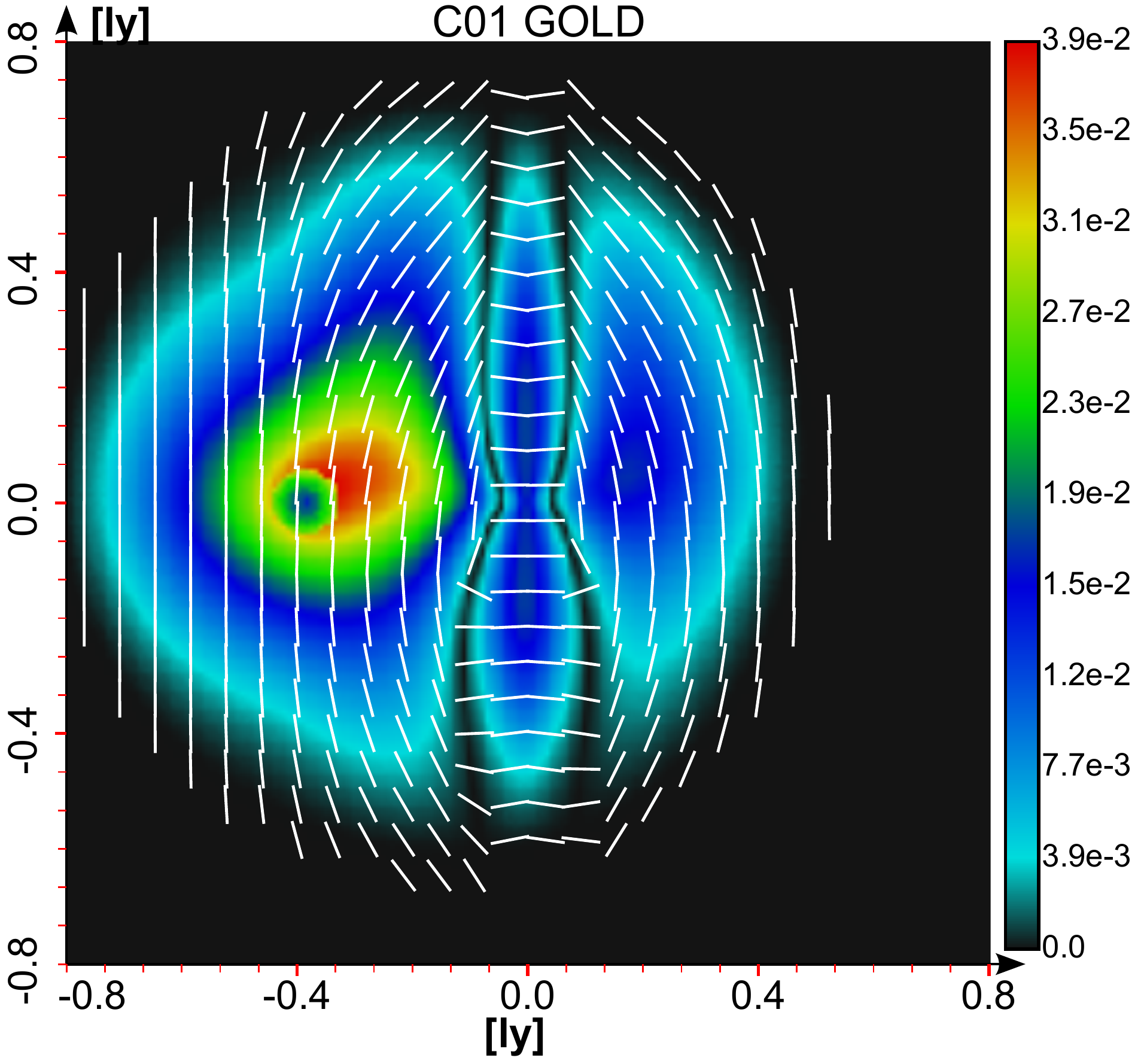}
                        \end{center}
                \end{minipage}
                        \begin{minipage}[c]{0.49\linewidth}
                        \begin{center}
                                \includegraphics[width=1.0\textwidth]{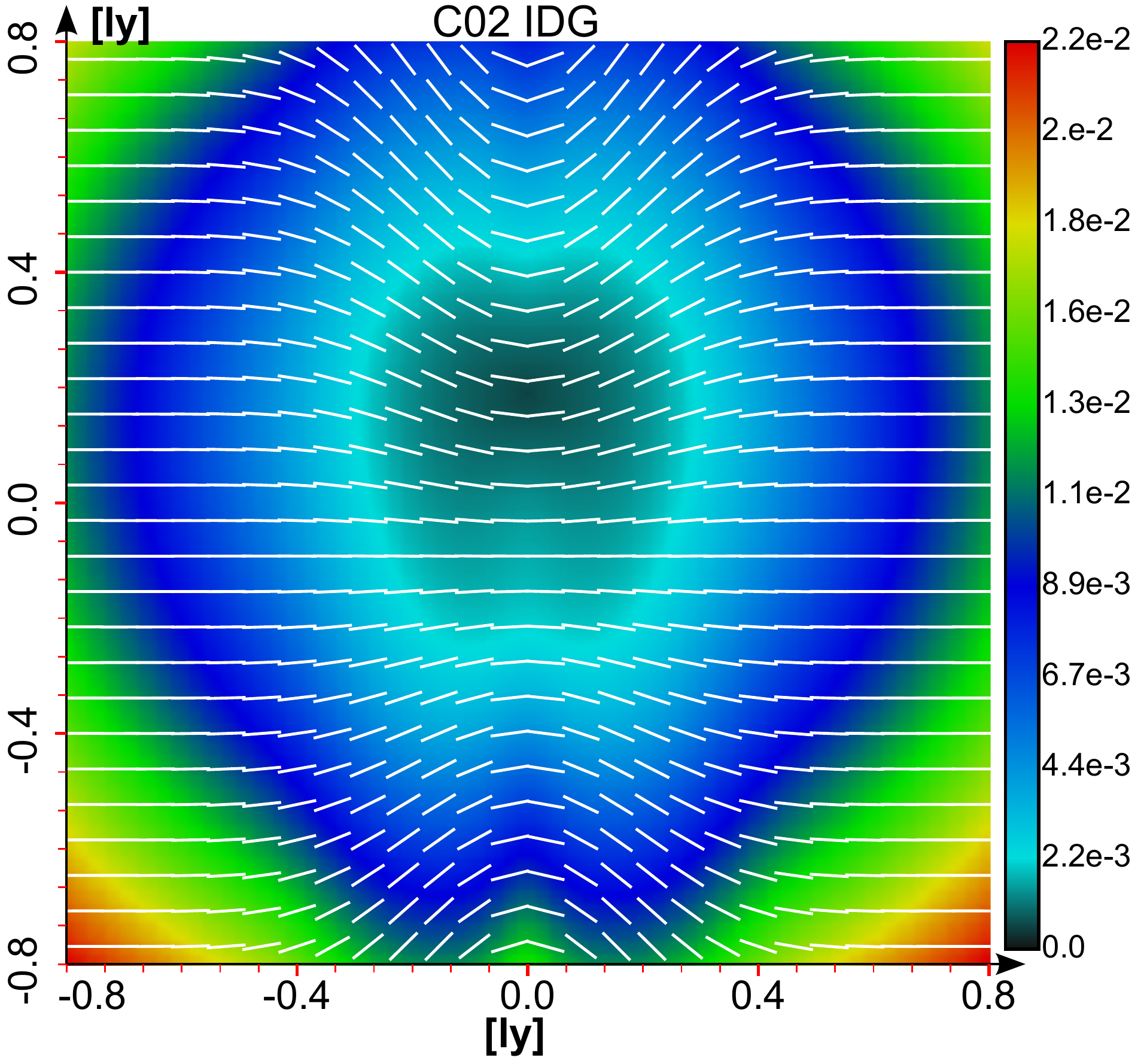}\\
                                \includegraphics[width=1.0\textwidth]{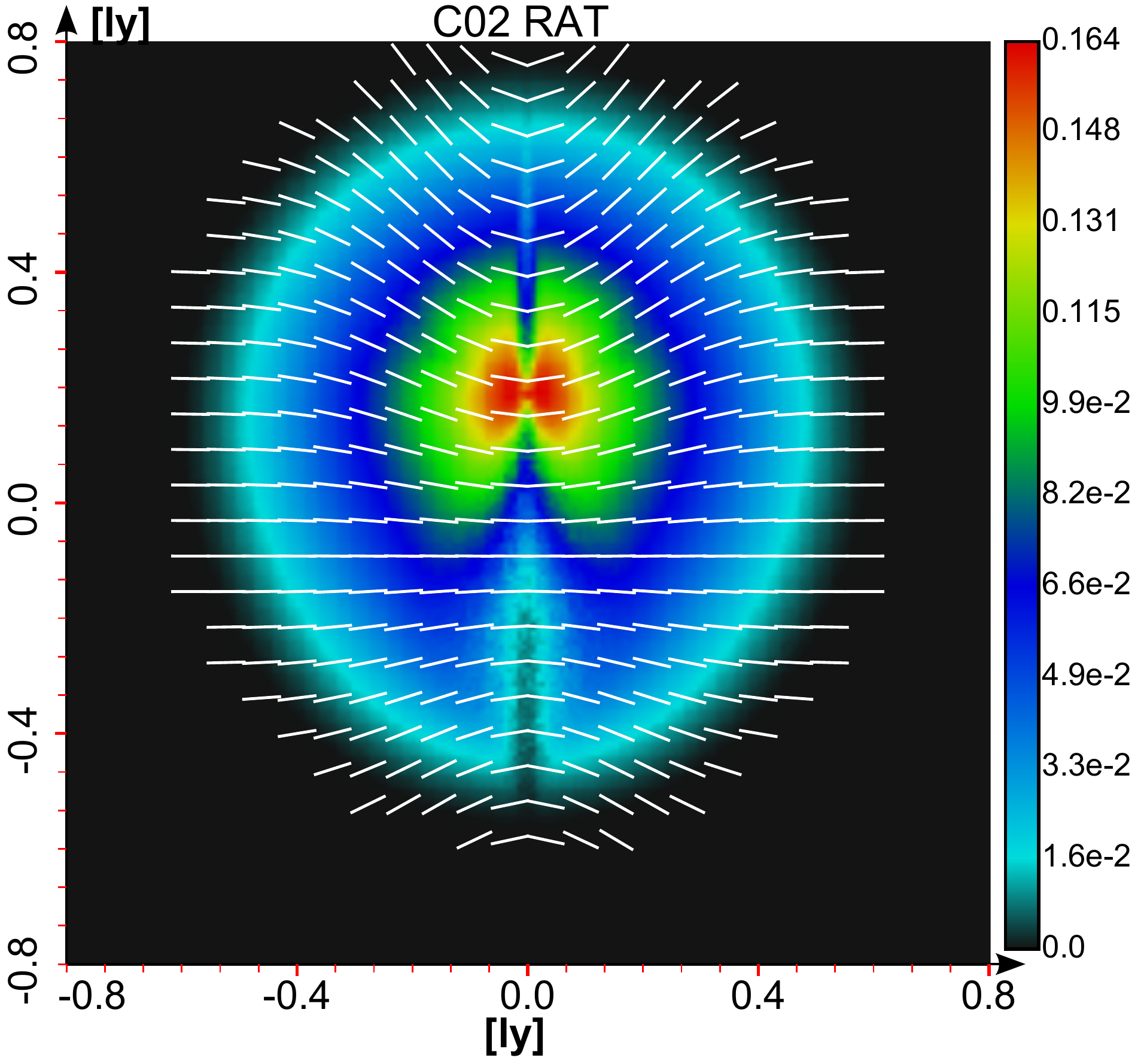}\\
                                \includegraphics[width=1.0\textwidth]{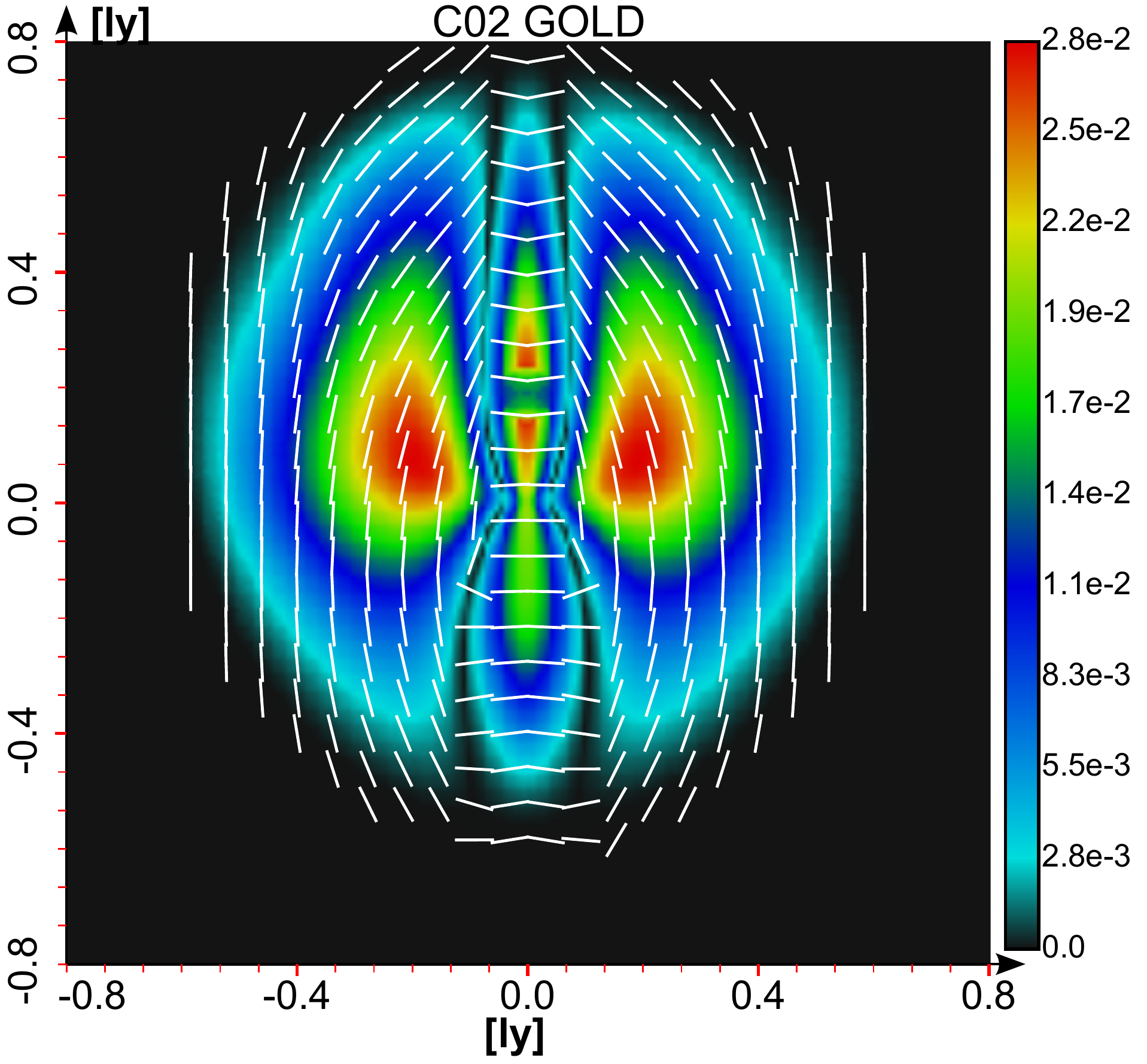}
                        \end{center}
                \end{minipage}
\caption{\small Maps with pattern of linear polarization overlaid with normalized orientation vectors at an exemplary wavelength of $\lambda= 723\ \rm{\mu m}$. We applied an offset of $90^{\circ}$ to the orientation vectors to match the projected hourglass magnetic field morphology. The cloud model (C01) is in the left column and the model (C02) is in the right column for dust grains aligned with the IDG mechanism (top row), RAT mechanism (middle row), and GOLD mechanism (bottom row).}
\label{CloudPolMap}
\end{figure}
\begin{figure*}[]
                \begin{minipage}[c]{1.0\linewidth}
                        \begin{center}
                                \includegraphics[width=0.85\textwidth]{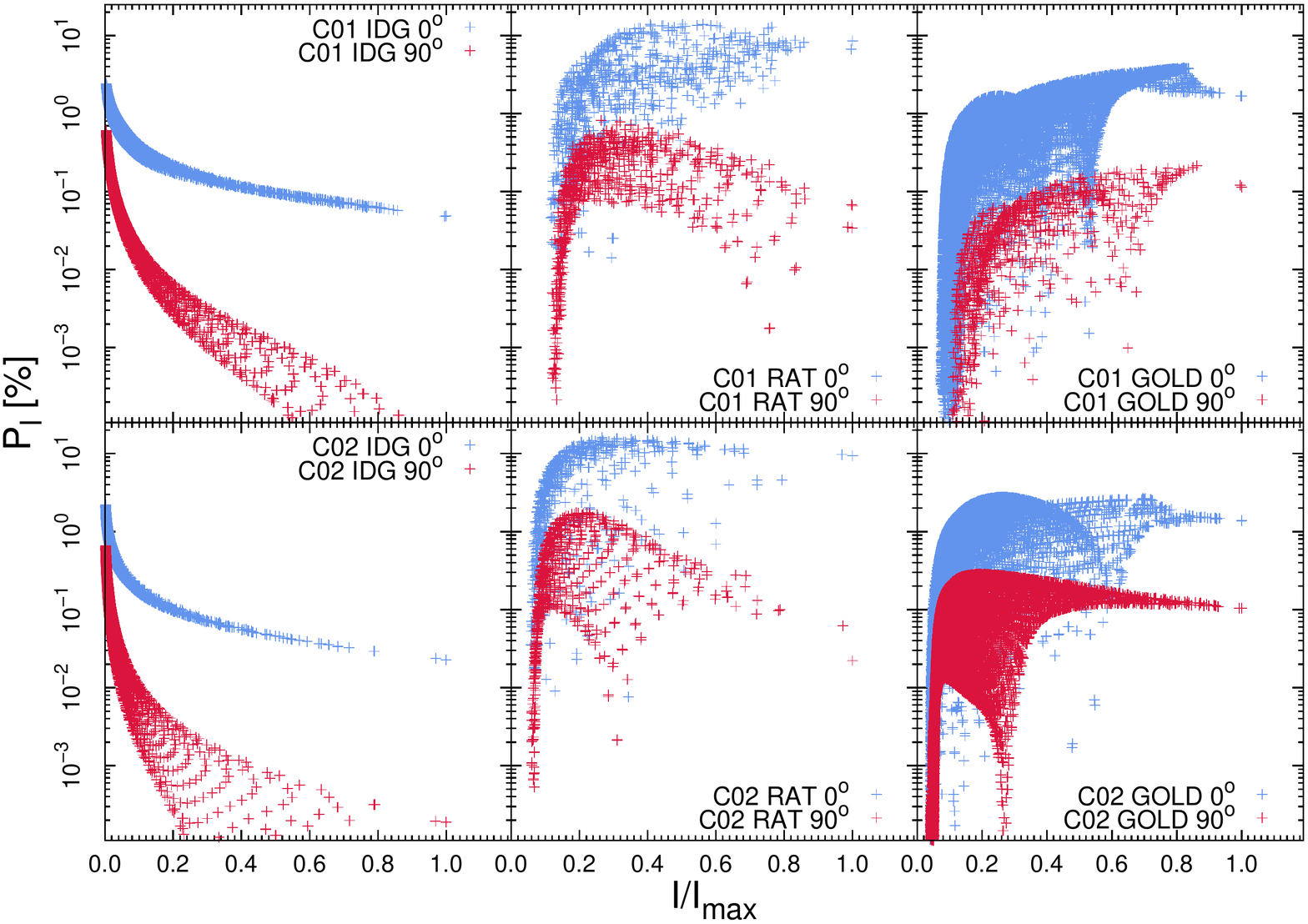}
                        \end{center}
                \end{minipage}                                  
\caption{\small Plots of the degree of linear polarization of the $C01$ model (top row) and $C02$ model (bottom row) as a function of normalized intensity (P-I relation). The results with IDG alignment (right column), RAT alignment (middle column), and GOLD alignment (left column). Both models $C01$ and $C02$ (blue dots) were rotated by $90^{\circ}$ around the y-axis of Fig. \ref{fig:globModel} (red dots).  }
\label{CloudPI}
\end{figure*}
Since we assume a constant supersonic velocity field, GOLD alignment alone barely depends on temperature and density. However, a dependency exists because of internal alignment (see Eq. \ref{eq:goldJ}) and the Mach-limit (see Eq. \ref{eq:mach}). As with RAT alignment, GOLD alignment is also limited by the local  magnetic field strength (see Eq. \ref{eq:larm}). Since the internal alignment mechanism is sensitive to high temperatures, this manifests itself as a sphere of completely randomized dust grains in the surrounding regions $B$ (Fig.\ref{fig:Cross} top right panel)  and $A$ (Fig.\ref{fig:Cross} bottom right panel) around the star. By comparing Fig.\ref{fig:Cross} (left panels), Fig. \ref{CloudPolMap} (left panels), and Eq. \ref{eq:gold}, we can see that the dominant parameter maintains the angle $\alpha$ between gas stream $\vec{v}$ and magnetic field direction $\vec{B}$. The areas of minimal grain alignment appear along the transition where GOLD alignment changes the sign of the Rayleigh reduction factor, which is shown by red lines in Fig.\ref{fig:Cross} (middle panel).\\
The resulting maps of linear polarization for the cloud models $\rm{C01}$ and $\rm{C02}$ are presented in Fig. \ref{CloudPolMap}. The normalized orientation vectors of linear polarization have an offset angle of $90^{\circ}$ to be parallel with the projected magnetic field direction. For the IDG mechanism (Fig. \ref{CloudPolMap} top row), the center of depolarization moves with the position of the star because of the change in the dust temperature distribution, leading to a nearly rotationally symmetrical pattern in both cases.\\
The polarization pattern for RAT alignment in the middle column of Fig. \ref{fig:Cross} follows the regions of highest radiation. Therefore, a high polarization coincides with the position of the star. The degree of linear polarization is reduced in the $\rm{C01}$ model along the regions of minimal grain alignment efficiency $Q_{\rm{\Gamma}}(\epsilon)$. The same effect can be detected in the polarization map of the $\rm{C02}$ model. However, here this effect is less evident leading to a butterfly-like pattern of linear polarization in the surrounding area of the star. Both, IDG and RAT alignment mechanisms resemble the underlying hourglass field morphology. The GOLD alignment in turn, can change the alignment angle along the line of sight and this, subsequently, changes the polarization direction.\\
A correlation between intensity $I$ and polarization $P_{\rm{l}}$ was recognized in \cite{2001ApJ...561..871H} following the power law $P \propto  \left(I / I_{\rm{max}} \right)^{-\alpha}$ where $\alpha$ is a fit parameter.
This finding is consistent in the literature of observations \cite[see][]{2002ApJ...574..822M,2003ApJ...598..392L,2005ApJ...631..361C}. Since the $P- I$ relation is hyperbolic in nature, we expect to find a flat tail when $I / I_{\rm{max}}$ reaches unity, making it an ideal test for  different grain alignment theories. Therefore, we calculated the increasing depolarization towards lower intensities for our cloud models $\rm{C01}$ and $\rm{C02,}$ making use of the polarization maps of Fig. \ref{CloudPolMap}. Here, each data point presents a different line of sight. We also calculated the $P - I$ relations for our cloud models rotated by $90^{\circ}$. The $P - I$ relation of the IDG mechanism shown in Fig. \ref{CloudPI} (left column) is consistent with observations. Since the anisotropy in radiation and grain alignment is directly correlated and stellar radiation is the dominant source of radiation in both cloud models, the $P - I$ relation in the RAT case is less definitive. In contrast to the IDG alignment, the $P - I$ relation of RAT alignment starts with an increase in polarization up to a point of $I/I_{\rm{max}} \approx 0.1$ and then remains  almost constant at its maximum value. However, the RAT alignment mechanism matches the expected $P - I$ relation curvature in agreement with observations for larger fractions of $I/I_{\rm{max}}$.\\
The GOLD alignment shows an exceptional behavior. In contrast to observational findings, the degree of linear polarization remains constant in the non-rotated case and increases even slightly for the rotated case. Data points show a broader scatter, compared with other alignment mechanisms.\\
The characteristic appearance of each grain alignment theory in the maps of linear polarization  is often discussed to account for observational data \citep[see][for review]{2015ASSL..407...59A}. So far, RAT alignment is considered among the most promising theories of grain alignment. The observational fact that grains are better aligned in close proximity to a bright star is a strong indication of RAT alignment \citep[][]{2011PASJ...63L..43M}. A direct correlation between grain alignment and  direction, as well  as the strength of the radiation field, was also confirmed in \cite{2010ApJ...720.1045A} and \cite{2011A&A...534A..19A} next to the star HD 97300 in the Chamaeleon I star-forming region. These observational findings are in accordance with the results of the synthetic linear polarization maps that were calculated with RAT alignment, as presented in this section.
\subsubsection{Discussion}
The velocity-dependent flip in the orientation of polarization vectors is characteristic of GOLD alignment and may help to determine areas with supersonic gas streams by observations. From the observational point of view, a flip in polarization angles was reported along the direction of molecular outflows \citep[e.g.][]{1998ApJ...502L..75R,2006ApJ...650..246C}. However, the rather unusual pattern of linear polarization, as presented in this section, can so far not be confirmed by observations. This may be a result of the limited set of parameters of the  cloud models that were considered. The two main limiting factors are velocity and magnetic field strength (see Eq. \ref{eq:gold} and Eq. \ref{eq:larm}), which were assumed to be constant for simplicity. Recent progress on the field of mechanical grain alignment also indicates that  dust-grain alignment may also be efficient in subsonic environments \citep[][]{2007ApJ...669L..77L}.\\
The IDG alignment has still its place in accounting for the observed alignment of small dust grains \cite{1992ApJ...385L..53C,2014ApJ...790....6H}. The maximum values in linear polarization are in good agreement with observations \citep{2005AA...430..979G,2011ApJ...732...97D} for all the applied alignment mechanisms. Taking the maximal degree of linear polarization as a measurement of the prevailing alignment mechanism, the polarization is  clearly dominated by IDG in the outer parts followed by RAT near the star.  Higher linear polarization degrees can easily be achieved for RATs by increasing the free parameter $f_{\rm{high-J}}$ and the upper cut-off $a_{\rm{max}}$ in the grain size distribution of the applied dust models. However, exact values of these parameter are currently speculative. In contrast to the IDG and RAT alignment, models considering GOLD alignment hold the lowest degree of linear polarization, which indicates that the observable net polarization of the cloud would lack any flip in polarization angles. This is also consistent with the behavior of the calculated $P-I$ relations. Both RAT and GOLD alignment show an unexpected distribution of linear polarization. This uncertainty and the deviations in calculated $P-I$ relations may be resolved with more sophisticated models from dedicated MHD simulations in forthcoming parameter studies.

\section{Summary and outlook}
\label{sect:sum}
We have presented POLARIS (\textbf{POLA}rized \textbf{R}ad\textbf{I}ation \textbf{S}imulator), a newly developed 3D MC-RT radiative transfer code that was designed to perform self-consistent dust heating calculations and polarization simulations. The main novelty of the POLARIS code is the calculation of linear and circular polarization maps considering dichroic extinction, thermal re-emission, birefringence, and scattering on dust grains that are partially aligned with the direction of the magnetic field. In contrast to previous MC codes, POLARIS  calculates the impact of imperfect grain alignment on light polarization by considering the major classes of state-of-the-art dust grain alignment theories (alignment owing to paramagnetic relaxation (IDG), alignment because of radiation-dust interaction (RAT), and mechanical alignment (GOLD) as a result of gas streams).\\
To calculate the dust temperature distribution, we use the combined techniques of continuous absorption and immediate re-emission. The code is openMP parallelized and RT simulations are performed in an adaptive octree cartesian grid, which allows for high performance and the consideration of MHD simulations as input data. Further advanced optimization techniques (ray-tracing, wavelength range selection, forced first scattering, peel-off technique, and modified random walk) enable us to perform complex 3D MC-RT simulations, including polarization on dust grains in a reasonable time. The code considers various sources of radiation to cover a broad variety of astrophysical problems.\\
The predictive capability of the POLARIS code was tested to run in specific complex environments resulting from MHD  simulations, as well as in case studies with analytical density, temperature, velocity, and magnetic field distributions. Here, we applied a method for combining the dust temperatures. We tested the dust heating algorithm of POLARIS in a complex MHD collapse simulation. The resulting distribution of combined temperatures matches simultaneously the compressed dust distribution and the positions of the considered sources of radiation.\\
The capability to create polarization maps according to the RAT alignment mechanism was tested in circumstellar disk models that were consistent with theoretical predictions. Here, we modeled the disk with an analytical function and applied a toroidal magnetic field geometry. The disk was heated by a central star. Stellar radiation and thermal dust re-emission lead to an anisotropy in the radiation field. This resulted in characteristic layers of dust grain sizes in the disk with larger sizes towards the mid-plane. This is in agreement with RAT alignment theory and verified the predictive capability of of the POLARIS code.\\
Linear polarization greatly depends on the dominant alignment mechanism. Each mechanism has an effect, in a unique way, on the local physical parameter leading to a characteristic pattern of linear polarization. We tested this scenario in an analytical model of two Bonnor- Ebert spheres with their centers close to each other, a constant velocity field, and an hourglass-shaped magnetic field morphology. We heated the dust with a star as the radiation source to calculate the anisotropy in the radiation field for RAT alignment. Finally, we derived synthetic maps of linear polarization, which considered the IDG, GOLD, and RAT grain alignment mechanism. The resulting pattern of linear polarization resembled the physical parameter in accordance with applied grain alignment theories and showed the expected PI-relations known from observations.\\
In forthcoming papers, we will present the code extensions for state-of-the-art  radiative line transfer, including Zeeman effect. Furthermore, we will apply POLARIS to interpret observational data by calculating the characteristic polarization pattern in MHD outflow simulations. Here, we  follow questions about the potential of multi-wavelength polarization measurements to identify distinct sub-structures of a larger magnetic field morphology.\\
The code is continuously in development to keep up with scientific progress, especially in grain alignment theory. POLARIS is available only on request. However, sample files, scripts and documentation are publicly available from \url{http://www1.astrophysik.uni-kiel.de/~polaris/}.

\appendix
\section{Benchmark}
\label{apAA}
\begin{center}
\begin{figure*}[]
\begin{center}
        \begin{minipage}[c]{0.15\linewidth}
                        \begin{center}
                                \includegraphics[width=0.9\textwidth]{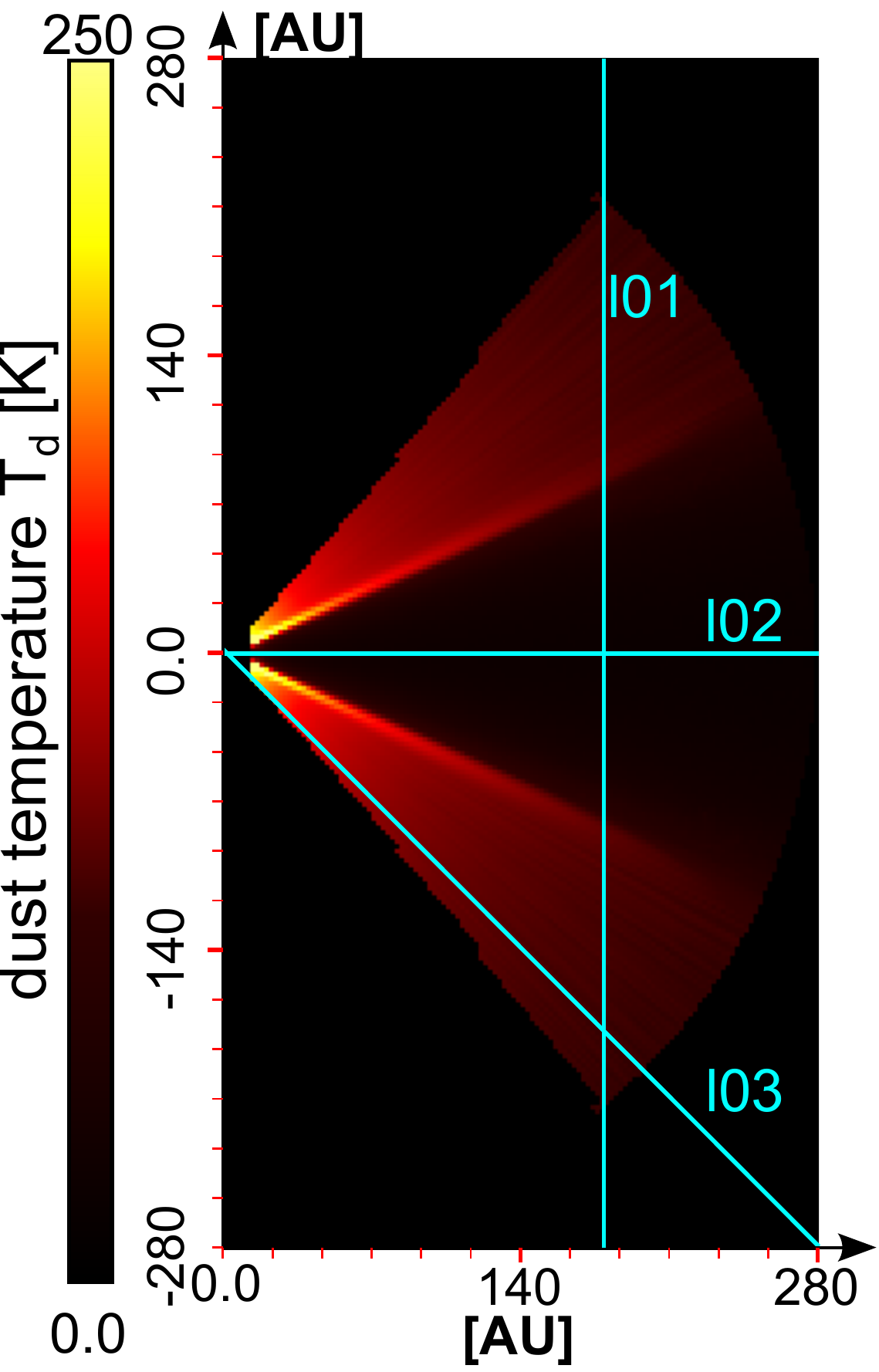}
                        \end{center}
                \end{minipage}
        \begin{minipage}[c]{0.27\linewidth}
                        \begin{center}
                                \includegraphics[width=1.0\textwidth]{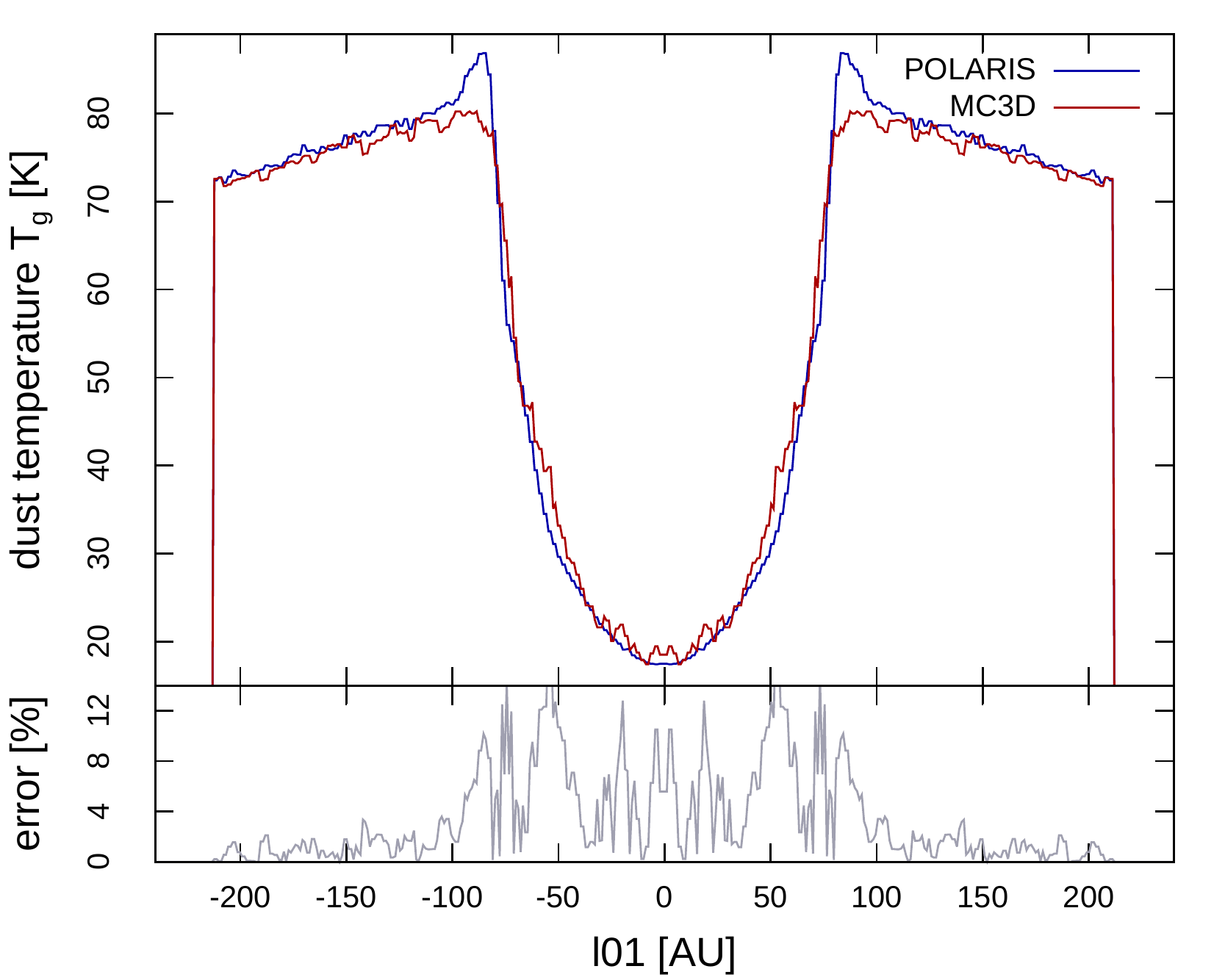}
                        \end{center}
                \end{minipage}
                                \begin{minipage}[c]{0.27\linewidth}
                        \begin{center}
                                \includegraphics[width=1.0\textwidth]{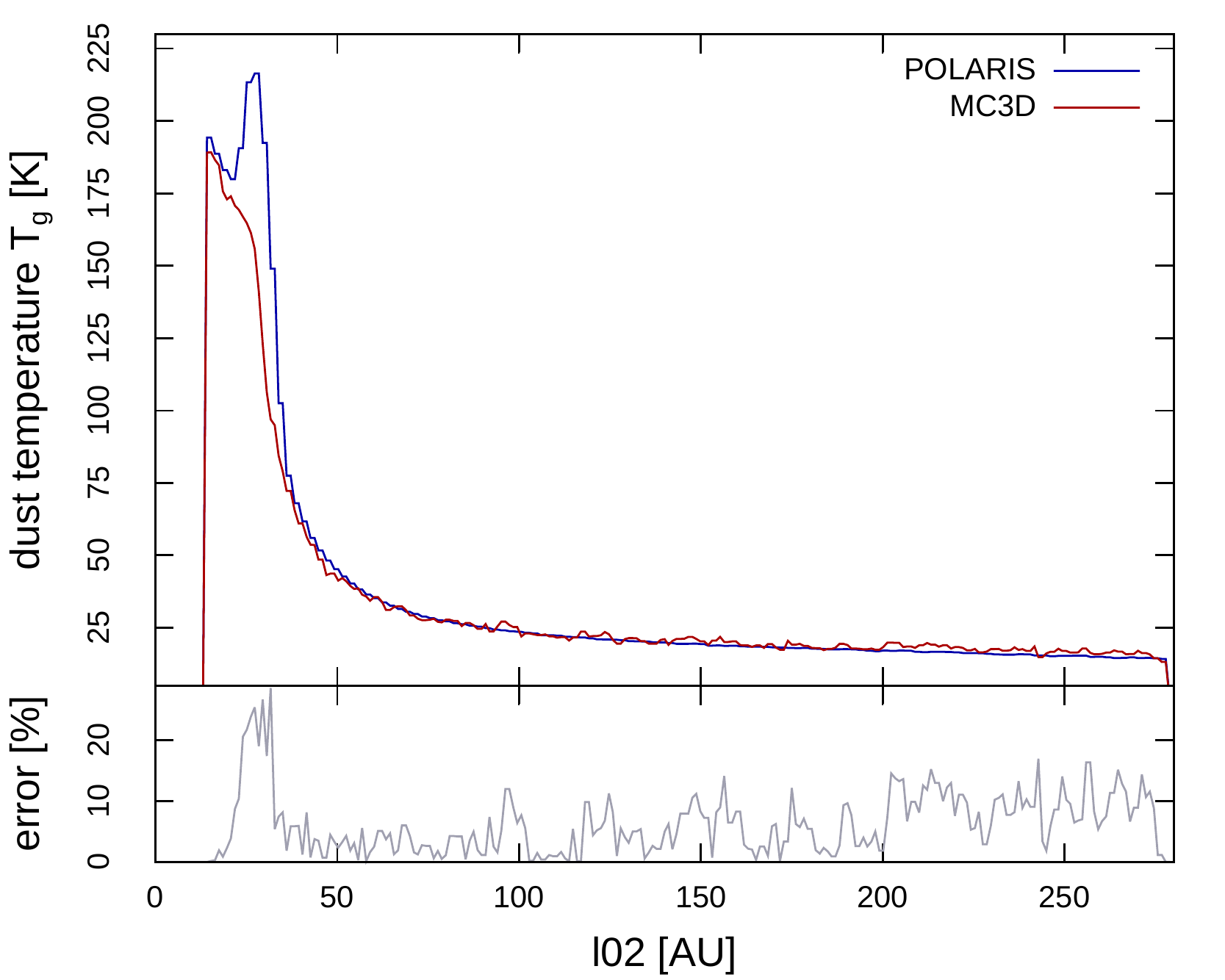}
                        \end{center}
                \end{minipage}  
                \begin{minipage}[c]{0.27\linewidth}
                        \begin{center}
                                \includegraphics[width=1.0\textwidth]{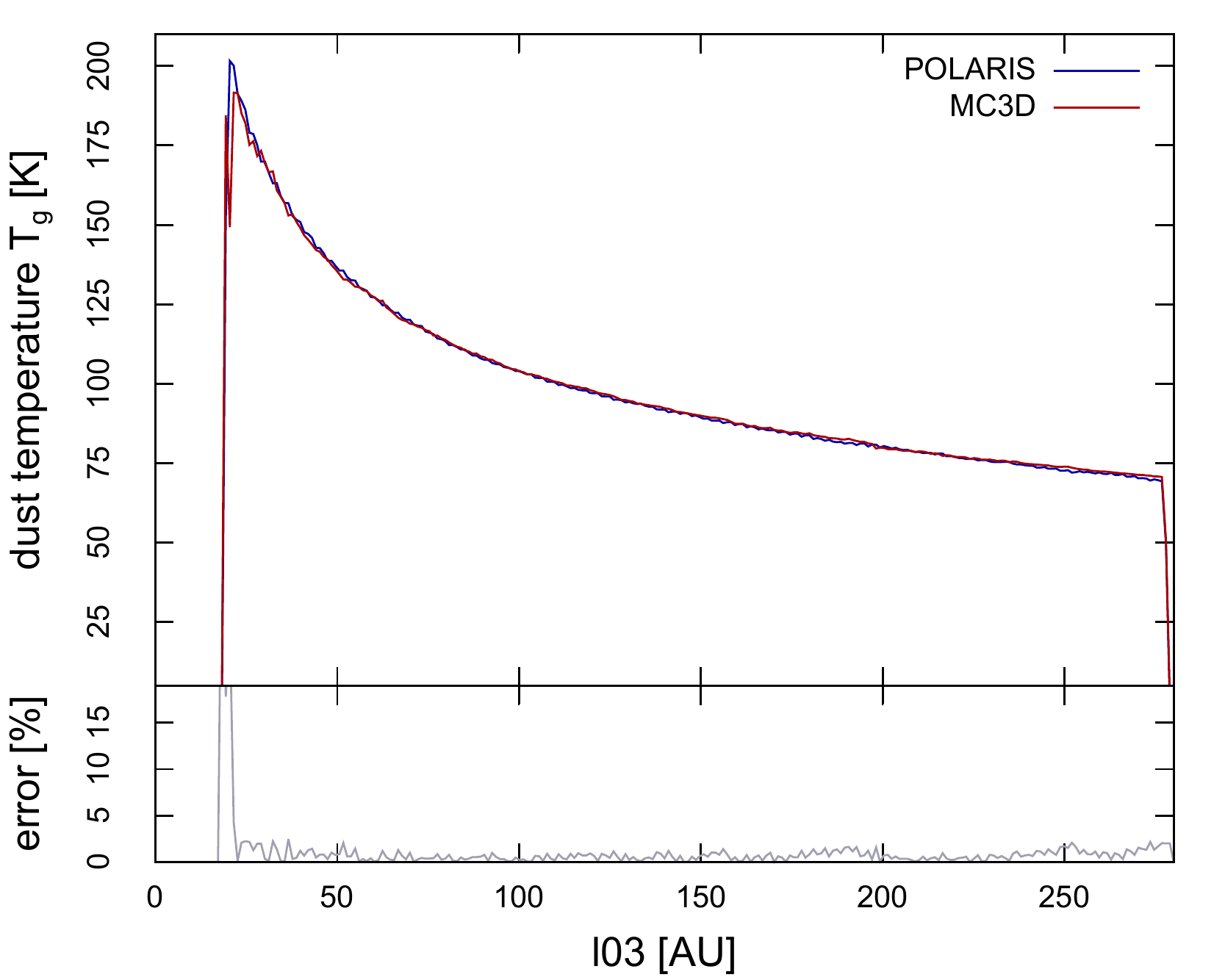}
                        \end{center}
                \end{minipage}
\end{center}
        
\caption{\small Resulting dust temperature  $T_{\rm{d}}$ for the $D03$ disk model (see Table \ref{tab:1}). The outer left panel shows the temperature distribution in a plane perpendicular to the mid-plane of the disk. 
The plots show the temperature along the edge of the disk, perpendicular to the mid-plane (left), along the mid-plane of the disk (middle), and the distribution along the disks surface (right). The red lines show the results from the MC3D code, the blue ones the results from the POLARIS code, and the gray lines show the errors between both codes.}
\label{fig:D03Temp}
\end{figure*}

\begin{figure*}[]
\begin{center}
        \begin{minipage}[c]{0.15\linewidth}
                        \begin{center}
                                \includegraphics[width=1.0\textwidth]{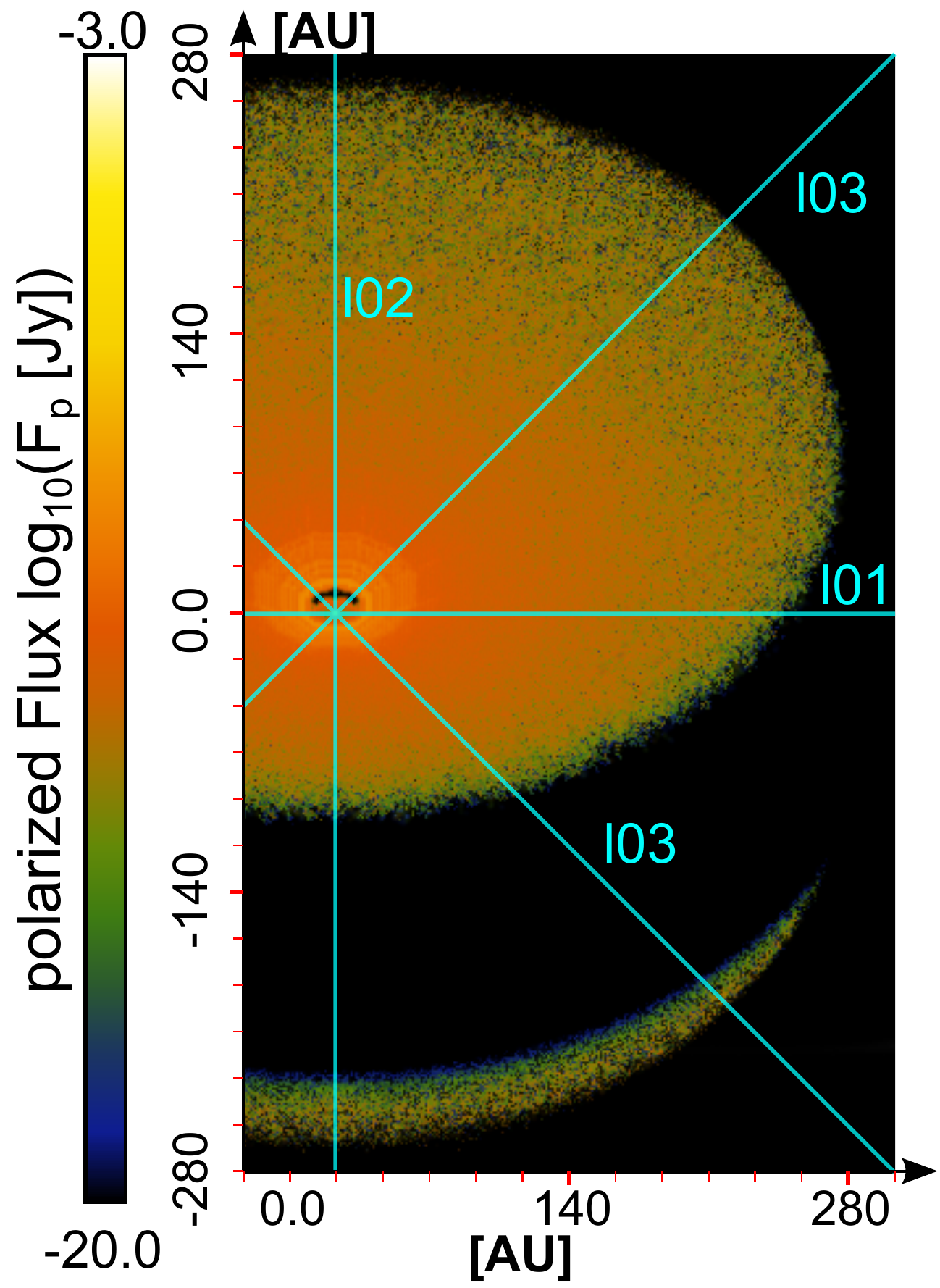}
                        \end{center}
                \end{minipage}
        \begin{minipage}[c]{0.27\linewidth}
                        \begin{center}
                                \includegraphics[width=1.0\textwidth]{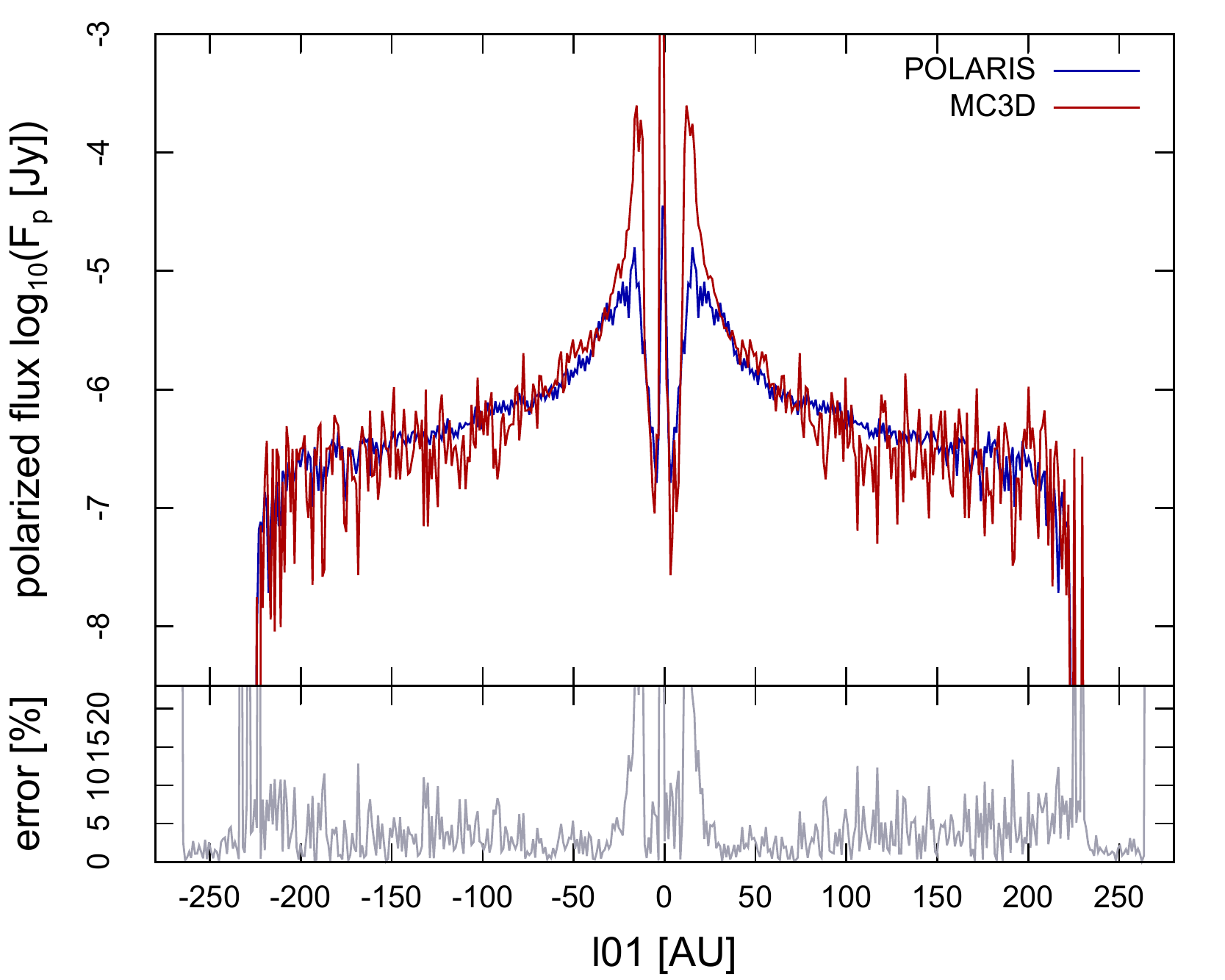}
                        \end{center}
                \end{minipage}
                                \begin{minipage}[c]{0.27\linewidth}
                        \begin{center}
                                \includegraphics[width=1.0\textwidth]{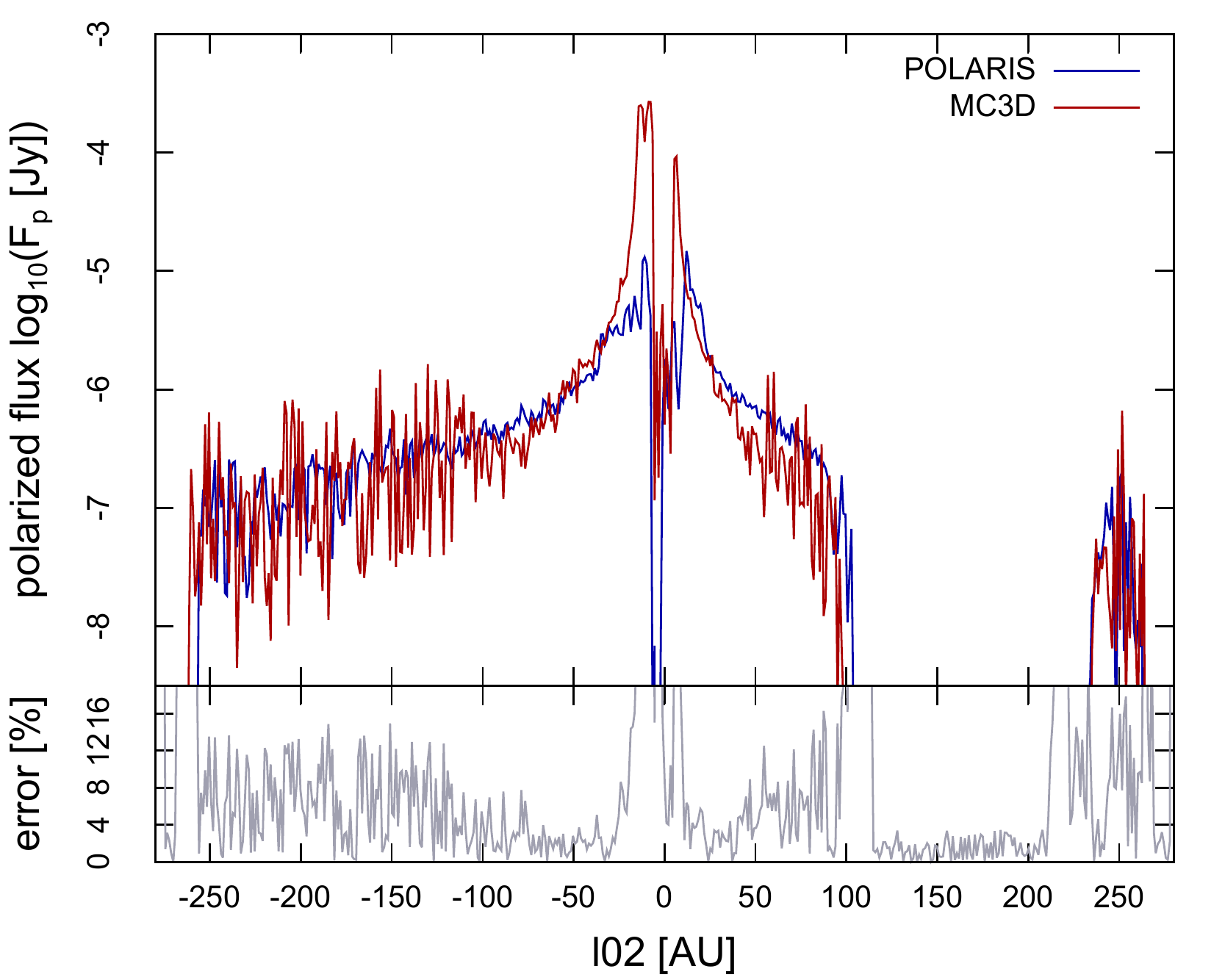}
                        \end{center}
                \end{minipage}  
                \begin{minipage}[c]{0.27\linewidth}
                        \begin{center}
                                \includegraphics[width=1.0\textwidth]{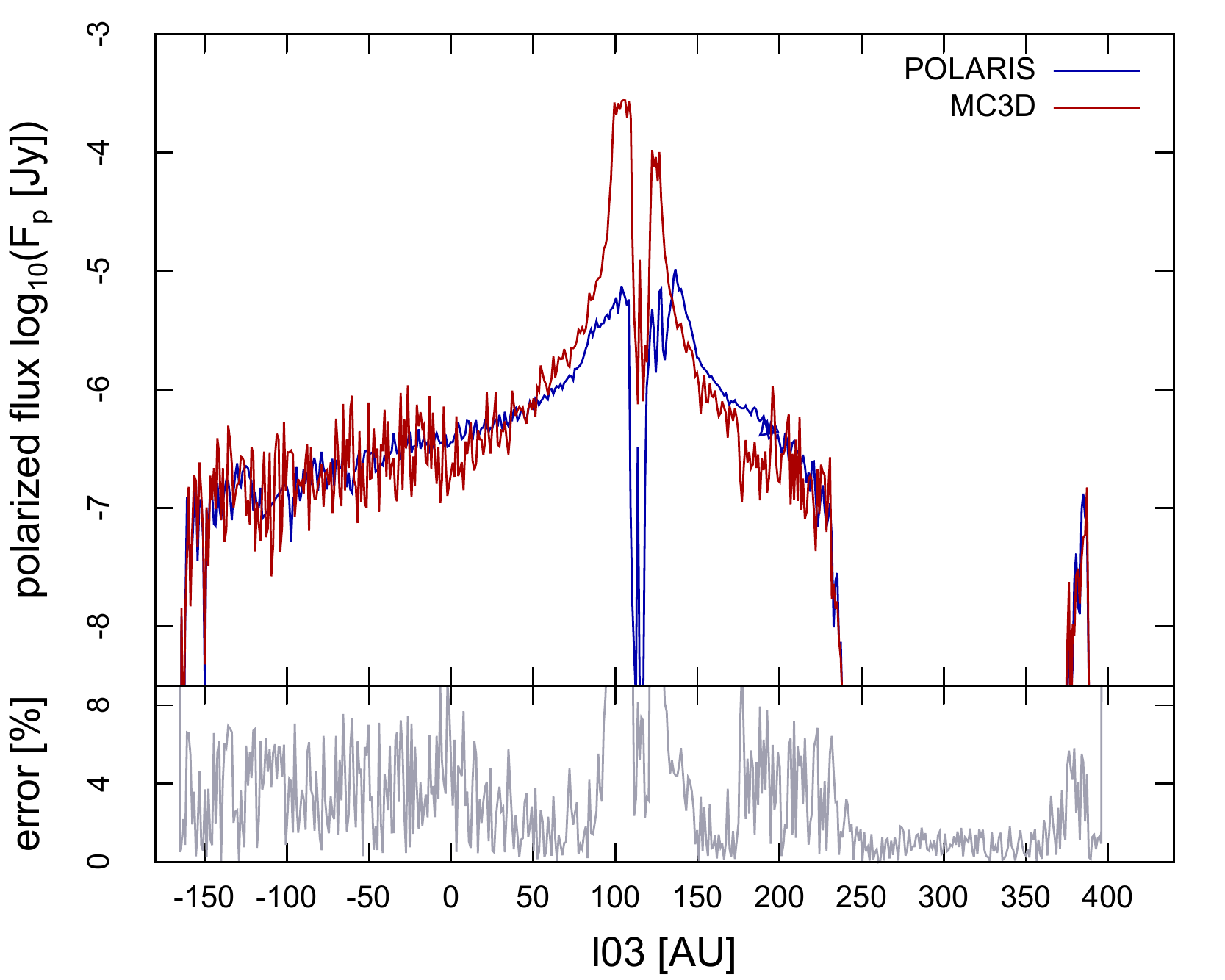}
                        \end{center}
                \end{minipage}
\end{center}
        
\caption{\small Resulting distribution of the polarized flux $F_{\rm{p}}$ for the $D03$ disk model (see Table \ref{tab:1}). The outer left panel shows the polarized flux distribution as a result of scattered stellar radiation under an inclination angle of $i=45^{\circ}$ and a wavelength of $\lambda = 730\ \rm{nm}$. The plots show the polarized flux through the center of the disk along the horizontal direction (left), along the vertical direction (middle), and the bisecting line (right). The red lines show the results from the MC3D code, the blue ones the results from the POLARIS code, and the gray lines show the errors between both codes.}
\label{fig:D03Int}
\end{figure*}

\begin{figure*}[]
\begin{center}
        \begin{minipage}[c]{0.15\linewidth}
                        \begin{center}
                                \includegraphics[width=0.9\textwidth]{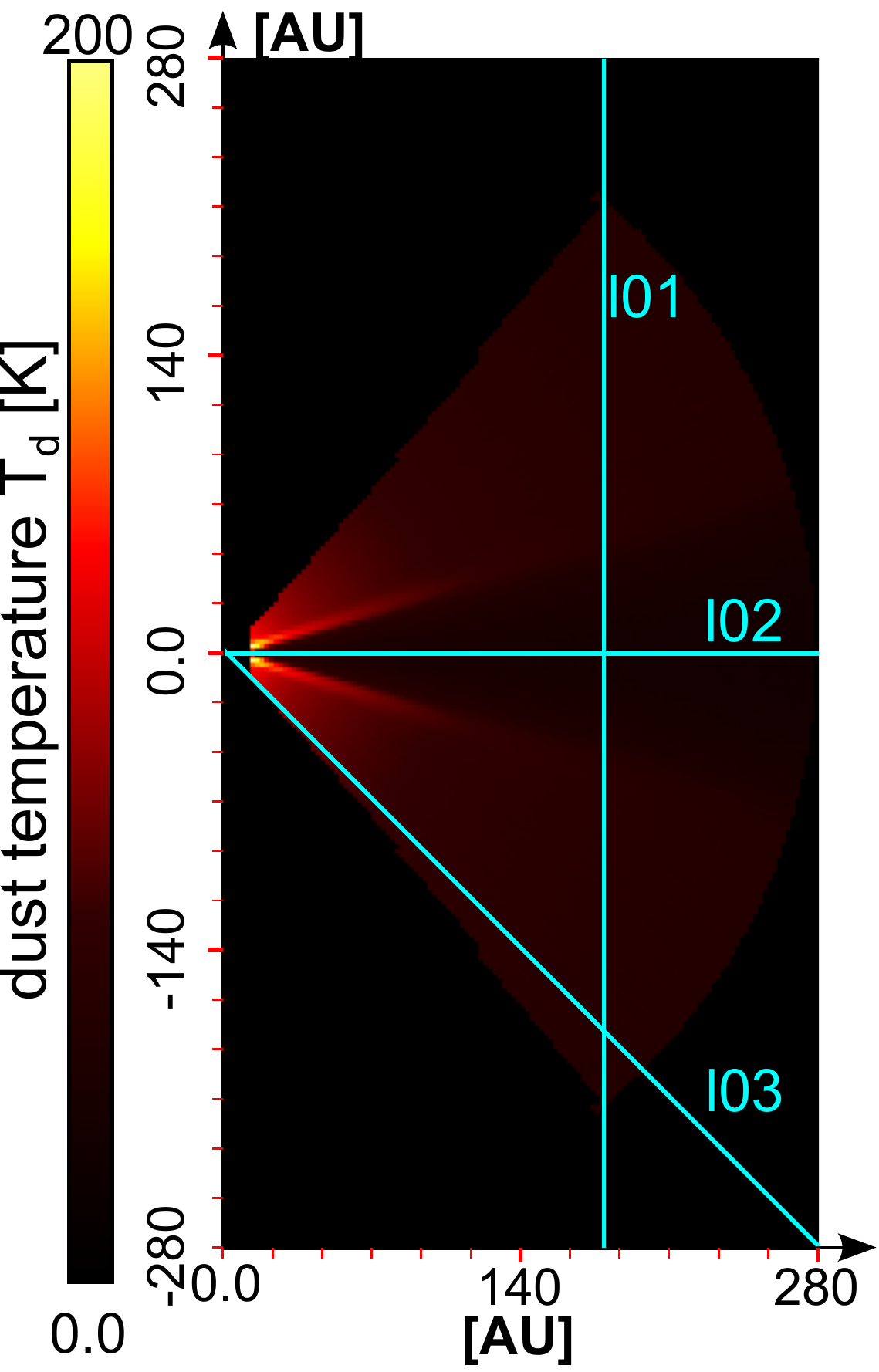}
                        \end{center}
                \end{minipage}
        \begin{minipage}[c]{0.27\linewidth}
                        \begin{center}
                                \includegraphics[width=1.0\textwidth]{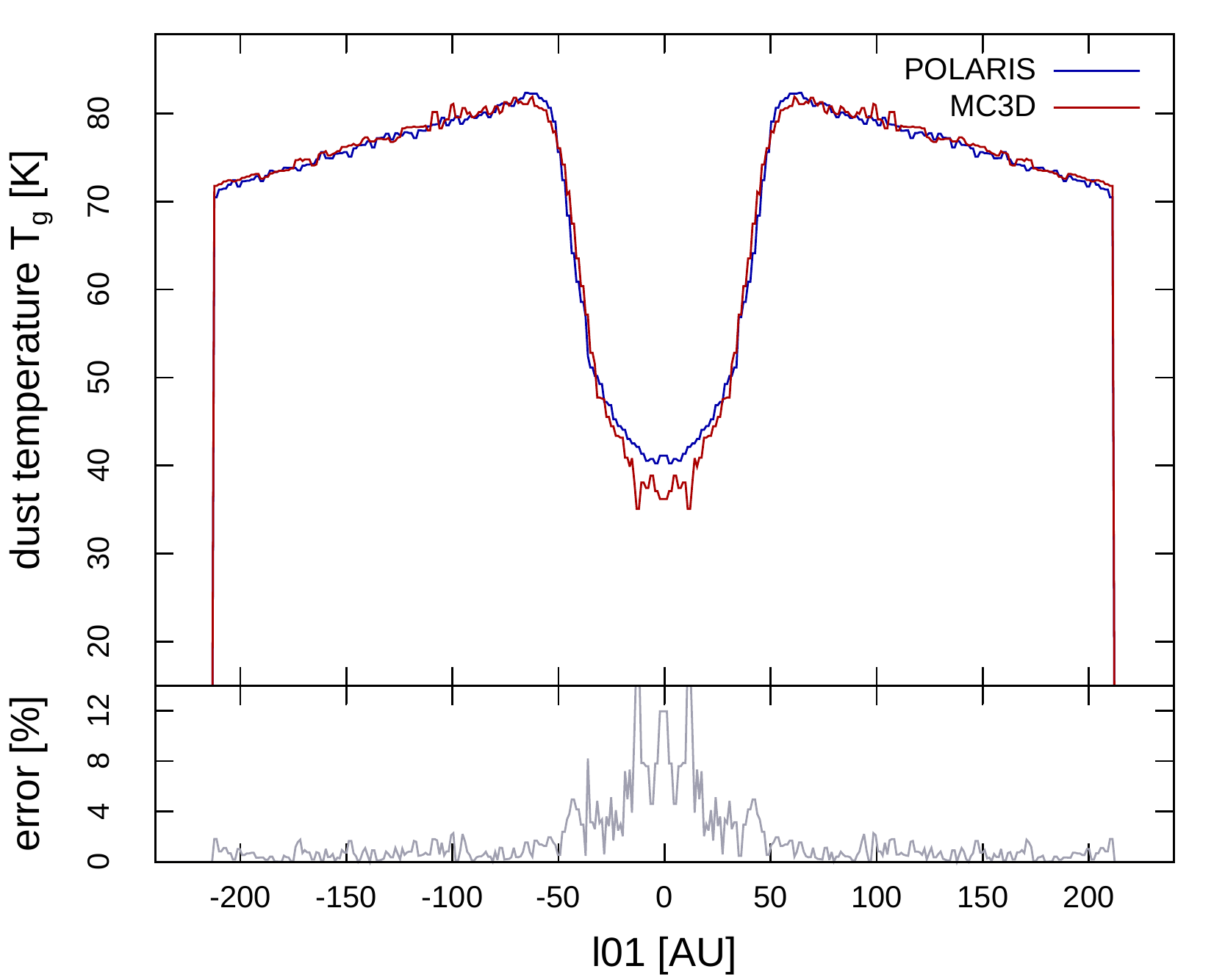}
                        \end{center}
                \end{minipage}
                                \begin{minipage}[c]{0.27\linewidth}
                        \begin{center}
                                \includegraphics[width=1.0\textwidth]{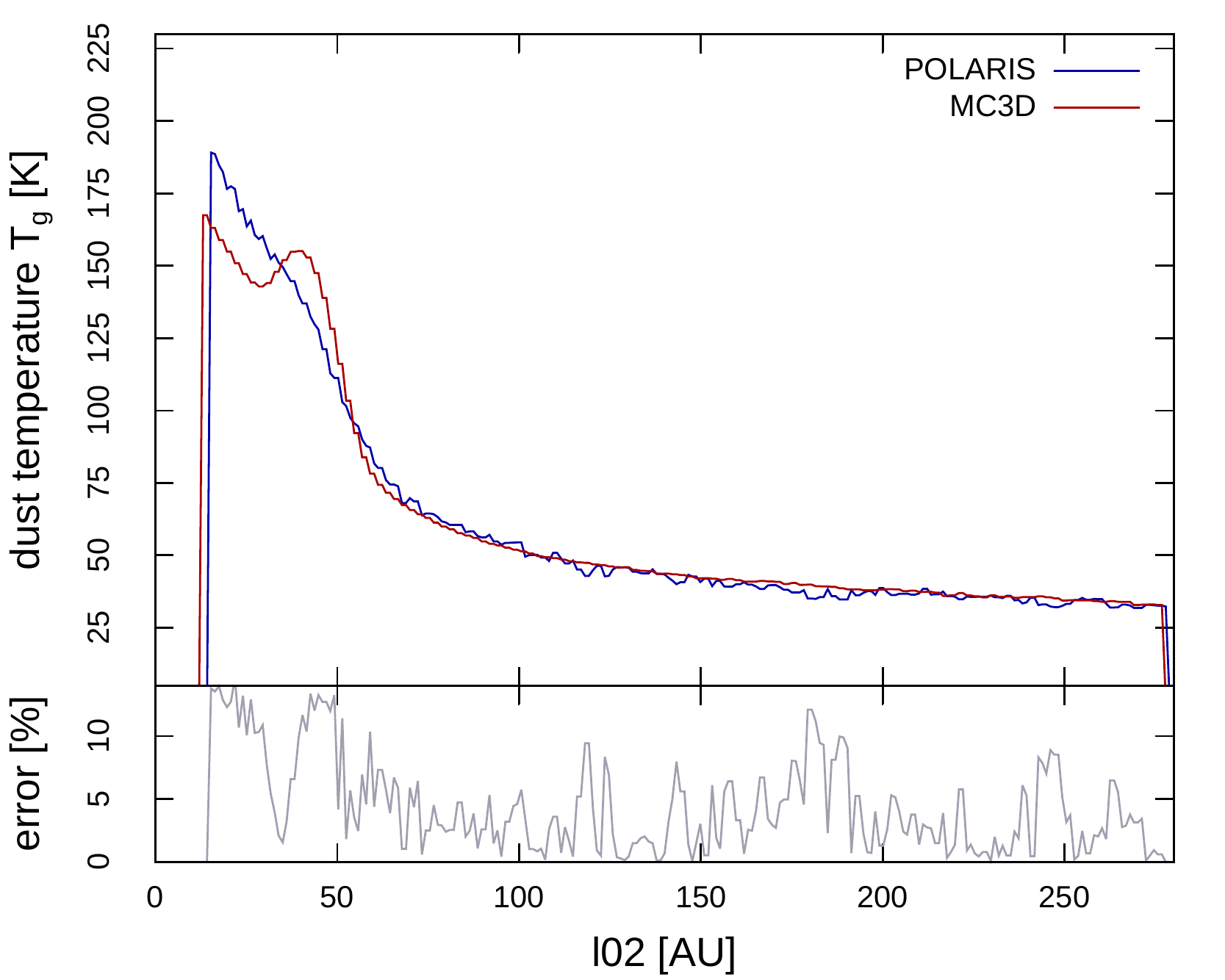}
                        \end{center}
                \end{minipage}  
                \begin{minipage}[c]{0.27\linewidth}
                        \begin{center}
                                \includegraphics[width=1.0\textwidth]{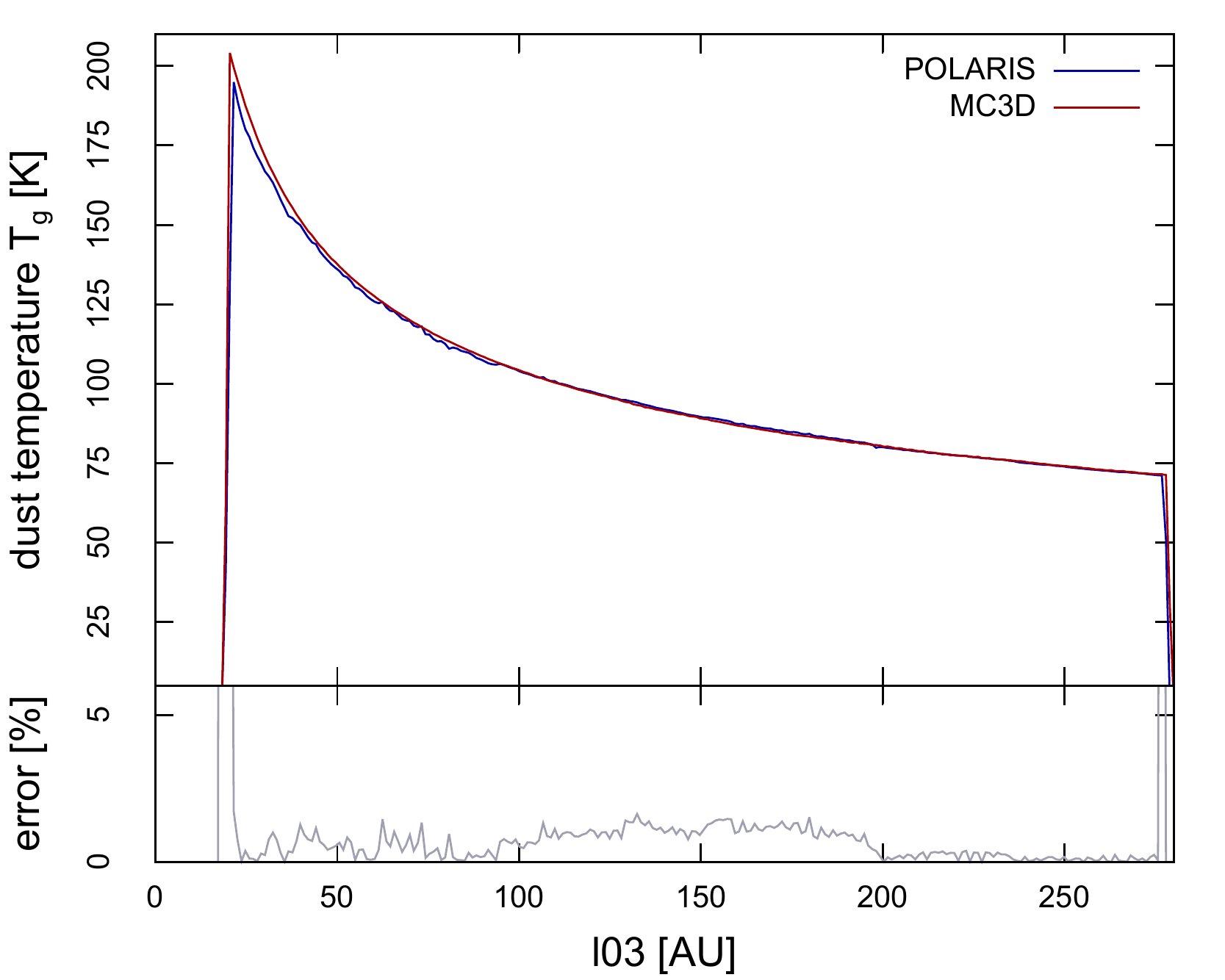}
                        \end{center}
                \end{minipage}
\end{center}
        
\caption{\small The same as Fig. \ref{fig:D03Temp} for the $D06$ disk model.}
\label{fig:D06Temp}
\end{figure*}

\begin{figure*}[]
\begin{center}
        \begin{minipage}[c]{0.15\linewidth}
                        \begin{center}
                                \includegraphics[width=1.0\textwidth]{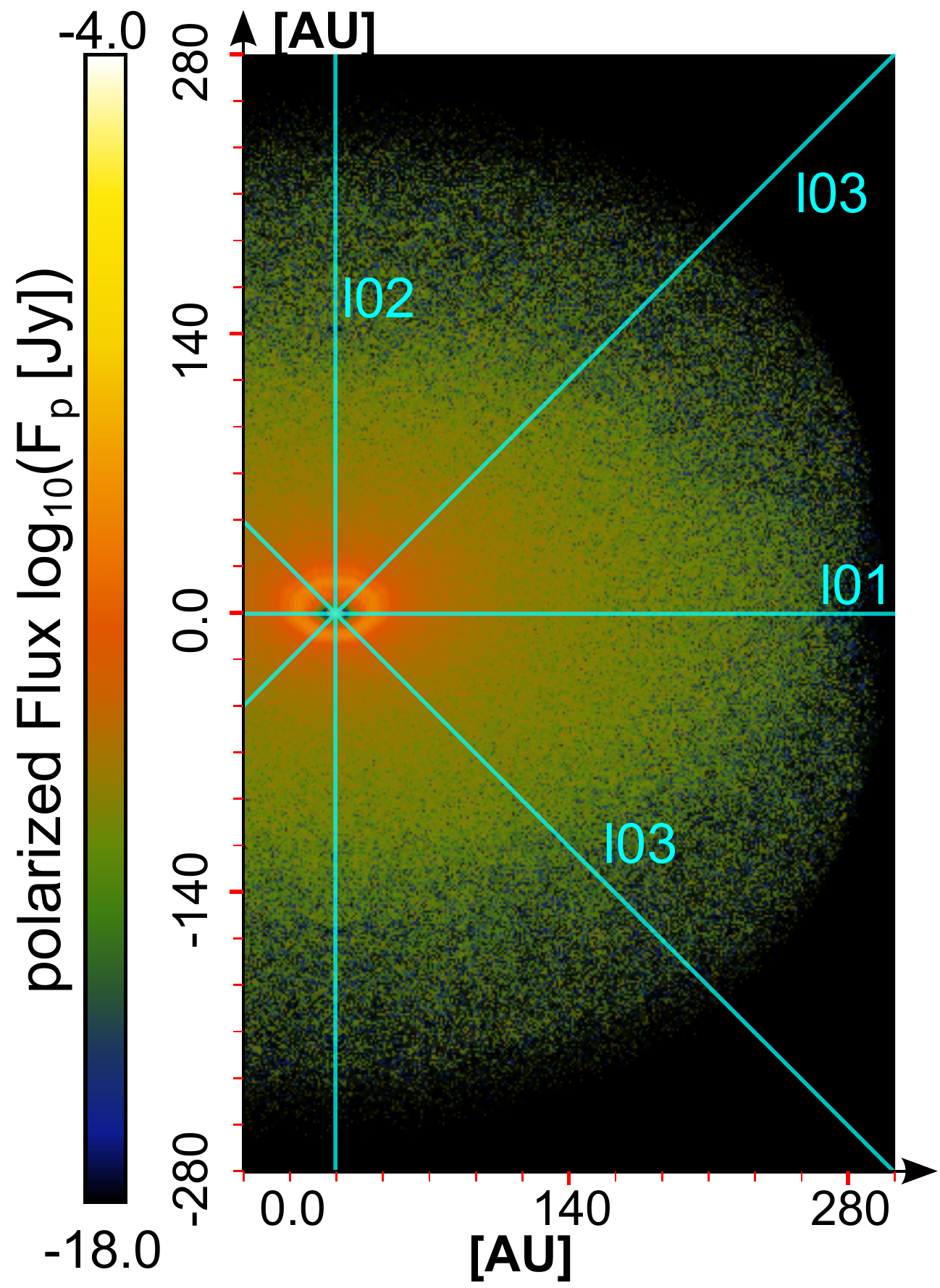}
                        \end{center}
                \end{minipage}
        \begin{minipage}[c]{0.27\linewidth}
                        \begin{center}
                                \includegraphics[width=1.0\textwidth]{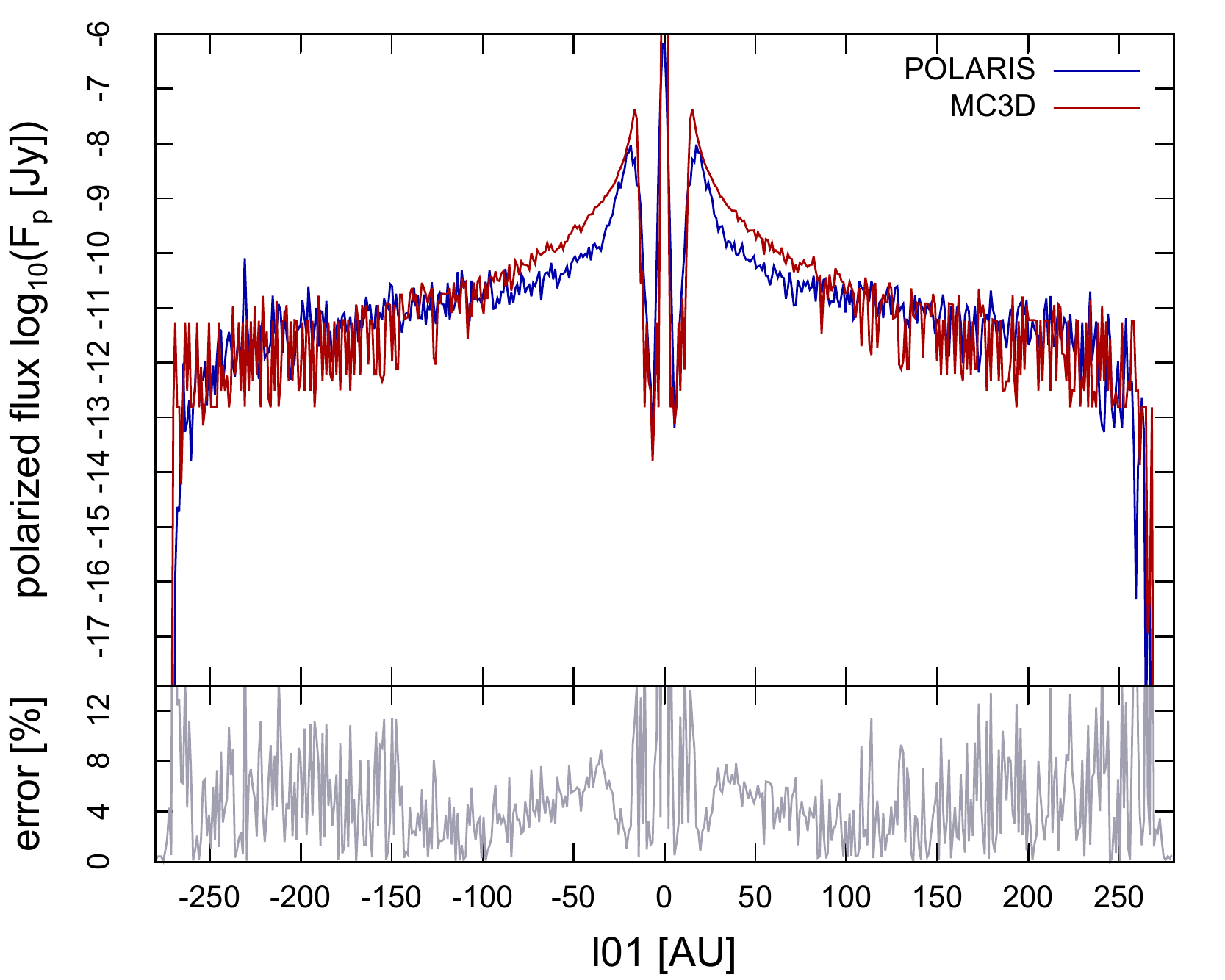}
                        \end{center}
                \end{minipage}
                                \begin{minipage}[c]{0.27\linewidth}
                        \begin{center}
                                \includegraphics[width=1.0\textwidth]{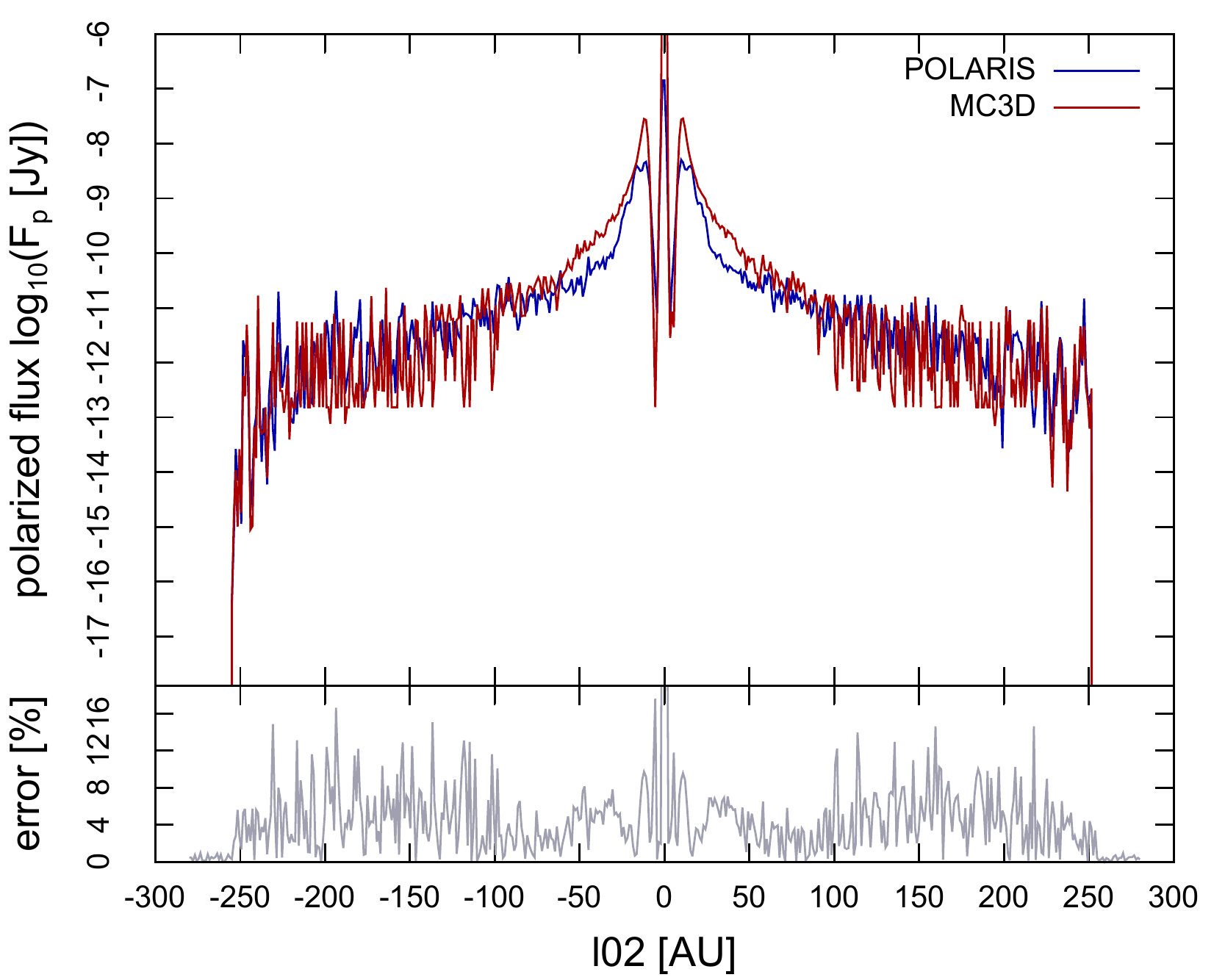}
                        \end{center}
                \end{minipage}  
                \begin{minipage}[c]{0.27\linewidth}
                        \begin{center}
                                \includegraphics[width=1.0\textwidth]{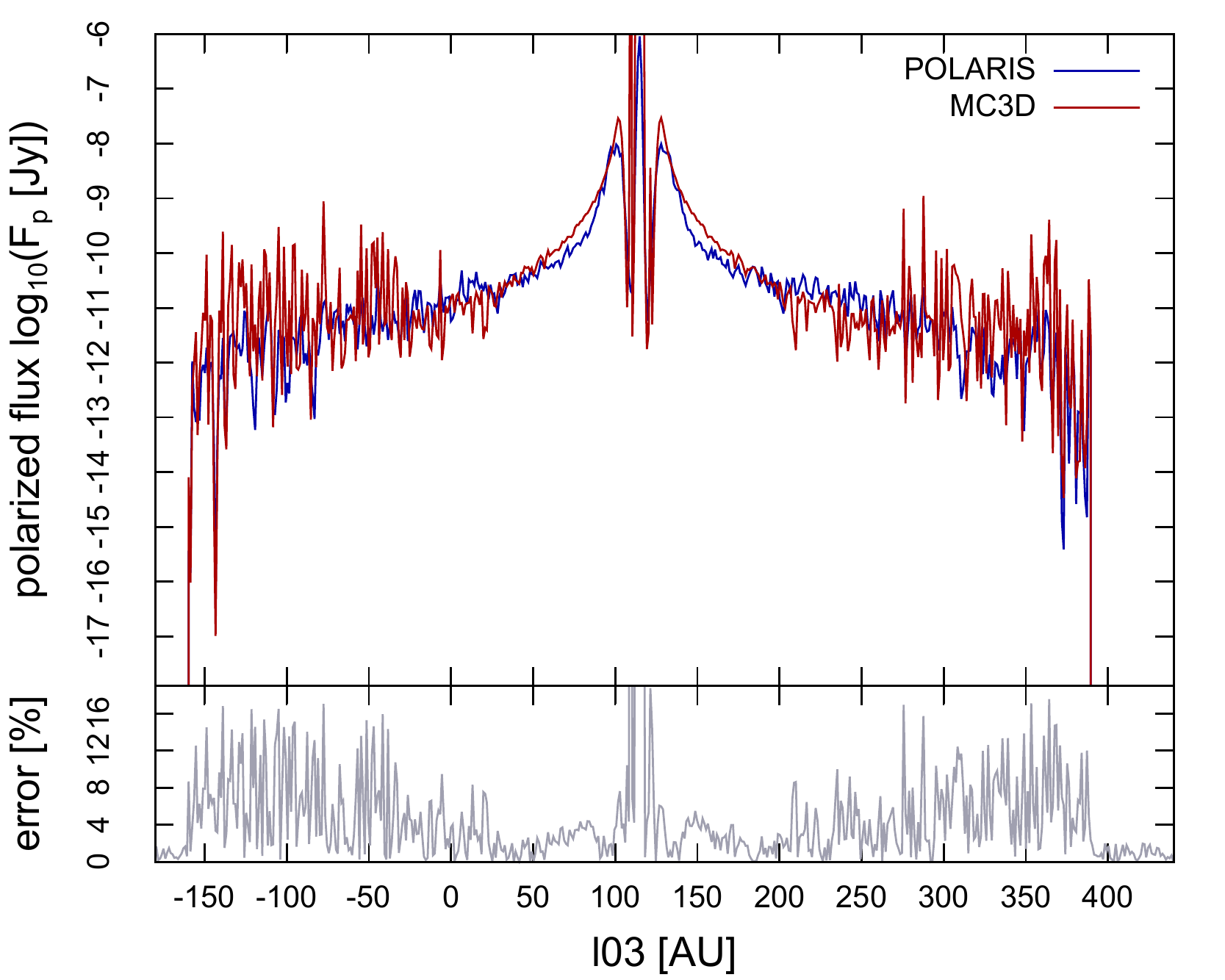}
                        \end{center}
                \end{minipage}
\end{center}
        
\caption{\small  The same as Fig. \ref{fig:D03Int} for the $D06$ disk model.}
\label{fig:D06Int}
\end{figure*}
\end{center}
To make predictions about the impact of imperfectly aligned dust grains on polarization measurements, the POLARIS code goes beyond previous approaches in this field. Its treatment concerning polarization as a result of dichroic extinction and thermal re-emission, combined with state-of-the-art dust grain alignment mechanisms is unique in that way. Therefore, essential features cannot be tested because of the absence of standardized benchmark tests. However, dust heating and polarization calculations of scattered light are implemented in many codes, but the POLARIS code is limited, so far, by its implemented octree grid geometry, making it inapplicable to benchmarks such as \cite{2009A&A...498..967P}. Although, POLARIS is a completely new line of development, we used the established and well-tested MC code MC3D \citep[see][]{2003CoPhC.150...99W} as a reference for light scattering and dust heating. Here, we confirm the accuracy of POLARIS by comparing the results of the disk models $D03$ - $D06,$ presented in Sect. \ref{setupDisk}, with the results of MC3D. To provide identical test cases for both codes, we converted the standard MRN MC3D dust catalog with spherical dust grains in a POLARIS database file format. We processed the MC3D output of the density distribution to create an octree grid that is required by POLARIS. Since, the dust temperature in low density areas amounts to unrealistically high values, and MC3D does not control  the sublimation temperature of the applied dust materials, we considered cells with a density of $n_{\rm{d}} < 10^{-20} \rm{m^{-3}}$ to be empty. For the MC calculations of the dust temperature distribution $T_{\rm{d}}$ and the polarized flux $F_{\rm{p}}=P\times F_{\rm{\lambda}}$, because of scattering, we used the parameter of Table \ref{tab:1}. The deviation between the results of both codes was quantified with an error of $e=100 |x_1-x_2|/\max(|x_1|,|x_2|)$, where $x$ stands for dust temperature and polarized flux, respectively.\\
The dust temperature distribution of the disk model $C03$ resulting from the MC3D code and POLARIS code is shown in Fig. \ref{fig:D03Temp}. In the outer regions, the temperature distributions match well, the difference in temperature is much higher at the inner edge of the disk. This difference is a result of the available grid geometries implemented in both codes. While MC3D performed the RT calculations on a spherical grid with smaller cells towards the center, POLARIS uses a Cartesian octree grid with a constant minimal cell size. The geometrical difference in both coordinate systems also results  in a different shaped inner border of the disk. Because of this, POLARIS  overestimates the temperature in the inner regions. For the same reason the dust temperature differs in the $C06$ disk model shown in Fig. \ref{fig:D06Temp}.\\
Finally, we compare the resulting polarized flux in Fig. \ref{fig:D03Int} for the $C03$ model and in Fig. \ref{fig:D06Int} for the $C06$ at a distance of $140\ \rm{pc}$ to the observer. While the overall trend along the selected lines matches quite well for both codes, the region of deviation remains near the inner radii in all disk models. The higher resolution in the center region of MC3D's grid leads to a layer of higher dust density at the inner edge. Subsequently, more light is scattered towards the observer. In comparison to MC3D, the synthetic images of the polarized flux POLARIS have a higher S/N because of the implemented optimizations of the forced first scattering algorithm and the peel-of technique (see Sect. \ref{optim}). However, in all test cases, the results of POLARIS fit  those of MC3D within the inherent limitations of applied grid geometries.

\section{Internal alignment equation}
\label{apA}
Numerical integration is one of the challenges for performing MC RT simulations in reasonable time. Here, we present the approximation we applied for the calculation of internal alignment. While the analytical solution for IDG, Gold, and RAT alignment are available (see Sect. \ref{sect:dust}), the internal alignment still requires  numerical integration. However, the distribution function can be rewritten as
\begin{equation}
        f(\zeta) \approx \exp \left(- \alpha \left[1+\delta\sin^2(\zeta)\right]\right)
,\end{equation}
with  $\alpha =J^2/2I_{\rm{||}}k_{\rm{B}} T_{\rm{d}}$ and $\delta = h -1$. The degree of alignment is determined by 
\begin{equation}
        \left\langle G(\zeta) \right\rangle = \frac{\int_{\rm{0}}^\pi{G(\zeta)\sin(\zeta)f(\zeta)d\zeta}}{\int_{\rm{0}}^\pi{\sin(\zeta)f(\zeta)d\zeta}}
\end{equation}
and also has an exact solution. This enables us to calculate the internal alignment with $x=\alpha\times\delta$

\begin{equation}
        \left\langle G\left(\cos^2(\zeta)\right)) \right\rangle(x) = \frac{e^x}{\sqrt{\pi x}  \times \text{erfi}\left(\sqrt{x}\right)}-\frac{1}{2 x}
        .\end{equation}
Here, $\text{erfi}$ is the complex error function, which can easily be pre-calculated and interpolated.\\
Obviously,   $\lim_{x \to +\infty}\left\langle G\left(\cos^2(\zeta)\right) \right\rangle(x)=1$ when internal alignment is taken into account. This is consistent with physics because in the supra-thermal regime ($J^2 >> J^2_{\rm{th}} \approx 2 I_{\rm{||}}k_{\rm{B}}T_{\rm{d}}$) and for disk-like dust grains ($\delta\rightarrow1$), internal alignment becomes irrelevant \citep[see][]{1996ASPC...97..425L,1997ApJ...484..230L}. 
For $x \rightarrow 0,$ the first order moment $\left\langle G\left(\cos^2(\zeta)\right) \right\rangle(x) $ remains positive and polarization vectors also cannot change their sign in IDG or RAT alignment as a result of imperfect internal alignment. 

\section{Combined Rayleigh reduction factor}
\label{apB}
As shown in this paper, several mechanism for grain alignment have been proposed over the decades. However, these mechanisms are not mutually exclusive and most are probably simultaneously at work inside the ISM. To investigate  this scenario, POLARIS enabled us to combine several grain alignment mechanisms in a single 3D MC-RT simulation. Following \cite{1995ApJ...451..660L}, we use \begin{equation}
        R\approx \frac{R_1 + R_2 + R_3 + R_1 R_2  + R_1 R_3 + R_2 R_3 + 3 R_1 R_2 R_3}{1 + 2 R_1 R_2 + 
 2 R_1 R_3 + 2 R_2 R_3 + 2 R_1 R_2 R_3}
\end{equation}
for the combined Rayleigh reduction factor. Here, we need to  emphasize that this approach remains a first approximation. It requires further research to investigate its validity and accuracy.

\begin{acknowledgements}
S.R. and S.W. thank the anonymous referee for providing useful feedback that improved this paper. We also wish to thank  colleagues Daniel Seifried for providing MHD data, Florian Kirchschlager for assistance with DDSCAT, Jan Philipp Ruge and Gesa H.-M. Bertrang for fruitful discussions about disk models and grain alignment, Florian Ober and Peter Scicluna for their contributions to radiative transfer and dust heating algorithms. We also wish to thank the students Yong Kyung OH and Jan Thiel for serving as code testers. For this project the authors S.R. and S.W. acknowledge the support of the DFG: WO 857/11-1.
\end{acknowledgements}

\bibliographystyle{aa}
\bibliography{./bibtex}
\end{document}